\documentclass[usegraphicx,useAMS,usenatbib]{mn2e}
\usepackage{graphicx}
\usepackage{ulem}
\usepackage{amssymb}
\usepackage{amsmath}
\usepackage{soul}
\usepackage{color}

\def\tilde{~}

\def\msun{\rm{\,M_{\odot}}}

\def\msunh{\rm{\,M_{\odot}{\it h}^{-1}}}

\def\kev{\rm{\,keV}}
\def\sev{\rm{\,keV}{cm^2}}

\def\kpc{\rm{\,kpc}}

\newcommand{\gr}{\kern 2pt\hbox{}^\circ{\kern -2pt K}} 

\def\Log10{{\rm~Log_{10}}}

\def\lcdm{$\Lambda$CDM~}
\def\om{\Omega_m}
\def\omb{\Omega_b}

\newcommand{\ltsima}{$\; \buildrel < \over \sim \;$}
\newcommand{\simlt}{\lower.5ex\hbox{\ltsima}}
\newcommand{\gtsima}{$\; \buildrel > \over \sim \;$}
\newcommand{\simgt}{\lower.5ex\hbox{\gtsima}}
\newcommand{\be}{\begin {equation}}
\newcommand{\ee}{\end {equation}}
\defcitealias{V16}{V16}
\defcitealias{ZuH11}{Z11}

\title[Cool core resiliency in galaxy cluster mergers]{
A study of cool core resiliency and entropy mixing 
in simulations of galaxy cluster mergers}

\author[R.~Valdarnini \& C. L. Sarazin]{
  R. Valdarnini,$^{1,2}${\thanks{E-mail: valda@sissa.it (RV)}
  C. L. Sarazin$^{3}$}
  \\
  $^{1}$SISSA --- Scuola Internazionale Superiore di Studi Avanzati, Via Bonomea 265, I-34136 Trieste, Italy\\
  $^{2}$INFN --- Iniziativa Specifica QGSKY, Via Valerio 2, I-34127 Trieste, Italy\\
 $^{3}$Department of Astronomy, University of Virginia,
530 McCormick Road, Charlottesville, VA 22904-4325, USA
}
\begin{document}

\pagerange{\pageref{firstpage}--\pageref{lastpage}} \pubyear{...}
\maketitle
\label{firstpage}

\begin{abstract}
        We present results from a suite of binary merging cluster simulations.
        The hydrodynamical cluster simulations are performed
        employing  a smoothed particle hydrodynamics (SPH) formulation in 
        which gradient errors are
        strongly reduced by means of  an integral approach.
      We consider adiabatic as well as radiative simulations, in which
        we include gas cooling, star formation and energy feedback from supernovae.
        We explore the effects of merging on the thermodynamic structure of the
        intracluster gas of the final merger remnant.
        In particular, we study how core entropy is generated during the merging
        and the stability properties of the initial cool-core profile
        against disruption.
         To this end, we consider a range of initial mass ratio and impact parameters.
         
          Final entropy profiles of our adiabatic merging simulations  are
         in good accord with previous findings \citep{ZuH11}, with cool-cores
         being disrupted for all of the initial merging setups.
         For equal-mass off-axis mergers, we find that a significant contribution to the final primary core entropy is due to 
hydrodynamic instabilities  generated by
        rotational motions, which are  induced by tidal torques 
         during the first pericenter passage.
        In radiative simulations, cool-cores
         are more
         resilient against heating processes;
         nonetheless, they are able to maintain their
         integrity only in the case of off-axis mergers with very unequal masses.
         We suggest that these results are robust against changes in the gas
         physical modeling, in particular to the inclusion of AGN thermal
         feedback.

         Our findings support the view that the observed core cluster morphology
         emerges naturally in a merging cluster context,
         and conclude that the merging angular momentum is a key
         parameter in shaping the thermodynamical properties of the
         final  merger remnant.

\end{abstract}
\begin{keywords}
Hydrodynamics --- methods: numerical
--- galaxies: clusters: general --- galaxies: clusters: intracluster medium --- X-rays: galaxies: clusters
\end{keywords}

%
%
\section{Introduction}
\label{sec:intro}

According to the  hierarchical scenario, the formation of structure
in the Universe proceeds under the action of gravity 
through merging and accretion of smaller structures. In this framework 
clusters of galaxies are the latest and most massive objects to be formed, 
with virial masses in the range $M \sim 10^{14}-10^{15} \msun$ \citep{Voit05}.

During their formation process, the gas is heated by adiabatic compression and 
shock-heating to higher temperatures. At virial equilibrium, about 
$\sim 90 \%$ of the baryons in a cluster will reside in the form of an 
 hot, X-ray emitting intracluster medium (ICM) at  temperatures $T\sim 10^7-10^8$ K.

Therefore, X-ray observations of the ICM
provide X-ray maps with which to
probe the spatial distribution 
of the cluster gas density, temperature and metallicities. 
Assuming hydrostatic equilibrium and 
spherical symmetry, these 
data can then be used to deduce the underlying dark matter (DM) distribution
and to determine cluster virial masses.

An accurate determination of cluster masses it is necessary in order to exploit
 the usefulness of clusters as cosmological probes, since at any given epoch 
 their number density  is a sensitive function of the background cosmological 
 model. This requires the clusters to be  dynamically relaxed, since otherwise
  cluster mass estimates will be prone to uncertainties.

  However, there is a large variety of observations indicating that galaxy clusters
  can be broadly classified into two categories: relaxed and un-relaxed 
  \citep[see][for a review]{Buote02}.
  The fraction of clusters exhibiting a disturbed 
  morphology grows with redshift and at the present epoch  can be even greater than
  $\sim 50\%$, depending on the adopted criterion used to measure the amount 
  of substructure  present in the cluster \citep{Buote02}.
   It is then fundamental to study the physics of cluster merging, not only
    in order to assess the status of the cluster dynamical equilibrium, but 
    because merging between substructures (or clusters themself) gives raise
     to a number of interesting physical processes \citep{Sar02,Molnar16}.

     During the merging process, collisions between substructures drive shocks into
     the ICM, heating the gas and injecting turbulent motions. These X-ray shocks 
      will boost X-ray luminosities, and leave a number of observational 
      signatures in the ICM, such as contact discontinuities (or cold fronts) in the gas 
      temperature, radio relics, relativistic electrons
      and other features \citep{Mark07,Feretti12}.

       Following the gas compression, mergers between clusters are also expected
       to drive star formation \citep{Fu99,Ro96,Roe14}, but with observations
       producing conflicting results. Some authors claim  an increase in the 
       star formation activity during mergers \citep{Bek10,St17}, while it is absent
       in other   merging systems \citep{Man17}.

       Moreover, major cluster mergers are the most energetic events since the Big-Bang,
       with energies $\simgt 10^{64}$ ergs. This renders these objects
       unique laboratories with which to study   dark matter models.
        Because of the collisionless nature of dark matter (DM), 
	 the position of gas and DM centers will be offset during a merging 
	 process. By contrasting X-ray and weak lensing data, it is possible
	 to derive upper limits on the cross section of self-interacting DM
	  \citep[][and references cited therein]{Molnar16}. From the 
	  Bullet cluster, \citet{Mar04}  were the first to put an upper limit
	   of the order of $\sigma_{\rm DM}/m_{\rm DM} \simlt 1 $ cm$^2$ g$^{-1}$ on the DM cross section per unit mass.
	  
	  Given the variety of physical phenomena and their non-linearity, 
	  N-body/hydrodynamical simulations are an indispensable tool with 
	  which to study merging of galaxy clusters.  Numerical simulations
	    aimed at studying cluster mergers have been performed either 
    in a cosmological framework \citep{Bu08,Pla09,Ras15,Ha17,Barnes18}, or
     by studying the collision between two clusters in isolation. 
     The binary merger simulations are implemented  by  first constructing two
    isolated gas+DM halos at equilibrium, and then the initial orbital trajectory
    is given by assigning  initial positions and velocities to the two halos.

    This method has the advantage that it allows the detailed study of a single 
     merging event.
     It also simplifies the interpretation of the simulation results
     because  the initial conditions are kept under control  and the simulation 
     can be contrasted with a specific observation. This approach has been 
     followed by many authors 
   \citep{Ro96,Ricker01,RT02,Poole06,MC07,Poole08,Mi09,Donnert17}.

   Specifically, idealized binary cluster mergers have been used to study 
   the merging configuration of the `Bullett cluster' \citep{Spr07,Mas08}, as well as
   that of `El Gordo' cluster \citep{Zh15,Zh18} and of other merging clusters
   \citep{Mac13,Molnar18,Ha19}.

   Simulated X-ray maps can be constructed to study cluster merging, for instance
by assessing the degree of relaxation of a specific system \citep{ZuH09}.
Moreover, the measured offset between X-ray and Sunyaev–Zel’dovich (SZ) maps 
allows the relative  velocity of the two merging clusters to be determined
 \citep{Molnar12,Zh14}. These limits in turn can be used to derive constraints on the
  assumed cosmological model.

  Another important topic in which mergers of galaxy clusters play an important role
  is in the study of DM properties. As previously outlined, major mergers are very
  energetic events in which a self-interacting DM (SIDM) is expected to exhibit 
   significant signatures. For this reason, merging simulations with a SIDM have
   been carried out by many authors \citep{Rob17,Kim17,ZuH19}, the simulations
   being aimed at investigating the impact of a SIDM on gas and DM properties 
   of the merged clusters.

 Finally, numerical simulations of merging clusters have been widely used to study
 the origin of the observed central properties of the cluster gas. 
 X-ray cluster surveys show
 that clusters can be divided
  into two categories according to  ICM central properties 
  \citep{Cav09,Joh09,Pratt10,McD13}: cool-core (CC) and non-cool core (NCC) clusters.
 CC clusters are characterized by a peaked X-ray emission, very short cooling times
  ($\sim$10\% of the Hubble time),  
   { central temperatures about $\sim 1/3$ of the virial ones and  radial entropy 
   profiles steeper in the core than those of NCCs \citep{Cav09}.}

   These short cooling times should induce a run-away cooling process that is not 
   observed;
    to balance radiative losses, some heating sources must be operating in the cluster 
    cores. This is the so-called `cooling flow' problem 
and various heating models have been proposed in the literature
to offset cooling and regulate the cooling flow 
\citep[see][and references cited therein]{Soker16}.
  
To  observationally define a CC cluster there are various criteria
     \citep{Barnes18}, which depend on the available data. However, there is
     some consensus that CC clusters are correlated with a regular X-ray
     morphology \citep{Chon12}, while this is not true for NCC clusters.
      The latter are often associated with a disturbed morphology \citep{Pratt10} 
      and exhibit a much flatter radial entropy profile than CC clusters. 

      These findings strongly suggest that  the CC/NCC dichotomy can be
      naturally interpreted in terms of the cluster merging histories. 
      In this framework, the population of NCC clusters originates as a consequence
      of major mergers that disrupt CC clusters.
      Conversely, the original 
      core morphology is preserved for relaxed clusters that have 
      not experienced a major merger recently.

      This scenario is an important issue for a better understanding of cluster
      formation and evolution, and  N-body/hydro  simulations of merging clusters
       have been widely employed 
  \citep{RT02,GO02,MC07,Poole08,Bu08,Pla09,ZuH11,Ras15,Ha17,Barnes18} 
  to address the survival of CCs during cluster mergers.

  \citet{GO02} used 2D idealized radiative  merging simulations 
 to conclude  that CCs do not survive major head-on mergers.
 Similar conclusions were reached for equal-mass mergers by \cite{RT02} and 
 \citet{Poole08}, but from
 their simulations the authors argue that CCs  are resilient to unequal-mass mergers
 if they are off-center.

These findings are in contrast with  those of \cite{ZuH11}. From a suite of 
idealized merging simulations,  performed over a range of different mass ratios
and impact parameters, the author finds that there is a significant degree of gas
mixing taking place during the mergings. This in turn leads to higher levels
of final entropy in the merger remnants and to CC disruption. Similar results 
were also obtained by \citet{Mi09}.

This discrepancy with previous simulations 
 \citep{RT02,Poole08} could be due to a number of causes, both physical and 
 numerical. For instance, both of the earlier authors performed their merging simulations
 using standard SPH, while \citet{Mi09} and \citet{ZuH11} employed 
the adaptive Eulerian mesh code FLASH. In terms of code capability to model 
fluid instabilities and gas mixing this could be 
a critical issue (see Section \ref{subsec:icstab}).
Moreover, the simulations of  \citet{RT02} and \citet{Poole08} incorporated 
radiative cooling, while the mesh runs were adiabatic. 

Early cosmological simulations \citep{Bu08,Pla09} have shown that there is a significant
connection between the presence of CCs and the cluster merging history.
In particular, \citet{Bu08} using cosmological simulations that included 
cooling as well as star formation and supernovae feedback, found that CCs are
resilient to late-time mergers.
More recently, \citet{Ras15} argued that CCs can be destroyed during late-time
mergers, with AGN feedback playing a key role in reducing overcooling and 
allowing CCs to be disrupted. This is contrast with the findings of 
\citet{Ha17}, for whom the low entropy levels exhibited by simulated CCs cannot be 
alleviated by AGN feedback. According to \citet{Ha17}, CC disruption 
depends critically on the  angular momentum of the merger.
 
Motivated by these considerations, we present in this paper a suite of hydrodynamical
simulations of merger clusters, aimed at investigating the resiliency of CCs
to cluster mergers. We perform a set of 
N-body/hydrodynamical  binary cluster merger simulations, with initial conditions
 spanning a wide range of mass ratio and impact parameters.

 We use an SPH code 
\citep[see][for a review]{Price2012}, based on a improved numerical scheme
 (see below). In a battery of hydrodynamical tests
 \citep[][hereafter \citetalias{V16}]{V16}, it has been
 demonstrated that this code  can be profitably  used in many astrophysical
 problems, without the shortcomings present in standard SPH.
 We perform both adiabatic  and radiative merger simulations,
  with the latters incorporating  radiative cooling as well as 
   star formation and energy feedback from supernovae.

   The initial conditions of our idealized merger simulations are set up
   as follows. For an isolated spherical halo composed of gas and DM initially in 
   equilibrium, we specify the radial DM density and gas entropy profiles.
To define halo parameters,  we use a \lcdm cosmology,  with 
$\om=0.3$, $H_0=70$ km s$^{-1}$ Mpc$^{-1}$, 
 and a baryon fraction of $f_b=\omb/\om=0.162$.
   For each of the gas and DM components, a particle realization of  positions 
   and velocities is then constructed, according to profiles computed 
   under the  assumption of hydrostatic equilibrium. 

   This procedure is used to construct both a primary and a secondary cluster, 
   the virial mass of the two being related by the merging mass ratio.
    To initialize the merger simulation, the particle positions and velocities 
    of the two halos are then shifted
    according to the initial orbital trajectory.

    Our initial condition set up is analogous to that implemented by 
 \citet{ZuH11} in his adiabatic merger simulation study. In particular, we adopt the 
 same range of collision parameters. 
 In the present study, we have purposely  chosen to adopt similar initial settings.
 This in order to compare with previous results on the effects of mergers on final 
 CC properties,  specifically when cooling is included in the simulations. 

 Moreover, the merger simulations of \citet{ZuH11} 
 were performed using  an adaptive mesh-based Eulerian code.
 For adiabatic simulations, it is then interesting to compare the entropy
 profiles of the final merger remnants against the corresponding ones 
 presented in  \citet{ZuH11}. This because the two sets of simulations have 
 been constructed by adopting very similar initial conditions, but the 
 codes used to perform the simulations are based on two completely different
 numerical hydrodynamical schemes.

 Our paper is organized as follows. In Section \ref{sec:sims} we describe our 
 hydrodynamical scheme, together with the method we use  to initialize halos in
 equilibrium and the orbital properties.
Section \ref{sec:results} is dedicated to the presentation of the results, in 
which we describe our findings from adiabatic and radiative simulations; 
a specific Section being dedicated to investigate the generation of entropy 
through mixing and shock-heating processes during the various merging phases. 
Finally, our main conclusions are summarized in Section \ref{sec:discuss}.

\section{Simulations}
\label{sec:sims}
%
The simulations are performed by employing an entropy conserving SPH 
formulation \citep{Price2012}.  To estimate first-order SPH derivatives, 
 the numerical scheme is improved
by using an Integral Approximation (IA) accompanied by a matrix inversion, 
thus  strongly reducing gradient errors in the momentum equation.  

The IA scheme was  originally proposed by   \citet{ga12}, and further tested
 in a variety of hydrodynamical tests \citep[][\citetalias{V16}]{ga12,Ross15}. The results of the tests
demonstrate that the new SPH formulation outperforms standard SPH and, in 
terms of accuracy, can be considered competitive  with other 
numerical hydrodynamic schemes \citepalias{V16}. 
In particular, with respect standard SPH, it is found 
 that  the IA scheme greatly improves the numerical modeling of subsonic 
turbulence \citep[\citetalias{V16},][]{V19}.
This aspect is particularly important  for the simulations presented here, 
in which a significant amount of turbulence is expected to be injected into the ICM 
during cluster collisions \citep{Sch17}.

We now outline the basic features of the hydrodynamical method
--- we refer the reader to  \cite{ga12} and \citetalias{V16}
for a comprehensive description of the IA method applied to SPH.
In what follows,  we will refer to the  SPH scheme described here as  integral SPH (ISPH).

\subsection{Numerical method}
\label{subsec:nummt}

In SPH, the fluid is described by a set of $N$
particles with mass $m_i$, velocity $\bf v_i$, density $\rho_i$
and specific entropy parameter $A_i$.\footnote{We use the convention of using
Latin indices to denote particles and Greek indices to denote
the three spatial dimensions.}
 The latter is
related to the particle pressure  by
$P_i=A_i\rho_i^{\gamma}=(\gamma-1) \rho_i u_i$, where $\gamma=5/3$ and
$u_i$ is the thermal energy per unit mass $u_i$.

At the particle position ${\bf r}_i$, the SPH gas density $\rho_i$ is given by
the summation 
 \begin{equation}
	 \rho_i=\sum_j m_j W(|{\bf r}_{ij}|,h_i)~,
    \label{rho.eq}
 \end{equation}
where the sum is  over neighboring particles $j$, and
 $W(| {\bf r}_{ij} |,h_i)$ is a kernel with compact support.
 We define $W_{ij} (h_i) \equiv W(| {\bf r}_{ij} |,h_i)$.
 For the simulations presented here, we use the  $M_4$ kernel \citep{Price2012},
 which is zero for  $|{\bf r}_i-{\bf r}_j| \geq  2 h_i$. 

In Equation  (\ref{rho.eq}) the sum is over a finite number of
particles $N_{nn}$, and the smoothing length $h_i$  
 is implicitly defined by the equation
   \begin{equation}
  \frac{4 \pi (2h_i)^3 \rho_i}{3}=N_{nn} m_i~,
    \label{hrho.eq}
   \end{equation}
which is solved  iteratively for each particle by setting 
$N_{nn}= 33$.

The SPH momentum equation in the IA framework reads
\begin{equation}
	\frac {d  {\bf  v}_{i,\alpha}}{dt}=-\sum_j m_j \left[
 \frac{P_i}{\Omega_i \rho_i^2}
	{\bf \mathcal A}_{\alpha,ij}(h_i) 
 +\frac{P_j}{\Omega_j \rho_j^2}
{\bf \mathcal {\widetilde {A}}}_{\alpha,ij}(h_j)
\right]~,
  \label{fsphw.eq}
\end{equation}
where  $\Omega_i$ is defined as
   \begin{equation}
   \Omega_i=\left[1-\frac{\partial h_i}{\partial \rho_i}
   \sum_k m_k \frac{\partial W_{ik}(h_i)}{\partial h_i}\right]~,
    \label{fh.eq}
   \end{equation}
and  the two terms ${\bf \mathcal A}_{\alpha,ij}(h_i)$ and
$ {\bf \mathcal {\widetilde A}}_{\alpha,ij}(h_j)$ are the IA generalization to the 
SPH derivatives 
 $ \nabla_i  W_{ij}(h_i) $ and $ \nabla_i  W_{ij}(h_j)$, respectively.

These IA terms are given by 
\begin{equation}
\begin{array}{lc}
{\bf \mathcal A}_{\alpha,ij}(h_i)= &
\sum_{\beta} {\bf \mathcal C}_{\alpha\beta } (i) \Delta_{\beta }^{ji} W(r_{ij},h_i)~, \\
{\bf \mathcal {\widetilde A}}_{\alpha,ij}(h_j) = & 
	\sum_{\beta} {\bf \mathcal C}_{\alpha\beta } (j) \Delta_{\beta }^{ji}  W(r_{ij},h_j)~.
\end {array}
  \label{iadf.eq}
 \end{equation}
	Here  ${\bf \Delta}^{ji}_{\alpha} =( {\bf r}^{j}- {\bf r}^i)_{\alpha}$
 and $ \mathcal{C}_{\alpha\beta } (i) $ are the elements of the  matrix
 $\mathcal{C}=\mathcal{T}^{-1}$ associated with the particle $i$. 
The inverse of this matrix is a $3 \times 3$ symmetric tensor  
 $\mathcal{T}$, which for   particle $i$ takes the form
\begin{equation}
	{\bf \mathcal T}_{\alpha\beta}(i)=\sum_j \frac{m_j}{\rho_j}    
	{\bf \Delta}_{\alpha} ^{ji} {\bf \Delta}_{\beta }^{ji}  W(r_{ij},h_i)~. 
    \label{iade.eq}
   \end{equation}

To properly handle shocks, the SPH momentum equation
 (\ref{fsphw.eq})
must be generalized to
include an artificial viscosity (AV) term:
 \begin{equation}
	 \frac {d {\bf v}_{i,\alpha}}{dt}=-\sum_j m_j
\Pi_{ij}
 {\bf {\mathcal {\bar A}}}_{\alpha,ij} ~,
  \label{dvdt.eq}
   \end{equation}
where  $\Pi_{ij}$ is the AV tensor and
   \begin{equation}
	   {\bf {\mathcal {\bar A}}}_{\alpha,ij}=\frac{1}{2}
 \left[
	 {\bf {\mathcal A}}_{\alpha,ij}(h_i) +
	   {\bf {\mathcal {\tilde A}}}_{\alpha,ij}(h_j)
\right].
  \label{asym.eq}
   \end{equation}
   
  We adopt here the Riemann-based
formulation proposed by \citet{mo97} to write the AV tensor as:
   \begin{equation}
\Pi_{ij} =
 -\frac{\alpha_{ij}}{2} \frac{v^{AV}_{ij} \mu_{ij}} {\rho_{ij}} f_{ij}~,
  \label{pvis.eq}
 \end{equation}
	where $\mu_{ij}= {\bf v}_{ij} \cdot
	{\bf r}_{ij}/|r_{ij}|$ if $ {\bf v}_{ij} \cdot {\bf r}_{ij}<0$ but zero
	otherwise,  ${\bf v}_{ij}= {\bf v}_i - {\bf v}_j$,
 $\rho_{ij}$ is the arithmetic mean of the two densities
 and
  $\alpha_{ij}=(\alpha_i+\alpha_j)/2$ is the symmetrized AV parameter.
 The signal velocity $v^{AV}_{ij}$ is estimated  as
   \begin{equation}
v^{AV}_{ij}= c_i +c_j - 3 \mu_{ij}~,
  \label{vsig.eq}
 \end{equation}
with $c_i$ being the sound velocity.
 The  symmetrized AV limiter $f_{ij}=(f_i+f_j)/2$, where
   \begin{equation}
	   f_i=\frac {|{\bf \nabla \cdot  v}|_i}
	   {|{\bf \nabla \cdot  v}|_i+|{\bf \nabla \times  v}|_i}~,
   \label{fdampav.eq}
 \end{equation}
 is introduced \citep{ba95} to suppress AV when in presence of strong shear flows.
  The individual particle
viscosity parameters $\alpha_i(t)$ are allowed to evolve in time according the
\citet{cul10} scheme, which is found to significantly reduce AV away from shocks
[\citealt[][V16: see equation (9) and following text]{cul10}].
The  $\alpha_i$'s can vary 
from  a minimum value $\alpha_{min}=0.01$ when  shocks  are absent,
 up to a maximum value $\alpha_{max}=1.5$.

\subsubsection{  Dissipative terms }
\label{subsec:visc}

In SPH,  the particle entropy $A_i$ is generated at a rate 

   \begin{equation}
  \frac {d A_i}{dt} =\frac{\gamma-1}{\rho_i^{\gamma-1}}
   ( Q_{AV} +Q_{AC}-Q_R )~,
    \label{aen.eq}
   \end{equation}
where the 
   $Q_{AV}$ term  refers to numerical viscosity effects
   \citepalias{V16}.
 $Q_{AC}$ is an artificial conduction (AC)
term and  $Q_{R}$ describes the effects of radiative cooling.
{ The latter is defined as $Q_R= \Lambda(\rho_i,T_i,Z_i)/\rho_i$, with
$\Lambda(\rho_i,T_i,Z_i)$ being the cooling function, 
 $T_i$ and $Z_i$ the particle temperature and metallicity, respectively.
Thus, $Q_R$ is the cooling rate per unit mass.}

The presence of the AC term is necessary 
 in  SPH simulations \citep{pr08} 
for treating contact discontinuities, 
such as when studying the growth of Kelvin--Helmholtz instabilities. 
 This term can be written as
   \begin{equation}
  \left ( \frac {d u_i}{dt} \right)_{AC} =
\sum_j \sum_{\alpha} \frac{m_j v^{AC}_{ij}}{\rho_{ij}}
	   \left[ \alpha^C_{ij}(u_i-u_j) \right ]  {\bf \Delta}^{ij} _{\alpha}
	   {\bf {\mathcal  {\bar A}}}_{\alpha,ij}/r_{ij}~,
  \label{duc.eq}
   \end{equation}
where $ v^{AC}_{ij}$ is  the AC signal velocity, {$\alpha^C_{i}$ is an AC 
parameter of order unity  and
 $\alpha^C_{ij} = (\alpha^C_i + \alpha^C_j)/2$ its symmetrized value.}

The form of the AC signal velocity depends on the problem under 
consideration \citep{Price18}. An appropriate choice in the 
presence of  self-gravity  is found to be \citep{wa08,VA12}
 \begin{equation}
	 v^{AC}_{ij} = |({\bf v}_i-{\bf v}_j)\cdot {\bf r}_{ij}|/r_{ij}~,
 \label{vsgv.eq}
 \end{equation}
which has been checked in several test cases
\citep{VA12} and is zero for a self-gravitating system at equilibrium.

%

The time evolution of  the AC parameter $\alpha^C_{i}$ is regulated
by a source term which is proportional to the Laplacian of the particle 
thermal energy,
and by a decaying term which quickly damps 
 $\alpha^C_{i}$ away from discontinuities.
A description of the  AC numerical settings  is given in \citet{VA12}.

Finally it is worth noting that incorporating an  AC term into the SPH 
thermal equation significantly improves a major shortcoming of classic SPH.
It is well-known \citep{Mi09} that the level of core entropies, found in non-radiative
standard SPH simulations of galaxy clusters, are well below those produced in similar
simulations using mesh-based codes. This difficulty is due to the Lagrangian
nature of SPH, in which subgrid diffusion processes are missed.
It is shown  that introducing an AC term  \citep{wa08,VA12},
 the levels of entropies found in galaxy cluster cores  are in much better 
agreement 
with those produced using mesh codes.

\begin{table*}
\caption{Initial parameters of the three test clusters used in the 
merging simulations.$^{a}$}
\label{clparam.tab}%
\centering
\begin{tabular}{cccccccccc}
\hline
Cluster & $M_{200}$ $[\msun]$  & $M^h_{DM}$ $[\msun]$ & $c_{200}$ & $r_{500}$ [Mpc]
  & $S_{500}$ $[\sev]$ & $S_0$ & $S_1$  & $f_{gas}$ \\
\hline
C1 &   $ 6 \times 10^{14}$ & $ 6.6 \times 10^{14}$    & 4.5  & 1.15& 931  & $5.3\times 10^{-3}$ & 1.75 & 0.1  \\
C2 &  $ 2 \times 10^{14}$  & $ 2.2 \times 10^{14}$ & 6.4  & 0.81& 466  & $5.3\times 10^{-3}$ & 1.84  & 0.086   \\
C3 &  $ 6 \times 10^{13}$  &  $ 6.6 \times 10^{13}$ & 7.0  & 0.54 & 211  & $6.4\times 10^{-3}$ & 2.1  & 0.074\\
\end{tabular}
\begin{flushleft}
{\it Notes.} {$^a$ Columns from left to right:
Name of the cluster model,
halo mass $M_{200}$ at the radius $r_{200}$,
total DM halo mass $M^h_{DM}$ within $r=2 r_{200}$, 
concentration parameter $c_{200}$,
cluster radius $r_{500}$ at which $\Delta=500$,
gas entropy $S_{500}$ at $r_{500}$,
gas entropy profile parameters $S_0$ and $S_1$(see the text),
and gas mass fraction $f_{gas}$ at $r_{500}$.}
\end{flushleft}
\end{table*}

  The  $Q_{R}$  term is present in those runs  for which 
radiative cooling is also included.
  For these simulations 
 the gas physical modeling  incorporates star formation and  energy feedback from supernovae
as well.
For the numerical aspects of the cooling implementation we refer to \citet{VA06}.

%
\subsection{Initial condition setup }
\label{subsec:icsetp}

For a variety of initial conditions, we perform N-body/hydrodynamical ISPH 
simulations to study the collisions between galaxy clusters.
Each cluster consists of an isolated spherical halo initially in equilibrium, 
composed of dark matter and gas particles.
The initial conditions of our idealized binary cluster mergers
are very similar, but not identical, to  those of
\citet[][hereafter \citetalias{ZuH11}]{ZuH11}.
We study collisions between a primary and a secondary cluster,
 with the primary mass  always set to  
 $M_{200} =6 \times 10^{14} \msun \equiv M_1$.
 Here  $M_{200}$ is the cluster mass within the  radius 
$r_{200}$. We define $r_{\Delta}$ as the radius  at which   the cluster  
mean density is $\Delta$ times the cosmological critical density $\rho_c(z)$:
 \begin{equation}
M_{\Delta}=\frac{4 \pi}{3} \Delta  \rho_c(z) r_{\Delta}^3.
 \label{mcl.eq}
 \end{equation}
In the following, we assume $z=0$ as the redshift at which $r_{200}$ is calculated.
For the secondary, with cluster mass $M_{200}\equiv M_2$,   
we consider three different  mass ratios $R=M_2/M_1=1:$1, $1:$3 and $1:$10.

For each collision with mass ratio $R$, we consider three different impact 
parameters $b$: a head-on cluster collision with $b=0$, and two off-axis
mergers with $b/r_{200}=0.3$ and $0.6$. Here $b$ is the impact parameter of the 
collision when the distance $d_{12}$ between the center of mass of the two clusters 
is $d_{12}=r^1_{200}+r^2_{200}$; see Figure 1 of \citetalias{ZuH11} for a geometric description
of the collision set up.  
The procedure we use to assign  initial separations and relative velocities
between the two clusters is described in Section \ref{subsec:ickin}.

\subsubsection{ Dark matter  halos }
\label{subsec:icdm}
We assume spherical symmetry for the initial dark matter (DM) and gas mass distribution.
For the DM density, we adopt an NFW  profile \citep{NA97}
\begin{equation}
\begin{array}{llll}
\rho_{DM}(r)&=&\dfrac{\rho_s}{r/r_s(1+r/r_s)^2}\, ,  &   0\leq r\leq r_{200} \, , \\
 \end{array}
 \label{rhodmins.eq}
 \end{equation}
where $c_{200}=r_{200}/r_{s}$ is the concentration parameter.
To avoid a divergent total mass, outside $r_{200}$ the DM density profile
 is suppressed exponentially \citep{Kaz04}
 up to a final radius $r_{max}=\xi r_{200} $  :
\begin{equation}
\begin{array}{lll}
\rho_{DM}(r)&=&\rho_{DM}(r_{200}) (r/r_{200})^{\delta} 
\exp{-\displaystyle{\left(\frac{r-r_{200}}{r_{decay}}\right)}} \, , \\
             & &~~   r_{200} < r < r_{max}~, 
 \end{array}
 \label{rhodmext.eq}
 \end{equation}
where $r_{decay}=\eta r_{200}$ is the  truncation radius, and the parameter 
$\delta$ is set by requiring the first derivative of the DM density profile
to be continuous at 
 $r=r_{200}$ 
 \begin{equation}
\delta=-\frac{r_s+3r_{200}}{r_s+r_{200}}  + \frac{r_{200}}{r_{decay}}~.
 \label{deltadm.eq}
 \end{equation}

For the runs presented here,  we set  the truncation parameters to
 the values 
 $(\xi, \eta) =(2,0.2)$; this choice   will be motivated in 
Section \ref{subsec:icstab}.
 For a cluster of given mass, the density profile
 is then specified by the parameter $c_{200}$. For our two test clusters with
 $M_{200} > 10 ^{14} \msun $  we set  the value of 
  $c_{200}$   using the CLASH $c-M$ relation of 
 \citet{gr16}
 \begin{equation}
c_{200}\simeq3.66 /(M_{200}/M_{\star})^{0.32}~,
 \label{cfit.eq}
 \end{equation}
where $M_{\star}=8\times10^{14}\msunh$.  For the  cluster $C3$ with 
  $M_{200}< 10^{14} \msun$ 
 we set  $c_{200}=7.03$.  This value is obtained by using the 
  following $c_{500}-M_{500} $  relation  for galaxy groups \citep{gast07,Sun09}
 \begin{equation}
c_{500}=3.96 /(M_{500}/10^{14} \msun )^{0.226}~,
 \label{cfitc.eq}
 \end{equation}
and solving numerically the halo profile to obtain $c_{200}$.
Table ~\ref{clparam.tab}  lists several initial  parameters of the three 
idealized clusters we use to construct our simulation suite.

A numerical realization of the DM density profile is then constructed 
by first evaluating 
{ the enclosed DM mass $ M_{DM}(<r)$ within the radius $r$}, which 
is normalized so that  it is equal to $(1-f_b) M_{200}$ at $ r_{200}$.
We subsequently invert $ q(r)=M_{DM}(<r)/  M_{DM}(<r_{rmax})=y$, where $y$ is a 
uniform random number in the interval $[0,1]$,  to obtain the radial particle 
coordinate $r$.
Finally,  Cartesian coordinates are assigned to the particle by randomly orienting 
the particle position vector $ {\bf r}$. 

\citet{Kaz04} showed that, for exponentially truncated NFW halos,
  particle velocities
are accurately determined if their energies are drawn from the 
system distribution function $f(\mathcal{E})$. For spherical symmetric systems
 \begin{equation}
\rho_h(r)= 4 \pi \int_0^{\Psi(r)} f(\mathcal{E})
\sqrt{2 [\Psi(r)-\mathcal{E}]} ~ d \mathcal{E}~,
 \label{rhof.eq}
 \end{equation}
where  $\Psi(r)=-\Phi(r)=-(\Phi_{h}+\Phi_g)$ is the relative gravitational 
potential and $\mathcal{E}= \Psi- v^2/2$ is the relative energy. Here 
the subscripts $h$ and $g$ denote the DM and gas components, respectively.

Equation (\ref{rhof.eq}) can be inverted \citep{Bin87} to give
 \begin{equation}
f(\mathcal{E})=\frac{1}{\sqrt{8} \pi^2}
  \left[
 \int_0^{\mathcal{E}} \frac{d^2 \rho_h}{d\Psi^2}
\frac {d \Psi}{\sqrt{(\mathcal{E}-\Psi)}}+
\frac{1}{\sqrt{\mathcal{E}}} \left(\frac{d \rho_h}{d \Psi} \right)_{\Psi=0}
\right]~.
 \label{fdist.eq}
 \end{equation}
The boundary term on the rhs of the equation is zero for any sensible choice of 
   $\Psi(r)$  and $\rho(r)$ \citep{Kaz04}. The second order derivative 
 $ {d^2 \rho_h}/{d\Psi^2} $ can be expressed as

 \begin{equation}
\frac{ d^2 \rho}{ d  \Psi^2} =   \left(\frac{r^2}{GM}\right )^2 \left [
\frac{ d^2 \rho}{ d r^2}  + \frac{d \rho}{d r} 
\left (\frac{2}{r} -\frac{4 \pi \rho r^2}{M} \right )
\right]~,
 \label{drho2.eq}
 \end{equation}
which has the advantage of avoiding numerical differentiation in the integral
(\ref{fdist.eq}), since $\rho$ is known analytically.

The function $ f(\mathcal{E}) $ is then evaluated numerically and its values 
tabulated over a grid of energies. For a given energy $ \mathcal{E}$ the value of $f$ is
obtained by interpolation.
For a particle at position $ {\bf r}$ with energy 
 $\mathcal{E}\in [0,\Psi(r)]$,  we randomly draw pairs
 $(f,\mathcal{E})$  and 
use an acceptance--rejection method
\citep{KD94,zemp08,Drakos17} to obtain the particle speed $v= \sqrt{2[\Psi(r)-\mathcal{E}]}$.
As for the particle position,  the direction of the  velocity vector is randomly
oriented.

These prescriptions for generating DM particle distributions at equilibrium are widely
used by many authors in numerical simulations of merging cluster galaxies, for which
initially DM halos are described by an exponentially truncated NFW profile.
However, a major drawback of the method introduced by \citet{Kaz04} is that 
the second derivative $ {d^2 \rho_h}/{d\Psi^2} $ is discontinuous at 
 $r=r_{200}$. This implies that the behavior of 
the function $ f(\mathcal{E}) $  can become inconsistent 
for certain values of $r_{decay}$ \citep{zemp08,Drakos17}, thus compromising halo 
stability over cosmological timescales.
\citet{zemp08} recommend the choice
 $\eta=r_{decay}/r_{200}=0.3$; here we show in Section \ref{subsec:icstab} that 
 by setting $\eta=0.2$ one can obtain sufficiently stable halos.

%

\subsubsection{ Baryonic halos }
\label{subsec:icgas}
We assume hydrostatic equilibrium to construct the cluster gas initial conditions.
Following \citetalias{ZuH11}, we initialize  gas density and temperature profiles
 by specifying analytically the initial cluster entropy profile.
{ Physically, this would be best-represented by giving the specific physical 
entropy per particle in the gas, $s(r)$.
However, both observations of the gas in clusters and previous simulations have 
instead utilized a related entropy parameter [written as $S(r)$ or $K(r)$].
To be consistent and allow easier comparison to observations and previous 
simulations, we will adopt this convention.

Thus, in the following we refer to the  ``gas entropy'' as this entropy parameter  
$ S \equiv k_B T/n_e^{2/3}$, where $k_B T$ is the gas temperature in keV 
and $n_e$ the electron number density.
From the Sackur-Tetrode equation \citep{Lan80}, it is easily shown that
$s = (3/2) k_B \ln ( S ) + {\mathrm const}$, where the constant is not important 
here.  Thus, it is straightforward to convert between $s$ and $S$.
However, one should be aware that differences in the entropy are exaggerated by 
$S$, since it is only its logarithm that enters into the physical entropy.}

CC clusters are observationally  characterized 
 \citep{Cav09,Pratt10,McD13,Ghi19} by dense, compact
cores with cooling times  shorter than $H_0^{-1}$. 
 A key feature of CC clusters is that of having 
a level of central entropy  below a threshold value $\simeq 60$ keV cm$^2$.

We then adopt for the gas entropy profile an observationally motivated 
 \citep{Cav09,Pratt10,Ghi19} functional form which consists of a power law behavior
and an entropy floor value:
 \begin{equation}
S(r)/S_{500} = S_0+S_1 \left (\frac{r}{r_{500}}\right )^{\alpha}~,
 \label{sprof.eq}
 \end{equation}
where  \citep{Ghi19}
 \begin{equation}
S_{500}\simeq  1963 
\left[ M_{500}/ ( 10^{15} h^{-1} \msun )\right]^{2/3} {\rm keV} \, {\rm cm}^2~.
 \label{snorm.eq}
 \end{equation}

%

For a given set of parameters $(S_0,S_1,\alpha)$, the entropy profile is then 
completely  specified and we numerically integrate the equations of hydrostatic 
equilibrium and mass continuity:
 \begin{subequations}
 \label{hydros.eq}
 \begin{eqnarray}
\qquad \frac{ d  P } {d r} & = & - \frac { G  M_{tot}(<r)}  {  r^2 } \rho_g
\, , \label{hydrosa.eq} \\
\qquad \frac { d M_g (<r)} {d r}&  = &  { 4 \pi r^2 \rho_g } \, . \label{hydrosb.eq}
 \end{eqnarray}
 \end{subequations}
The numerical integration of these equations is found more manageable if 
one integrates the temperature instead of pressure. The latter can be expressed 
as 
 \begin{equation}
P=\frac{\rho_g k_B T}{ \mu m_p}= K \rho_g ^{5/3}~,
 \label{pres.eq}
 \end{equation}
where $ K(r)=S(r)(\mu/\mu_e)^{2/3}/(\mu m_p)^{5/3}$
 and for the mean molecular weights  we assume
$\mu=0.59~,\mu_e=1.14$. Equations  (\ref{hydrosa.eq})  and (\ref{hydrosb.eq}) now read
 \begin{subequations}
 \label{thydros.eq}
\begin{eqnarray}
\frac { d T }{d \log r} &= & - \frac{2 \mu m_p }{5 k_B  } 
\frac{ G  M_{tot}(<r)}{r}
+\frac{3}{5} T \frac{d \log K} { d \log r} ~, \label{thydrosa.eq}\\
\frac { d \log M_g }{d \log r} &= &  \frac{ 4 \pi r^3 \rho_g } {M_g(<r)} \label{thydrosb.eq}~.
 \end{eqnarray}
 \end{subequations}

\begin{table}
\caption{\label{kinerr.tab}%
The simulation ID is defined according to the mass ratio $R$ and the
impact parameter $b$ of the collision, second column the corresponding ID 
of the \citetalias{ZuH11} runs.
The simulations start at $t_s=0$, the time $\tau^{num}$ is 
the simulation 
time at which the center of mass position and velocities of the two halos are  
closest to the corresponding \citetalias{ZuH11} initial conditions. 
$\varepsilon_x$ and $\varepsilon_v$ are the corresponding relative errors, 
$\tau^K$ is the analytic solution given by the Kepler's problem. 
}
\centering
\begin{tabular}{cccccc}
\hline
    ID & ID \citetalias{ZuH11}  & $\tau^{num}$[Gyr] & $\tau^K$[Gyr] & $\varepsilon_x$ & $\varepsilon_v$  \\
\hline
  $ R01b00$ & $S1$   & 3.81  & 3.73  & $1.06\times 10^{-2}$ &  0.115  \\
  $ R01b03$  & $S2$ &  3.94  & 4.17  & $2.3\times 10^{-2}$ &  0.09  \\
   $ R01b06$  & $S3$ & 4.81  & 5.10   & $4.1\times 10^{-2}$ &  0.13  \\
  $ R03b00$  & $S4$ &  2.81  & 2.73   & $1.4\times 10^{-2}$ &  0.09  \\
  $ R03b03$  & $S5$ &  2.87 & 2.99   & $1.2\times 10^{-2}$ &  0.08  \\
  $ R03b06$  & $S6$ &  3.31 & 3.41   & $2.8\times 10^{-2}$ &  0.11  \\
 $ R10b00$  & $S7$ &  2.32 & 2.32   & $3.7\times 10^{-2}$ &  0.06  \\
   $ R10b03$  & $S8$ &  2.44 & 2.53   & $1.03\times 10^{-2}$ &  0.08  \\
   $ R10b06$  & $S9$ &  2.75& 2.87   & $2.4\times 10^{-2}$ &  0.09  \\
\end{tabular}
\end{table}

To integrate  these equations it is necessary to specify two boundary conditions. 
 Our first condition is that $r(M_g=0)=0$, whilst the second requires that the halo gas mass 
 at $r=r_{500}$ yields a gas mass fraction $f^h_g=M_g/M_{tot}$ given by the measured
relation \citep{Sun09}  
 \begin{equation}
f_{gas} = 0.035h^{-3/2} ( M_{500} h  / 7 \times 10^{12} \msun )^{0.135}~.
 \label{fgas.eq}
 \end{equation}
To construct our gas density and temperature profiles we proceed as follows.
 For a given set of entropy parameters $ (S_0,S_1,\alpha)$,  we initially choose an arbitrary value  of $P(r=0)\equiv P_0$ which is used to compute $\rho_0$ 
and $T_0$. Equations (\ref{thydrosa.eq})  and (\ref{thydrosb.eq}) are then 
numerically integrated up to $r=r_{500}$ , after which the value of 
$f^h_g$ is contrasted with that of $f_{gas}$.
 We iterate the whole procedure in order to bracket the value of $P_0$ 
 until the quantity $|f^h_g-f_{gas}|/f_{gas} $ is below a certain threshold 
value ($\simlt 1 \%$).
When this condition is satisfied and $P_0$ is a root value, we propagate the 
solution outward to $r_{max}=2 r_{200}$.
 This normalization  procedure implies a gas mass fraction at $r= r_{200}$ that 
may differ from the cosmic value  $f_b$, and therefore that may
  not be entirely consistent with 
the normalization adopted in Section \ref{subsec:icdm} for the DM component,
the latter making use of the cosmic gas fraction $f_b$ to set the halo DM mass  
to $(1-f_b) M_{200}$ at $ r_{200}$.
However, we have verified that for all  the considered halos 
the difference between the two gas fractions at $r= r_{200}$  
 is very small (say $\simlt 1\%$).

The radius up to which the gas density profiles is continued beyond $ r_{200}$ 
coincides here with the DM  truncation radius. The choice of the gas truncation 
radius requires some care. Setting this radius to $ r_{200}$ would imply 
that gas particles at the halo edge will begin to flow outward,
owing to the absence of an external pressure. This effect implies 
 a steepening of the gas density profile and a mass leak, which can have a significant 
impact on the initial mass profile when  the collision timescale is large.

 To avoid this mass-leaking issue, an approach commonly employed
 \citep{Turner95,Ricker01,Poole06,MC07,Donnert14}
 consists of  surrounding the gaseous halo with a low-density, dynamically negligible, confining medium.
However, this procedure comes at the cost of adding a large number of SPH particles
to the simulations. We choose here a different approach by extending the gaseous halo 
beyond $ r_{200}$  and up to a maximum radius $r_{max}=2 \, r_{200}$.

The shell between $ r_{200}$  and  $r_{max}=2 \, r_{200}$  then acts as a buffer zone that
is able to keep the gas particles within $ r_{200}$ confined. 
Clearly the edge particles at $r\simeq 2 \, r_{200}$  will begin to flow outward, 
 leading to a steepening of the density profile. 
This steepening can be considered unimportant as long as collisions
between clusters occur on timescales much shorter than that necessary to 
modify the initial gas density profile at radii  $r\simlt  r_{200}$. 
It will be seen in Section \ref{subsec:icstab} that these conditions are always verified 
for the cluster collisions we consider.

We assume  $\alpha=0.93$ in Equation (\ref{sprof.eq}) and
 we report in Table ~\ref{clparam.tab}  the coefficients  $S_0$ and $S_1$, together 
with $S_{500}$, for each of the three test clusters.
   These coefficients
are the best-fit values of the entropy profile given 
by \citet{Ghi19} for their CC subsample of X-COP clusters (their Table 3, CC entry),
 but increased by about a factor of $ \simeq 30 \%$. 
 It has been found necessary to introduce 
this offset in the coefficients  to obtain a physical meaningful solution to 
Equations (\ref{thydrosa.eq})  and (\ref{thydrosb.eq}) up to $r_{max}$.
This is because, using the original coefficients, it is not possible to propagate
the numerical solution beyond $r_{200}$, the radial profile of $P(r)$ being 
characterized by a very steep decline with radius.  As a result, at radii
slightly beyond $ r_{200}$, the pressure becomes numerically consistent with zero.

By setting $r\simeq 10^{-2} \, r_{500}$,  we obtain entropy core values for our test clusters of $S/S_{500} \simeq 3 \times 10^{-2}$
from Table ~\ref{clparam.tab}.
For cluster C1 this gives $S(r=10^{-2} \, r_{500})\simeq 28$ keV cm$^2$, still below 
 the threshold value for CC clusters \citep{Cav09}.
These core values at $r\simeq 10^{-2} \, r_{500}$  
 are in the lower portion of the observed range of entropies 
for CC clusters; see Figure 6 of \citet{Ghi19}. 

Note that for cluster C3, the coefficient $S1$ has 
been further increased by $ \simeq 70\%$, with respect the best-fit value
of $1.35$ given by \citet{Ghi19}.
This is because all of the CC clusters of the X-COP sample used by the authors
have $M_{200} >6 \times 10^{14} \msun $, whereas here cluster C3 has 
 $M_{200}=6 \times 10^{13} \msun $.
From Figure 10 of \citet{Sun09} it  can be seen that a core entropy of 
    $S/S_{500} \simeq 3 \times 10^{-2}$ at  $r\simeq 10^{-2} r_{500}$  
 is well below the observed range of entropy values reported by 
 \citet{Sun09}  for their group sample.
Finally, for cluster C1 (C3)  we have at $r=10$ kpc 
an the entropy core value  of $ \simeq 25 \sev$ ($\simeq 12\sev$), 
  which is about 20\% higher than the entropy value   
found at the same radius for the corresponding test cluster of \citetalias{ZuH11}.

Once the initial gas density and temperature profiles  are computed for the range 
 $0\leq r\leq  r_{max}$, we use them to construct the initial  
particle configuration. To this end, we store the profiles on a very fine grid. 
This in order to obtain $\rho(r)$ and $ T(r)$ at a generic radius $r$
 using grid values. 

In SPH, a non trivial issue is the realization of the gas particle distribution
which must reproduce the required density profile $\rho(r)$. 
From Equation (\ref{rho.eq}), one can see that in SPH the density at a given 
particle position depends on the masses and relative positions of nearby particles.
This in turn implies that any specific realization of particle positions which 
satisfies the prescribed profile $\rho(r)$, must be found by solving simultaneously
for all the particle positions. This is a difficult task and several methods have been 
devised to solve the problem.

A relatively simple approach, which starts from a uniform distribution and solves 
 Equation (\ref{rho.eq}) by varying particle masses, cannot be applied in SPH. 
This is because numerical instabilities are found 
 to arise \citep{Mo06} when large mass contrasts are present between particles.
Similarly, a random realization of particle positions generated from the specified 
$\rho(r)$, as we did in Section \ref{subsec:icdm} to construct the DM  particle positions, 
 cannot be used here. The noise induced by Poisson sampling implies  the development of
large fluctuations which in turn  quickly perturb the initial equilibrium profile.
 
These difficulties have lead many authors to develop alternative methods to solve 
the problem of properly generating initial conditions in SPH. These methods 
can be summarized as follows:
lattice stretching \citep{he94,Ross07}, 
viscous damping  \citep{Wa07,Price07,Pakmor12,Price18}, 
relaxation \citep{Zu86,Nag88}, space partition based either on 
 tessellation \citep{Pakmor12,raskin16,Rein17} or 
 weighted Voronoi tessellations  \citep{Diehl12,vela18,Ar19}.

Initially, we implemented the method of \citet{Diehl12} to setup our SPH initial conditions.
The algorithm is based on a Voronoi tessellation in which particles are moved 
iteratively toward a relaxed configuration. However, during the merging simulations 
the halo profiles constructed according to this procedure were found to deviate 
from the initial equilibrium solution.  In several cases this happened on a
timescale shorter than that  occurring  between the start of the simulation and 
the cluster collision.

We interpret this behavior as a direct consequence of the entropy profile we use
to construct our initial conditions. For CC clusters, the average entropy profile
 implies an equilibrium solution with a very steep density profile toward the 
cluster center. As a result, it is intrinsically difficult to keep SPH gradient 
errors under control.
We have indirectly verified that is the case by running a halo in isolation for several gigayears;
the physical parameters of the halo were those of 
cluster C1. The DM and gas  particle realization were constructed following the 
procedure just described, but the parameters of the entropy profile were 
 the best fit values extracted from a subsample of simulated  NCC clusters
\citep{V19}. In such a case, the halo gas density profile was found quite stable 
 over the whole simulation period.

%

To solve this issue, we then adopt the following procedure. Initially, gas 
particle 
positions are obtained by a radial transformation applied  to the coordinates of 
a  uniform glass distribution of points.
This is a minimal noise configuration, and it is generated by applying 
to an initial Poisson distribution of points a reversed gravitational 
acceleration together with a damping force \citep{Wa07}. Particle positions are 
advanced until a low energy state is reached.

The radial transformation is such that the new 
gas positions are consistent with the desired initial gas mass profile. 
This profile is now much 
more stable than that obtained from a random realization (Section \ref{subsec:icstab}), 
but its stability properties are not yet sufficient to consistently satisfy  
the initial condition set up required by the merging runs studied here.

To further improve the stability properties of the gaseous halo, we add to the momentum
equation a time-dependent friction term \citep{Price07,Pakmor12}
 \begin{equation}
	 \frac {d {\bf v}_{i}}{dt}= -\frac{{\bf v}_i - {\bf V}_{cm}}{ \tau_{damp}} +
	 \left( \frac {d {\bf v}_{i}}{dt}\right)_{SPH}~,
 \label{fdamp.eq}
 \end{equation}
where the second term on the rhs is given by Equation (\ref{fsphw.eq}),
${\bf V}_{cm}$ is the halo center of mass velocity, and $\tau_{damp}$ is
a time-dependent damping time scale. This is written as 
 \begin{equation}
 \tau_{damp}=\frac { \tau_{dyn} }{\alpha_{damp}}~,
 \label{tdamp.eq}
 \end{equation}
where $\tau_{dyn}= 1/ \sqrt{4 \pi G \rho_{d}}$ is the local dynamical timescale, 
$\rho_d({\bf x}_i) $ is the particle DM density, and $\alpha_{damp}$ is a friction 
parameter which controls the strength of the friction.
The DM density $\rho_d$  must be calculated at run time according to an SPH prescription, 
but one can introduce a DM to gas density ratio: 
 $\beta_{dyn}=\rho_{d}/\rho_g$. The halo stability can then be exploited to avoid the 
calculation of the DM density by using $ \beta_{dyn} \rho_g$ in place of $\rho_d$ in Equation (\ref{tdamp.eq}).
As it will be seen from the plots of 
 Section \ref{subsec:icstab}, a conservative value for $\beta_{dyn}$ is obtained 
 by setting $\beta_{dyn}\simeq50$. It has been found that this choice also has an impact on the 
value of $\alpha_{damp}$. In principle $\tau_{damp}$  should be a small fraction of 
$\tau_{dyn}$, but with the adopted value of $\beta_{dyn}$ very stable halos are already
obtained when $\alpha_{damp}\simeq3 $.

Our merging simulations are then performed by using the generalized momentum equation
 (\ref{fdamp.eq}) as the simulations start, and switching off 
( $\alpha_{damp}=0$ )  the friction parameter
 at a simulation time which depends on the initial merging 
kinematics (Section \ref{subsec:ickin} ).
Initially, we set gas particle velocities to zero and temperatures are 
assigned by interpolating grid values. These are calculated from the numerical solution and the
interpolation is done according to the radial particle coordinates.
 
Finally, we set the mass of DM  and gas particles  according to  the 
scaling $m_d\simeq8 \times 10^8 \msun (M_{200}/2 \times 10^{14} \msun)$
and   $m_g=f_b m_d/(1-f_b))\simeq 0.16 m_d $, respectively.
These mass assignments are consistent with previous findings 
\citep{V19}, in which    ICM profiles  extracted from a
set of hydrodynamical cluster simulations were found numerically converged 
when similar settings were adopted for the particle masses.
However, as outlined before, in SPH numerical instabilities
can arise in the presence of very different gas particle masses. This implies that the 
simulation
numerical resolution  is enforced by the  smallest mass of the binary system. 
We thus write
 \begin{equation}
m_d\simeq8 \times 10^8 \msun (M^{(2)}_{200}/2 \times 10^{14} \msun)~.
    \label{mdark.eq}
 \end{equation}

For cluster C1, the total number of gas particles  
  ranges then from  $N_g^{(1)} \simeq 2 \times10^5 $ when 
the collision mass ratio is $R=1:$1, up to  $N_g^{(1)} \simeq 2 \times10^6 $  
when $R=1:$10.  For cluster $C3$  one has  $N_g^{(2)} \simeq 1.27  \times10^5 $.
The gravitational softening parameters of the  particles are set according
to the scaling $\varepsilon_i =15.8\cdot ( m_i/6.2\times10^8 \msun)^{1/3}\kpc$.
 Additionally, in some test cases we run high-resolution (HR) 
simulations  in which the particle masses are scaled down by a factor $4$,
 with respect the reference value given by equation (\ref{mdark.eq}).

%

\subsubsection{Initial merger kinematics}
\label{subsec:ickin}
To construct the orbits of our merging simulations,  we choose a Cartesian system of 
coordinates $ \{x,y,z\}$, with the center of mass of the two clusters being at 
the origin. The orbits are initialized in the $ \{x,y\}$ plane at $z=0$, with 
$ \{{\bf d}^{in},{\bf V}^{in}\}$  being the initial separation and relative 
velocity vectors, respectively. Thus, the initial $ \{x,y\}$  coordinates of the two cluster center of mass
 read $-( d^{in}_x, d^{in}_y)/(1+R)$  and
$( d^{in}_x, d^{in}_y)R /(1+R)$. Similarly, the velocity components are given by
$-( V^{in}_x, V^{in}_y)/(1+R)$  and
$( V^{in}_x, V^{in}_y)R /(1+R)$. 

As already outlined, our collision parameter space is the same as in \citetalias{ZuH11}. However, there
is here a significant difference in the initial condition setup of the halos.
For the reasons discussed in Sections \ref{subsec:icdm} and
\ref{subsec:icgas}, the initial DM and gas mass profiles are continued beyond
$ r_{200}$ and extended up to $ 2 r_{200}$.  This implies that, unlike in \citetalias{ZuH11}, 
 the relative initial separation cannot be set here to the sum of the two 
virial radii, but to twice its value: $d_{in}=2( r^1_{200}+r^2_{200})$.
As discussed in Section \ref{subsec:icstab}, this is to avoid a significant 
overlap at the start of the simulation between the 
mass profiles of the two halos, which in turn would soon put the profiles out of
equilibrium.

Our initial condition vectors $ \{{\bf d}^{in},{\bf V}^{in}\}$    at $t=0$ must then
be chosen  such that, at some later simulation time $t$, the orbit of the binary 
cluster system produces the initial conditions of \citetalias{ZuH11}. These consists of a
separation $d_{in}/2$ between the two cluster centers of mass, with a collision
impact parameter $b$ and relative infall velocity $ V \simeq 1.1  \sqrt{G M_{200}/r_{200}}$. 
The latter value is justified by cosmological simulations  \citep{Viv02}.

 In order to realize these settings, we adopt a procedure similar to that described in
\citet{Poole06}.
For a specified set of initial conditions
 taken from \citetalias{ZuH11},
we first approximate the two 
clusters as point-like and accordingly assign positions and velocities to the two points.
We tag this orbital status as occurring at the time $t_f$,
 i.e. the start of the \citetalias{ZuH11} simulations.
We now numerically solve
 Kepler's problem by seeking the time $t_i < t_f$ such that the separation between the 
two points is $d_{in}$. The orbital positions and velocities at $t_i$ then complete  the 
solution vectors   $ \{{\bf d}^{in},{\bf V}^{in}\}$. 

To account for tidal distortions we first run a DM only merging simulation, using
 as initial conditions the solution vectors 
  $ \{{\bf d}^{in},{\bf V}^{in}\}(t=0)$  previously determined. 
During the simulations we denote as 
  $ \{{\bf X}_{cl},{\bf V}_{cl}\}(t)$  the center of mass position and velocities of the
two halos. These vectors are contrasted with the \citetalias{ZuH11} initial conditions
  $ \{{\bf X}_{cl},{\bf V}_{cl}\}(\tau^K)$,  which the binary system must 
reproduce at 
the simulation time $t_s= t_f-t_i \equiv\tau^K$.  
To quantify the  deviations between the specified set of initial conditions and the 
 numerical solution we define the following norms
 \begin{equation}
\left\{
\begin{array}{lcc}
	\varepsilon^{(cl)}_x(t) & = & \left. \left\Vert {\bf X}_{cl}(t) -
	{\bf X}^Z_{cl}(\tau^K)\right\Vert \right/ 
	\left\Vert {\bf X}^Z_{cl}(\tau^K)\right\Vert ~ \\
	\varepsilon^{(cl)}_v(t) & = & \left. \left\Vert {\bf V}_{cl}(t) -
	{\bf V}^Z_{cl}(\tau^K)\right\Vert \right/ 
	\left\Vert {\bf V}^Z_{cl}(\tau^K)\right\Vert ~, \\
\end {array}
\right .
   \label{norm.eq}
 \end{equation}
with $cl=1,2$ being the halo index.

We define as  position  error $ \varepsilon_x(t)$   the maximum of the two
error norms: 
$\varepsilon_x(t)  =  MAX( \varepsilon^{(1)}_x(t) , \varepsilon^{(2)}_x(t) )$, the 
velocity error $ \varepsilon_v(t)$  being similarly defined.
These errors are computed and saved at run times  $t_m=\tau^k + m \Delta t$, 
centered around $\tau^K$. We set  the  grid spacing to $\Delta t= 1/16$ Gyr and
$m$ is an integer ranging between $-20$ and $20$. 
Finally, the simulation time at which our constructed set of error values 
has a minimum, is identified as the simulation time $t=0$ in the corresponding 
merging run of \citetalias{ZuH11}. We label as $\tau^{num}$ this solution time obtained numerically.
Table \ref {kinerr.tab} lists the values of $\tau^{num}$  and $\tau^{K}$  for each
of our merging runs, together with the notation we use to label the different
simulations. 
Note that the difference $\tau^{num}-\tau^{K}$ is smallest for head-on collisions,
while it is largest ($\simeq 0.3$ Gyr) for the $R=1:$1 off-axis merger with $b=0.6$.
The simulation time $t_s$ here is then related to that of \citetalias{ZuH11} by 
the relation:
 \begin{equation}
t^Z=t_s -\tau^{num}~.
   \label{time.eq}
 \end{equation}

Our merging simulations are performed up to a simulation time
$t_s=\tau^{num}+t_{fin}$, where $t_{fin}=10$ Gyr.
We analyze simulation results when $t^Z \geq0$, at epochs spaced by $ 1$ Gyr.
To ease comparisons between our results and those of \citetalias{ZuH11},
 hereafter we will always use the simulation time $t^Z$, which
we  abbreviate as $t$. 
 
The damping factor  $\alpha_{damp}$ in Equation (\ref{fdamp.eq}) 
is  set to zero when $t\geq0$, but with some exceptions (see later). This
guarantees that at $t=0$ our halo entropy profiles possess the correct
radial behavior.
Our merging simulation suite is constructed by performing both adiabatic and 
radiative
simulations. For adiabatic runs we consider all of the collision parameter space, 
consisting of nine different merging simulations. For radiative simulations, we run only a limited number of mergers because of
the high computational cost of the simulations.

\subsection{Stability tests}
\label{subsec:icstab}
As already discussed in Section \ref{subsec:icdm}, the stability of spherically 
symmetric DM halos with an 
exponentially truncated NFW profile depends critically of how initial 
particle velocities
are assigned. According to \citet{Kaz04}, the long term halo evolution
is significantly affected if the  particle velocities are initialized using the local
Maxwellian approximation. By contrast,  much more stable halos are obtained when the initial particle energies are consistently extracted from the 
equilibrium distribution function $f(\mathcal{E})$.
However, the choice of the halo truncation parameters 
  $\xi=r_{max}/r_{200} $ and  
  $\eta=r_{decay}/r_{200}$ is  not entirely arbitrary.
In particular, an overly sharp truncation  ($ \eta \simlt 0.1$)  can 
lead to instabilities in the halo evolution \citep{zemp08,Drakos17}.
{ The solution is  use increase the truncation radius ($ \eta \simeq0.3 $), truncating 
the atmosphere more smoothly, } as 
already done in some merging runs \citep{Zh14}.

To validate our choice of the truncation parameters ($\xi,\eta$)  we studied the evolution
over cosmological timescales of three isolated DM halos. All of the halos
have $M_{200}= 6 \times 10^{14} \msun$, but their initial density profiles have
different truncation parameters $(\xi,\eta)$.  The three pairs of values 
we consider are $(\xi,\eta)=(3,0.3),(1.4,0.1)$ and $(2,0.2) $. 
We refer  to the corresponding halo realizations as DMa, DMb and DMc,
 respectively.
 Initial particle position and velocities are initialized  according to the procedures described 
in Section \ref{subsec:icdm}, and we use Equation (\ref{mdark.eq})
to set the DM particle mass to 
$m_d\simeq2.4  \times 10^9 \msun $.
The number of DM particles $N_p$ then ranges from 
 $N_p \simeq 3 \times 10^5$    for the   DMb halo, up to 
 $N_p \simeq 3.8 \times 10^5$    in the case of  the DMa halo.

\begin{figure*}
\centering
\includegraphics[width=0.95\textwidth]{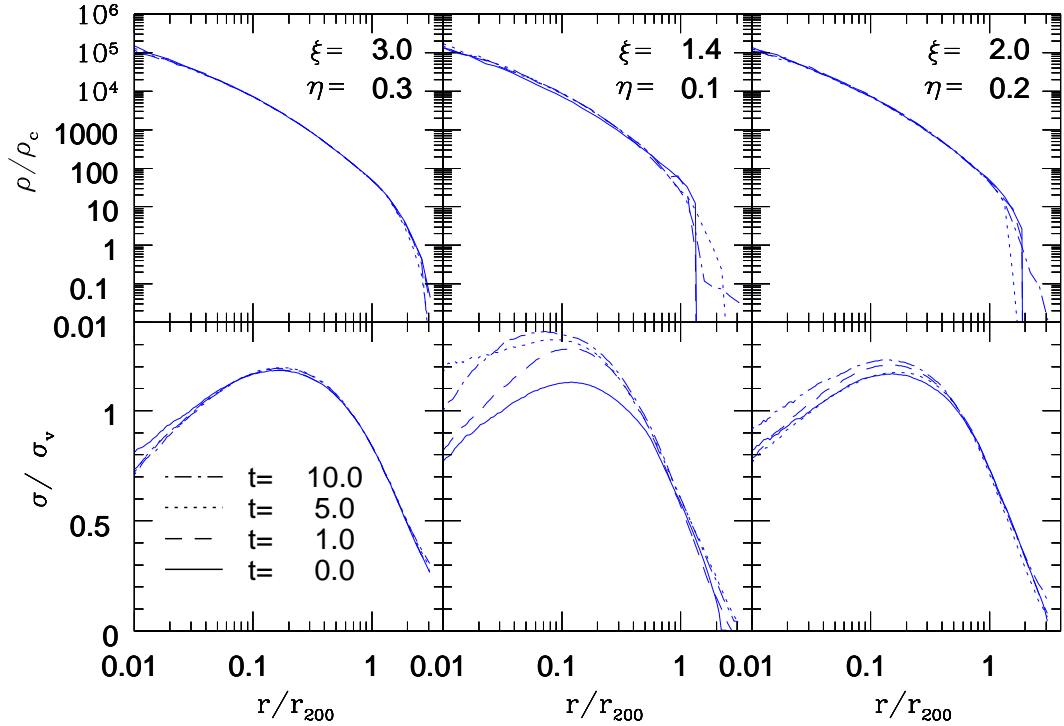}
\caption{ Evolution of the density (top panels)   and velocity dispersion  
(bottom panels) profiles for three different isolated DM halos 
(left to right). 
The halos are initially in equilibrium, with the particle distributions
realized according to the procedures described in Section \ref{subsec:icdm}. 
 All of the halos have  $M_{200}=6 \times  10^{14}\msun$,
 but their initial conditions differ in the choice of the 
truncation parameters $(\xi, \eta)$ (see text).
 Density is in units of the cosmological critical density and  velocity
dispersion in units of 
$\sigma_{200}= \sqrt{G M_{200} /r_{200}} \equiv \sigma_v$.
Different line styles refer to different epochs, 
 as indicated in the bottom left panel, where the time is in Gyr.
 (Note that the times decrease from top to bottom in the legend in this and subsequent figures.)
  \label{fig:dmhalo}}
\end{figure*}

Figure \ref{fig:dmhalo} shows the time evolution
of the density and velocity dispersion profiles for the three halos at 
four different time slices:  $t=0,~1,~5$  and $10$ Gyr.  
For the range of initial conditions we consider, the time span 
$t_{\rm hit}$ occurring 
between the start of the simulation and the direct hit between the primary and
 secondary cluster  cores (see later) 
 ranges from $\simeq$ Gyr, in the case of head-on
collisions, up to $\simeq8$ Gyr for the  $R10b06$ initial merging  configuration.
The different epochs displayed in Figure \ref{fig:dmhalo} have been chosen with 
the criterion of covering the whole range of time spans $t_{\rm hit}$.

As it can be seen, the best stability properties are exhibited by the DMa halo.
For this halo, both density and 
 velocity dispersion profiles are quite stable up to $\simeq 10$ Gyr.  This 
is in accord with previous findings \citep{zemp08,Drakos17}, and confirms 
 that setting $\eta=0.3$ is the safest choice when stability is an issue. 
However, this choice of $r_{decay}$ requires the continuation of the DM halo 
beyond  $r_{200} $ and up to  $r_{max}\simeq3 r_{200} $, if one wants  
 to avoid an abrupt truncation in the density profile.
This choice of the truncation parameters $(\xi, \eta)$ in turn implies that 
 the simulations will begin with  a significant overlap between the two
DM halos, if the  initial separation between the two clusters is chosen
to be $d_{in}= r^1_{200}+r^2_{200}$, as in \citetalias{ZuH11}.
These initial settings then put the initial mass profiles of the two halos
out of equilibrium, and it is not clear what the impact of these initial conditions is on ICM properties during cluster mergers \citep{MC07}. 

We choose here  to put the initial separation between the two cluster center 
of mass to the value $d_{in}=  \xi(r^1_{200}+r^2_{200})$, this choice
of the initial setup  being  clearly  advantageous because 
the merging runs are then performed with halos initially at equilibrium.
To avoid very large values ($\simgt10$ Gyr) of the simulation time 
$\tau^{num}$, when the two clusters orbital parameters best approximate 
the \citetalias{ZuH11} initial conditions, we decided here to use a value of $\xi$ 
smaller than that of the  DMa halo ($\xi=3$).

Figure \ref{fig:dmhalo} shows that the stability properties of the
  DMc halo, having $(\xi,\eta)=(2,0.2) $,  are much better than those of 
the DMb halo with $(\xi,\eta)=(1.4,0.1)$. There is some evolution in the 
density profile beyond $r_{200}$, but the velocity dispersion profile is
much more stable than those of the DMb halo.  To set up initial conditions 
for our DM halos, we thus adopt as truncation parameters the pair of values
 $(\xi,\eta)=(2,0.2) $. 

{ This choice is motivated by the criterion of having the DM halos
be as stable as possible, but without extending them very far beyond $r_{200}$.}
It must 
be stressed that this choice  does not necessarily imply that 
final gas profiles are  significantly affected by using the pair 
$(\xi,\eta)=(1.4,0.1)$. For instance, Figure 2 of \citetalias{ZuH11} shows little 
evolution in
 the  gas profiles of an isolated halo, although the DM component is 
initialized by setting   $\eta=0.1$.

We now investigate the stability properties of halos which contain both
DM and gas. As for the DM only tests, we always set the  halo mass 
 at $r_{200}$  to $M_{200}= 6 \times 10^{14} \msun$. 
 We setup the gas density and temperature 
profiles according to the procedure described in Section 
\ref{subsec:icgas}, the entropy profile parameters being those of cluster C1. 
All of the  halos have then the same physical parameters and analytical
profiles. The total DM and gas halo masses at 
  $r_{max}=2 r_{200} $ are then 
 $M^h_{DM}  \simeq 6.6 \times 10^{14} \msun $ and  
 $M^h_{gas}  \simeq 9.2 \times 10^{13} \msun $, respectively.  
Accordingly, from  Equation (\ref{mdark.eq}) the number of DM (gas) particles is
 $N_p\simeq 2.7 \times 10^5 ~(  2 \times 10^5)$.

We initially  consider two different particle realizations of the initial gas 
density profile.  The first halo (RN) has gas particle positions drawn
from a uniform random distribution. This is the simplest approach to 
realize the desired density profile, but for the reasons discussed in 
Section \ref{subsec:icgas}, the stability of its gas profiles 
can be considered very poor.
Thus, we use the profile evolution of this halo 
realization as a benchmark, against which to assess the stability properties 
of other procedures.
For the second halo (GL) gas positions are obtained by 
transforming the radial coordinates of a  glass-like configuration  
 of points. The transformation is consistently done by numerically solving for the radial coordinate of each particle that satisfies the requested mass profile.

\begin{figure*}
\centering
\includegraphics[width=0.95\textwidth]{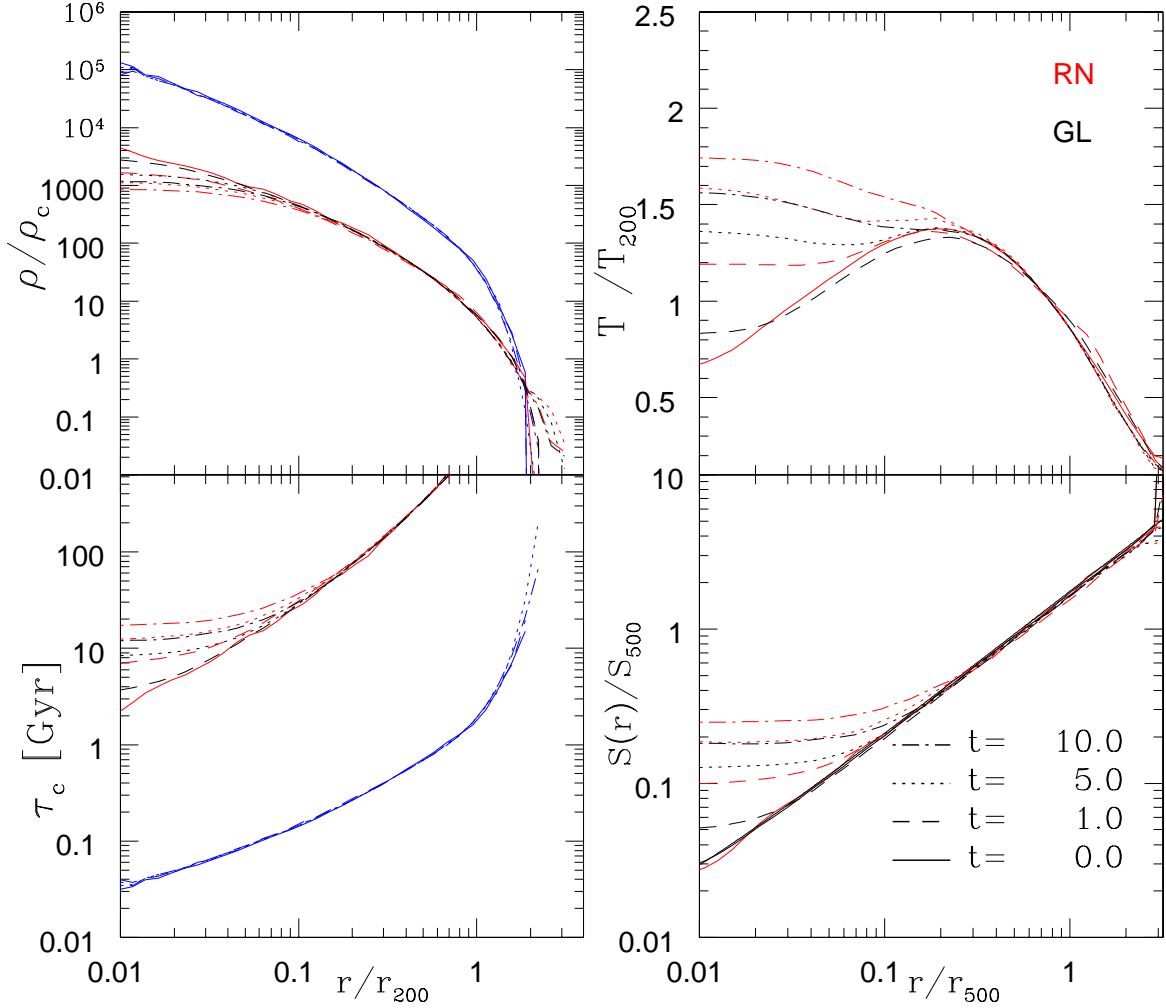}
\caption{
Time evolution of several gas profiles for two isolated gas+DM halos.
Both halos have the same physical parameters and
 $M_{200}=6 \times  10^{14}\msun$.  
  The initial gas density and temperature profiles 
are  derived from the entropy profile as given by Equation \ref{sprof.eq},
and the parameters $S_0$ and $S_1$ are those of cluster C1 
(see Table \ref{clparam.tab}).
From the top left going clockwise: the gas density, temperature, entropy and 
	cooling time $\tau_c$
(equation~\ref{tc.eq}).
As in Figure \ref{fig:dmhalo},  different line styles refer to different 
epochs.
The two halos differ in the realization of the  gas particle distribution
used to model the initial gas density profile
$\rho_g^{(in)}(r)$.
We consider an initial setup in which gas particle positions are randomly drawn (RN) to  obtain $\rho_g^{(in)}(r)$, whilst in the  other setup  
the gas positions are obtained by radially stretching a uniform glass point 
distribution (GL).
 In each  panel, red (black) lines are for the RN (GL) run.
 In the left panels, the blue lines refer to the halo DM density profile 
(top) and to the local dynamical time $\tau_{dyn}(r)$ (bottom).
For the sake of clarity these profiles are shown only for the RN run.
Similarly, the gas  profiles of the GL halo at $t=0$  are 
not depicted.
Entropy is normalized to $S_{500}$, as given by Equation  (\ref{snorm.eq}), and 
 $T_{200}$ is the mass-weighted temperature within $r_{200}$.
In the bottom right panel the solid black line indicates the analytical 
entropy profile (\ref{sprof.eq}).
  \label{fig:ichaloa}}
\end{figure*}

 For the two halo realizations, Figure \ref{fig:ichaloa} shows the time evolution of the gas density, temperature, and entropy profiles.  The temperature 
is in units of $T_{200}$, the mass-weighted temperature within $r_{200}$, and
entropy in units of $S_{500}$.
Additionally, we also show the radial profiles
of the dynamical time in gas and the cooling time
(defined in equation~\ref{tc.eq} below).

As expected, the plots clearly show the very poor stability properties 
of the RN halo.
On the other hand, there is some improvement when using glass-like initial
 conditions. The entropy profile $S_{GL}(r)$ exhibits a better stability, 
with deviations from the initial reference profile systematically smaller 
than in the RN case. At $t=1$ Gyr there are small deviations in the 
very inner region ($ r\simlt 0.02 r_{500}$), and at $t=10$ Gyr the 
profile $S_{GL}(r)$ is similar to that of $S_{RN}(r)$ at  $t=5$ Gyr.

 
These results demonstrate that in order to improve the profile stability of our
SPH particle realization, one must resort to more sophisticated methods.
As outlined in Section \ref{subsec:icgas}, the use of a relaxation method
\citep{Diehl12} was found to improve the profile stability, but not 
in a very significant way with respect the GL run.
Motivated by previous findings \citep{Price07,Pakmor12}, in 
order to keep the initial configuration in equilibrium we then add a 
time-dependent damping force to the SPH momentum equation.  The procedure  and
the parameter settings are described in the previous Section.

\begin{figure*}
\centering
\includegraphics[width=0.95\textwidth]{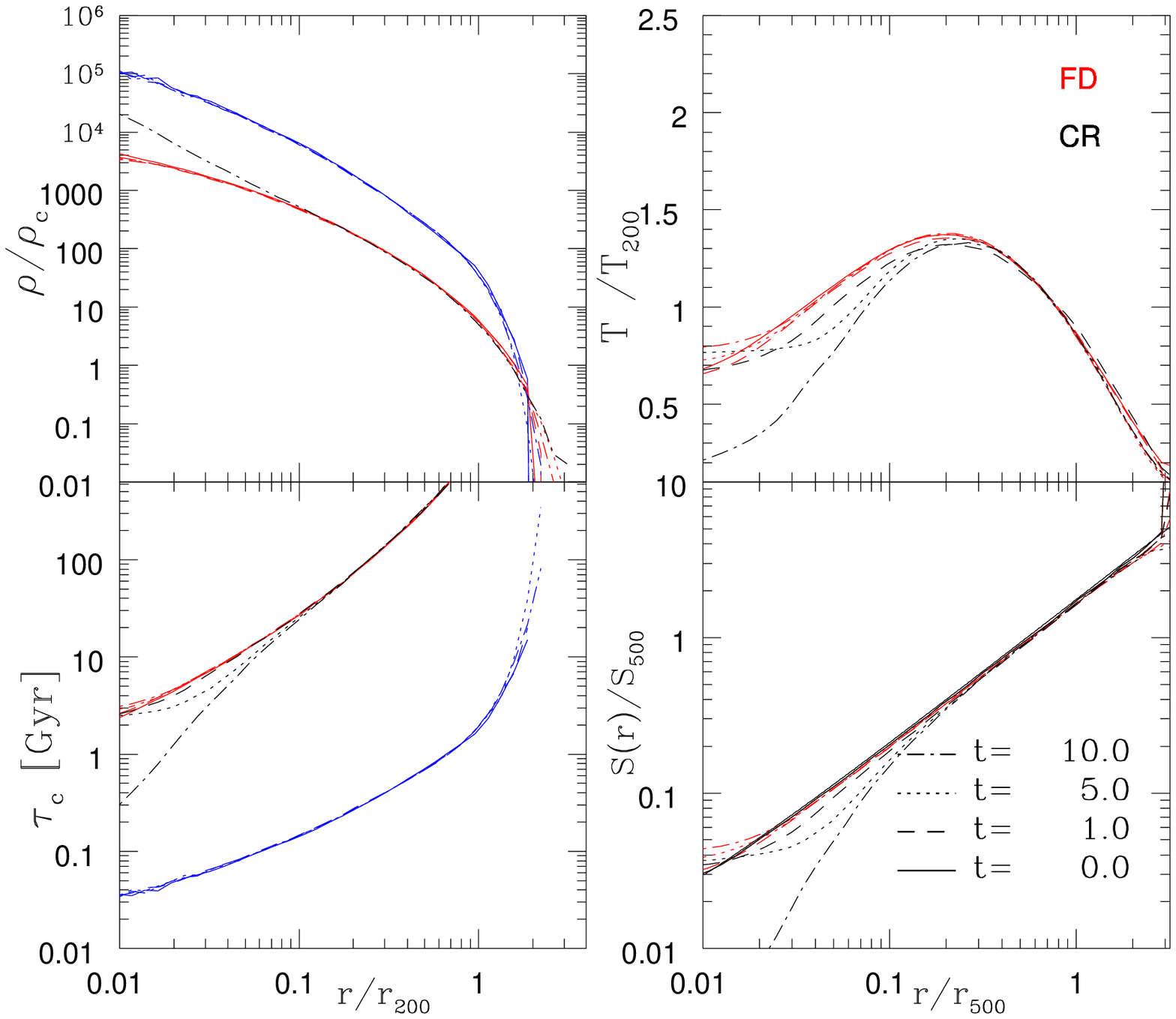}
\caption{As in Figure \ref{fig:ichaloa}, time evolution for two 
isolated gas+DM halos. The FD profiles refer to  an halo
initialized as the GL halo of Figure \ref{fig:ichaloa}, 
but with a friction term (\ref{fdamp.eq}) added to the SPH 
momentum equation. The CR halo has the same initial conditions of the RD 
halo, but its gas evolution is followed by including radiative cooling
in the SPH equations  and without the presence of the damping 
term (\ref{fdamp.eq}). For the sake of clarity, in the top left panel 
the gas density profile of the CR halo  is shown only for $t=10$ Gyr.
  \label{fig:ichalob}}
\end{figure*}

For this test case, which we label as FD,   Figure \ref{fig:ichalob} shows 
the time  evolution of the different gas profiles. The meaning of the 
different panels and lines being the same of Figure \ref{fig:ichaloa}.
The profile evolution clearly indicates that the damping method is very good 
in maintaining  the stability of the initial SPH particle realization, 
and in turn the gas profiles.
We have verified that this behavior holds for clusters C2 and C3 as well.

 Accordingly, we implement this setup procedure to construct stable gas profiles. The initial particle positions are extracted from a
 uniform glass-like distribution, as for the GL halo.  The hydrodynamic
SPH force equation is then generalized in Equation (\ref{fdamp.eq}) to 
incorporate a friction term.  The latter is present from the start of the 
simulation ($t_s=0$) up to the time when the binary system has reached 
 the optimal configuration aimed at reproducing the \citetalias{ZuH11} initial conditions 
($t_s=\tau^{num}$). After this epoch ($t=t_s-\tau^{num}\geq0$) the friction 
term is switched off ($\alpha_{damp}=0$) in the momentum equation.
With these settings, we can consistently 
compare our simulation results with those of \citetalias{ZuH11}, having realized the 
 same cluster orbital and gas profile initial conditions

However, it must be stressed that in some mergers the clusters will come 
in contact having  higher core entropies  than  those initially 
specified. {We define the time span $t_{\rm hit}$ as that occurring between the start of 
the simulation and when the two clusters cores collide or interact strongly.  For our 
head-on mergers ($b = 0$), we find that this is well approximated by the epoch when 
the distance between the two cluster centers of mass is smaller than $r^1_{200}$.  
However, for our offset mergers with $b = 0.3$ or $0.6$, we find that the secondary 
core passes by the primary core without being significantly affected during the first 
pericentric passage.   After the secondary reaches the apocenter, it falls more directly 
into the primary core.   Thus, this second encounter is nearly head-on, and we therefore 
apply the same definition as for $b=0$ to this second encounter.  Empirically, we find 
that this timescale does approximate the time when the secondary core is significantly 
affected.}
For example, for the $R10b06$ merging run 
one has $t_{\rm hit}\simeq  8$ Gyr. 

In general, we find that there is some small evolution in the inner 
($r\simlt 0.1 \, r_{500} $) level of initial ($t=0$) entropy 
 of the primary cluster when $t_{\rm hit}\simgt 4$ Gyr. 
This shows that long term stability in the 
initial  profiles is not always achieved, even after the application of a 
friction term to the motion of the SPH particles. 
To assess the impact of this behavior on the final ($t=10$ Gyr) entropy
profile  of the merged clusters,  we performed some of our merging simulations 
with the friction term still active up to $t=t_{\rm hit}$. 
These runs will be discussed in detail later;
unless otherwise stated in the following, we will 
discuss merging simulations in which 
 the friction term $\alpha_{damp}$ is set to zero when $t\geq0$.

Finally, we also show in Figure \ref{fig:ichalob}  the time evolution of the 
different gaseous halo profiles when the SPH entropy equation (\ref{aen.eq})
incorporates radiative cooling.
Following \citetalias{ZuH11},
we adapt the bremsstrahlung cooling time approximation:
\begin{equation}
\tau_{c}\simeq28.7 \, {\textrm{Gyr}} \left ( \frac{S}{100 \, \textrm{keV} \, \textrm{cm}^2}\right) ^{1/2}
\left( \frac{n_e}{10^{-3} \, \textrm{cm}^{-3}} \right)^{-2/3}.
\label{tc.eq}
\end{equation}
The test runs with radiative cooling are indicated as CR in the panels. 
As for the FD runs, the initial conditions are 
the same as for the GL halo, but here the damping term is absent in the momentum 
equation (\ref{fdamp.eq}). These settings allow us to assess the impact of
radiative cooling on the thermal evolution of an isolated halo initially in 
equilibrium.
The cooling time profiles in Figures~\ref{fig:ichaloa} and \ref{fig:ichalob}
show that the condition 
$\tau_c >> \tau_{dyn}$ is always satisfied at all radii,
thus suggesting that radiative processes are
not very important dynamically
(e.g., motions induced by cooling will be very subsonic).

The results indicate that  in the halo inner regions 
($r\simlt 0.1 \, r_{500} $), radiative losses become significant on time scales 
$t^h_{rad}\simgt 5$ Gyr, in accordance with the range of cooling times 
$\tau_c(r)$ 
displayed by the CR halo in the bottom left panel.
From Table \ref{kinerr.tab} one can see that the condition 
$t^h_{rad}>> \tau^{num}$  is not always satisfied, this in turn implies 
that  for some merging runs with cooling the entropy profile at 
$t_s=\tau^{num}$ ($t=0$)  will not satisfy the prescribed initial conditions.

To construct the initial setup for the merging simulations 
with cooling, we then  proceed as follows. 
The simulations are performed up to $t_s=\tau^{num}$ as in the adiabatic case, 
with the friction
term present and,  in particular, the cooling term $Q_R$ in equation  
 (\ref{aen.eq})  switched off.  This guarantees that both adiabatic and
radiative  simulations will start at $t=0$ with the same profiles.
After this  epoch the radiative merging runs are performed with the
 $Q_R$ term now present in Equation (\ref{aen.eq}) and 
the damping term   switched off.

Note that the previous discussion about the appropriate level of core entropy when 
$t_{\rm hit}$ is large (say $ \simgt 5$ Gyr), is not relevant here.
This is because the time evolution of the CR halo profiles demonstrate that core heating due to numerical effects is subdominant with respect radiative 
losses.

%

%
\section{Results}\label{sec:results}

In this Section, we present our main results from the simulations we performed.
We first discuss results from adiabatic simulations and subsequently those 
obtained from the cooling runs. Our findings are
qualitatively discussed 
 in light of the impact on the final entropy profiles of the different 
merging processes we consider.

\subsection{Adiabatic runs} 
\label{sec:ad}

\begin{figure*}
\centering
\includegraphics[width=0.95\textwidth]{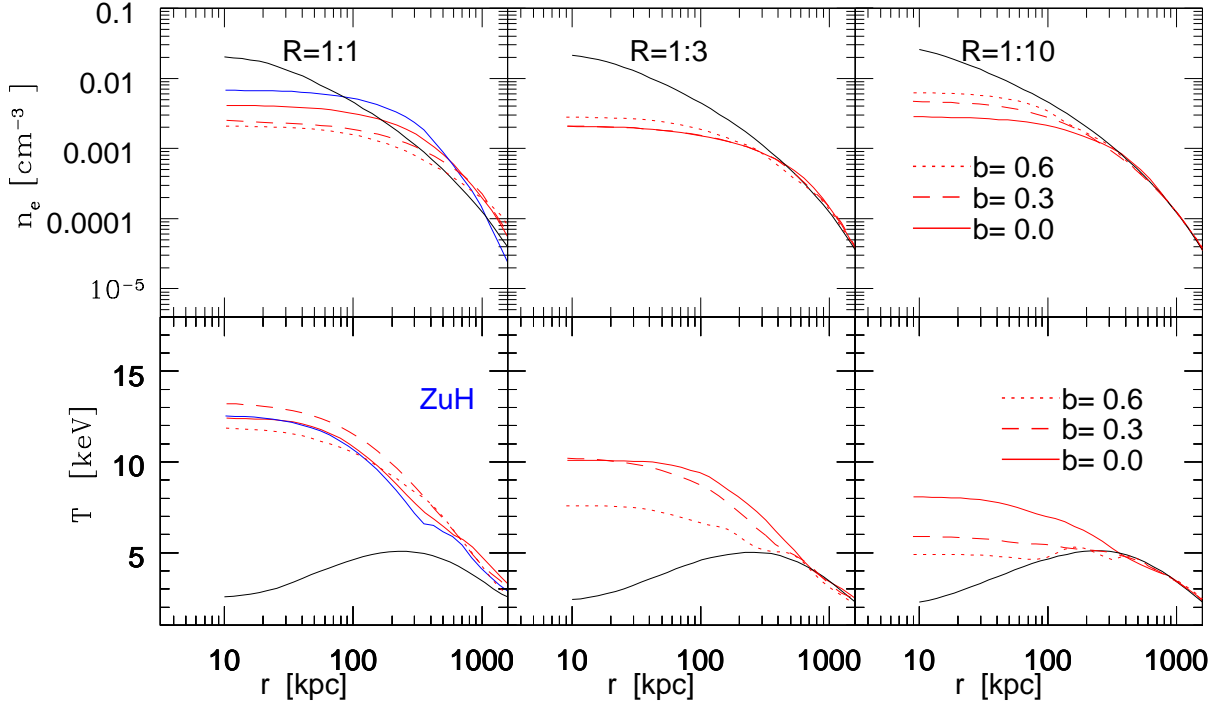}
\caption{Final gas density (top panels) and mass-weighted temperature
(bottom panels) profiles at $t=10$ Gyr.
From left to right, the mass ratios are $R=1:1, 1:3$ and $1:10$. 
Each bottom panel refers to the same  cluster mass ratio $R$ as the 
corresponding top panel.
 In each panel, the different line 
styles refer to merging runs with different impact parameters, while
 the solid black line shows the initial profile at $t=0$ Gyr for the more 
massive subcluster. 
The origin is centered at the location of the gas density 
peak of the final merged system.
 The solid blue lines in the left panels are the  profiles extracted 
at $t=10$ Gyr 
from a merging run with $R=1$:1 and $b=0$, but with the  initial entropy 
profile being given as in \citetalias{ZuH11}.
 \label{fig:adrho}}
\end{figure*}

For the nine merging simulations we show in Figure \ref{fig:adrho}
the final gas density and temperature  profiles of the resulting merging clusters.
The plots are depicted at $t=10$ Gyr, an elapsed time since the start
of the collision which should be sufficiently large to guarantee a relaxed state
for all of the considered merging configurations.

{The  radial profiles are calculated for each radial bin
by spherical averaging the extensible physical quantities.
These are total electron number for the average electron density, total thermal energy
divided by 3/2 of the total particle number for the gas temperature, and total
physical entropy (adding up the specific physical entropy per particle $s$ for all the
particles, dividing by the total number of particles, and converting to the average
entropy parameter $S$ as discussed as discussed at the start of Section  
\ref{subsec:icgas})
 for the entropy parameter.}

As in \citetalias{ZuH11}, for each physical quantity  we have subdivided the plots by showing 
in each panel of Figure \ref{fig:adrho} profiles extracted from merging runs 
with the same mass ratio but different impact parameters.
This layout is common also to the other Figures, and allows
a better comparison with previous findings.

 The radial behavior of the final density profiles depicted in 
Figure \ref{fig:adrho} exhibit the common feature of a flattened density 
($ n_e \simeq 2\times 10 ^{-3} \, {\rm cm}^{-3}$) at cluster 
radii $r \simlt 300$ kpc.
This flattening is in sharp contrast with the initial density profiles, which are 
constructed so as to reproduce that of cooling flow clusters. The 
initial profiles   steadily increase toward the cluster centers and 
 have  much higher central densities
 ($ n_e \simeq 2 \times 10 ^{-2} \, {\rm cm}^{-3}$).

Similarly, the final temperature profiles no longer show the initial 
inversion and steadily increase toward the cluster centers.
 There is a wide range of central temperature values, from $ \simeq 5 \kev$
up to  $ \simeq 15 \kev$, depending on the mass ratio $R$ and impact 
parameter $b$ of the merging simulation.

These findings strongly suggest that the initial cool-core cluster
configurations do not survive the impact on the gas of the processes 
 that occur during the collisions. This issue is central to the paper 
and will be addressed later, when discussing the final entropy profiles.

It is instructive to compare the profiles of Figure \ref{fig:adrho} 
with the corresponding ones shown by \citetalias{ZuH11} (Figure 15 and 16 of his paper).
There are strong similarities, but also interesting differences. 
In general, both density and temperature profiles   have the same 
 radial behavior as the corresponding profile of \citetalias{ZuH11}. In particular, for a 
specific mass ratio $R$, the hierarchy of the profiles at any given radius $r$ 
is always reproduced. This is reassuring because it validates our setup 
procedure and the code we are using.
  
Nonetheless, when contrasted against \citetalias{ZuH11} values, 
the  central temperatures  are found  smaller by a factor
lying in the  range $ \simeq 20 -30 \%$.
Differences in the final profiles of thermodynamic quantities  between
our simulations and those of \citetalias{ZuH11}  can be attributed to a number of 
reasons. To be specific, the largest impact will be caused by
differences in the setup of the initial cluster 
kinematic and physical parameters, and by the different numerical hydrodynamical 
schemes used to perform the simulations.
The latter can be significant, and in order to pin down its 
impact it is necessary to reduce as much as possible the effects of differing 
initial conditions.

To this end, we use as reference the merger with $R=1:$1 and $b=0$.
This merging configuration has the advantage of having a very short
collision time ($t_{\rm hit} \simeq 2 $ Gyr), so that differences between 
our initial orbital settings and those of \citetalias{ZuH11} are minimized. In what
follows, we will  refer to this simulation in brief as $R1b0$.

For a better comparison of our merging simulations with those of \citetalias{ZuH11},
 we perform a head-on merging simulation with mass ratio $R=1:$1 and
initial conditions
constructed as follows:  The  DM halo of each of the two clusters has a mass of
 $M_{200} = 6 \times 10^{14} \msun$, equal to that of cluster C1  in 
 Table ~\ref{clparam.tab}, but we set the cluster radius to 
 $r_{200}\simeq 1.55$ Mpc. This value is that reported in Table 1 of \citetalias{ZuH11}
for his cluster C1, and its a bit smaller ($\simeq 10\%$) 
than our corresponding  value ($r_{200}\simeq 1.76$ Mpc)\footnote{This difference is due to an error in the reported
value of $r_{200}$, J.A. ZuHone private communication}. Note that the
concentration parameter is the same for the two clusters ($c=4.5$).
It must be stressed that this small difference in the cluster radii has a significant
impact when comparing final results, such as entropy profiles. This is because
 small differences in the reference radius $r_{500}$ or $r_{200}$ 
(Equation \ref{sprof.eq} or Equation 1 of \citetalias{ZuH11}) induce differences in the 
 initial gas entropy at the same physical radius, which in turn imply 
 much larger differences in the final entropy profiles.
 
To construct our DM density profile we truncate the cluster at 
  a final radius $r_{max}=\xi r_{200} =1.4  r_{200}$,  and adopt a
decaying radius  $r_{decay}=0.1 r_{200}$ (as in \citetalias{ZuH11}). 
These radii are smaller than 
those adopted in our initial conditions: $(\xi, \eta) =(2,0.2)$, 
but for this merging configuration the collision time is very short   
and the considerations of Section \ref{subsec:icdm} 
about DM stability can be considered secondary.
Moreover, as in \citetalias{ZuH11} we initialize  our cluster center of mass positions
 with a relative initial separation  of $d_{in}=( r^1_{200}+r^2_{200})$.

Finally, our initial  gas  profiles are constructed according to the 
procedures described in Section \ref{subsec:icgas}, but truncating 
the profiles at $r=r_{200}$ and using Equation 1 of
\citetalias{ZuH11} with the same parameters $S_0$ and $S_1$ to specify the initial 
entropy profile $S(r)$. In the following, we refer to this simulation as 
ZuH and we will use it as our reference run against which to contrast 
 our simulation results with those of \citetalias{ZuH11}. The initial physical settings and
 kinematics of the ZuH simulation are  now identical to those of simulation S1
in \citetalias{ZuH11}, so that differences between the final thermodynamic profiles of the two 
runs can be entirely attributed to 
 the different numerical schemes used to perform the
simulations.

In the left panels of
 Figure \ref{fig:adrho},  the solid blue lines indicate the 
density and temperature profiles of this simulation extracted at $t=10 $ Gyr.   
The difference between these profiles and the corresponding ones 
 of the $R1b0$ run (solid red lines) can then be interpreted as 
originating  from  the 
different settings in the initial conditions between the two simulations.

A visual inspection shows that the difference in the ZuH density profile
and its \citetalias{ZuH11} counterpart S1 is minimal. Both of the profiles have
 the same central density
($n_e\simeq  10^{-2}$ cm$^{-3}$ at $r=10$ kpc) 
and a knee at the same radius $ r\simeq 300$ kpc.
Similarly, the  temperature profiles are also in accord.
 The left bottom panel of Figure \ref{fig:adrho}  shows a central temperature of $\simeq 13$ keV for the ZuH run, 
whereas in \citetalias{ZuH11} the central temperature  of the S1 run is $T \simeq 14$ keV.
These agreements  strongly  suggest the validity of the 
 the hydrodynamic  code used here to carry out the simulations.
We postpone further discussion of this topic to later when  we address the radial behavior of the final entropy profiles.

The final profiles of the $R1b0$ run can also be contrasted with 
 the corresponding ZuH profiles 
in order to assess the impact of 
different initial conditions and collision parameters on the final merged cluster.
In particular, Figure \ref{fig:adrho}  shows that 
the ZuH temperature profile is in good accord with the profile of its parent 
simulation, whilst it can be seen that 
in the inner cluster region ($r\simlt 300$ kpc)
the density profile $n_e(r)$ is higher than that of $R1b0$ 
by about a factor $\sim$ two and 
has a steeper  decline with radius  at $r\simgt 500$ kpc.

This difference in the final density radial behavior is a consequence 
 of two distinct effects. In the ZuH simulation, the initial entropy profile 
is the same as that of \citetalias{ZuH11}, and from Figure 15 (left panel) of \citetalias{ZuH11} it is
easily seen that this leads to a much steeper initial density profile
than that of the $R1b0$ run.  This initial difference  is not destroyed
during the merging phases and still has an impact on the density profiles
at $t=10$ Gyr. On the other hand,  at large radii the initial  ZuH 
density profile is truncated at $r_{max} =r_{200}$, half the value 
of the $R1b0$ simulation. 
As already discussed in Section \ref{subsec:icgas}, 
this implies a significant leakage of gas particles in the cluster outer regions during the merger.
Thus, the final gas density at large cluster radii will
be smaller than in the $R1b0$ run.

\begin{figure*}
\centering
\includegraphics[width=0.95\textwidth]{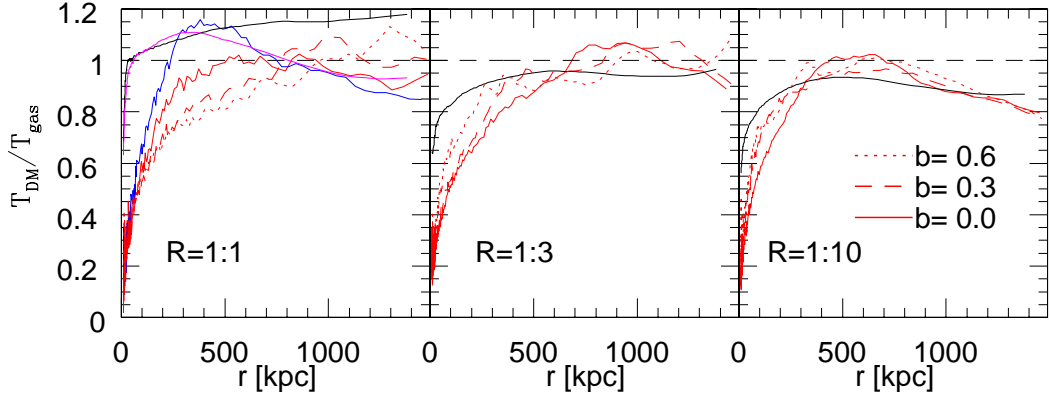}
\caption{Final profiles of the DM to gas temperature ratio $T_{DM}/T_{gas}$.
 The meaning of the  symbols is the same 
as in Figure \ref{fig:adrho}; 
the solid black lines show the initial profiles at $t=0$ Gyr.
In the left panel, the solid magenta line refers to a merging run with
$R=1:$1 and $b=0$, but having the DM halos initially truncated at 
$r_{max}=1.4 \, r_{200}$ instead of $r_{max}=2 \, r_{200}$. The blue line shows the final  profile of the ZuH run (see text).
\label{fig:adteq}}
\end{figure*}

These findings demonstrate that final differences between the gas cluster
 profiles of our simulations and those of \citetalias{ZuH11} can be entirely interpreted 
in terms of the adopted initial entropy profile. This will be confirmed
later when studying the radial behavior of the entropy profile.

\begin{figure*}
\centering
\includegraphics[width=0.95\textwidth]{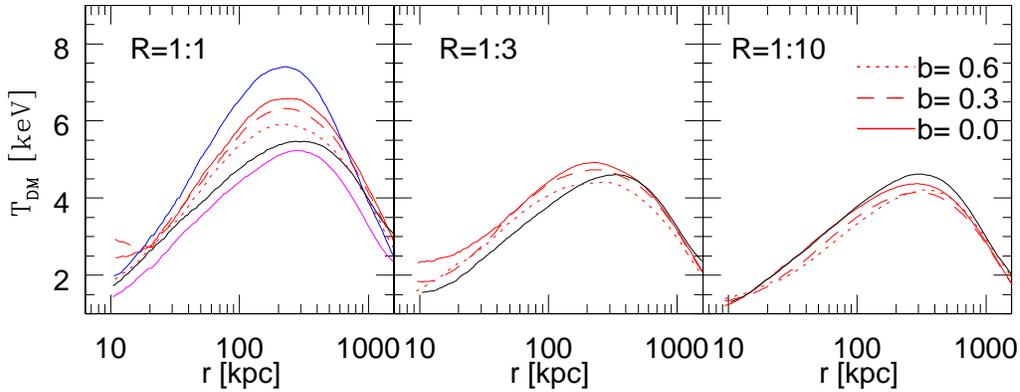}
\caption{Final profiles 
 at $t=10$ Gyr of the DM temperature $k_B T_{DM}=\mu m_p \sigma^2_{DM}/3$.
 The panel presentation, color coding, and line styles are the same as in Figure \ref{fig:adteq}. 
The solid magenta line in the left panel is the
 initial DM temperature profile of the 
merging simulation with the smaller truncation radius.}
 \label{fig:adsdm}
\end{figure*}

However, the approach used here to initialize cluster dark matter particle orbits differs in 
several ways from that of \citetalias{ZuH11}.  As can be seen in Figures 
 \ref{fig:adteq} and \ref{fig:adsdm}, this has an impact on the final DM velocity dispersion $\sigma_{DM}$ in several runs. 
For ease of comparison with the gas temperature $T_{gas}$ and previous 
findings \citepalias{ZuH11}, we introduce the DM temperature $T_{DM}$:
 \begin{equation}
k_B T_{DM}=\mu m_p \sigma^2_{DM}/3.
  \label{tdm.eq}
 \end{equation}
 
Final profiles of the DM to gas temperature ratio  
$\kappa(r)\equiv T_{DM}/T_{gas}$ are shown at $t=10$ Gyr in 
Figure \ref{fig:adteq} for the 
different merging simulations. In accord with \citetalias{ZuH11} (Figure 19), the ratio 
$\kappa(r)$ is of order unity ($ \simeq 0.9$) at all 
cluster scales.
The only exception is in the innermost cluster regions
($\simlt 300$ kpc) where the $\kappa(r)$'s tend to zero.
This is expected, since baryons in the core will raise their entropy through
mixing processes with post-shocked high-entropy material.

However, Figure \ref{fig:adteq} shows that there is a significant difference 
 between the initial ratio  $\kappa(r)$  of the head-on merger with $R=1:$1 
and the others. In fact, for the $R1b0$ merger run the initial $\kappa$ is
systematically higher by $\simeq 20$\%  compared to the other simulations.
This is in sharp contrast with the corresponding profile in Figure 19 of \citetalias{ZuH11}, 
which does not exhibit such a feature and whose behavior is in line with 
the others simulations.  

We argue that this difference can be interpreted  as originating from
the adopted initial conditions.
At variance with \citetalias{ZuH11}, we initially set the center of mass of our clusters
separated by a distance $d_{12}=2(r^1_{200}+r^2_{200})\equiv d^{in}$
The merging simulation time $t=0$ is then defined  when $d_{12}=d^{in}/2$.
This procedure then implies that at $t=0$, the two clusters have already 
had time to interact. For the gas component, the impact on the initial profiles 
of this interaction is negligible (Figure \ref{fig:adrho}), 
but for the DM halos one expects some amount of heating and an 
increase in the DM velocity dispersion. 
The strength of this effect will be weaker as the mass ratio $R$ gets higher.

 Figure  \ref{fig:adsdm} shows the radial profiles $T_{DM}(r)$, corresponding 
to the ratios depicted in Figure \ref{fig:adteq}. The left panel  ($R=1:$1)
shows that the initial profile $T_{DM}(r)$   (solid black line) is  
a bit higher ($\simeq 20\%$) than the initial profiles displayed in the other
two panels   ($R=1:$3 and  $R=1:$10), thus confirming the previous 
reasoning. In fact, this effect is significant only when $R=1:$1.

To demonstrate the correctness of this interpretation, we ran an additional
 merger simulations. As with the $R01b00$ run, we study a head-on merger 
with both cluster masses being $M_{200} =6 \times 10^{14} \msun $. At variance with
the initial condition setup described in Section \ref{subsec:icdm}, here we 
truncate the DM halos at a cut-off radius $r_{max}= 1.4 \, r_{200}$. 
We then perform the simulation and study the  final $\kappa(r)$ and 
$T_{DM}(r)$ profiles. If this heating effect depends on the cut-off radius 
 $r_{max}$, then at any given radius the final profiles of this simulation 
should approach the profiles of simulations with lower 
mass ratios. The profiles are shown (solid magenta lines) in 
 Figures \ref{fig:adteq} and \ref{fig:adsdm}, and confirm these expectations.   

Finally, a comparison with Figure 18 of \citetalias{ZuH11} shows that the final 
 profile $T_{DM}(r)$ of the ZuH run (solid blue line, Figure 
 \ref{fig:adsdm})  is in accord with the corresponding profile
of simulation S1.  Moreover, at $ r\simeq 200$ kpc the ZuH profile has 
a  peak value ($\simeq 7.2$ keV) 
that is about 15\% higher than  the peak
 of   the $R01b00$ run at the same location.  This offset between the two runs 
in the peak of the final DM velocity dispersion is interpreted as 
originating from the differences in the adopted initial conditions. 
In particular, for simulation $R01b00$ the two clusters have 
initial radii of $r_{200}\simeq 1.76$ Mpc, whilst initially 
$r_{200}\simeq 1.55$ Mpc for the ZuH simulation.

\begin{figure*}
\centering
\includegraphics[width=0.95\textwidth]{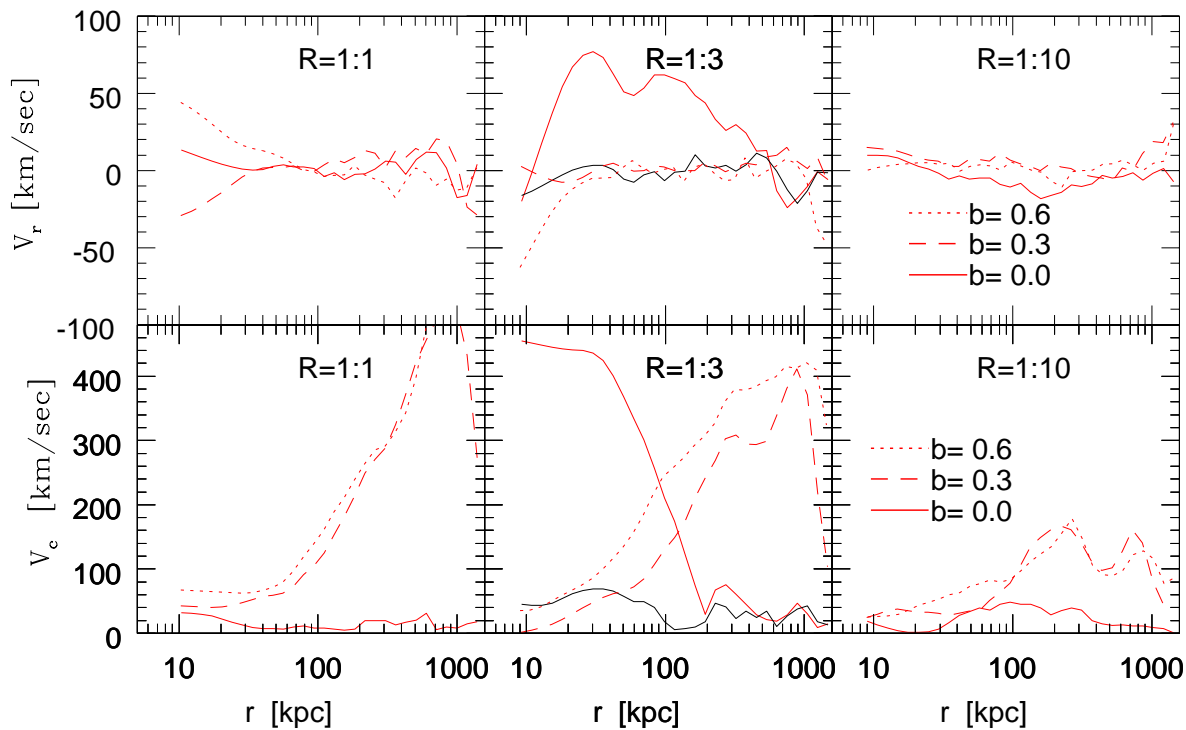}
\caption{Mean gas radial  ($V_r$, top panels) and 
 circular ($V_c$, bottom panels) cluster velocities profiles at $t=10$ Gyr
for mergers without gas cooling.
 The meaning of the  symbols is the same 
as in  Figure \ref{fig:adrho}.
The solid black lines in the middle panels refer
 to the profiles of the $R=1$:3 and $b=0$ run evaluated at  $t=11$ Gyr.
 \label{fig:advmrb}}
\end{figure*}

We show in Figure  \ref{fig:advmrb} the mean radial 
($V_r$)  and circular ($V_c$) gas  velocities profiles at $t=10$ Gyr.
The latter  is defined at the cluster radius $r$ as 
$V_c(r)=\sqrt{ \bar{v}_{\phi}^2(r)+{\bar v}_{\theta}^2(r)}$, 
where $\bar{v}_{\phi}$ and ${\bar v}_{\theta}$ are the mean 
azimuthal and polar velocities, respectively.
The profiles of Figure  \ref{fig:advmrb} can be contrasted with the 
corresponding profiles in Figures 21 and 22 of \citetalias{ZuH11}. 
All of them exhibit a radial behavior which is in accord with their \citetalias{ZuH11} 
counterparts, with the only exception being the head-on $R=1:$3 merger
(simulation S4 of \citetalias{ZuH11}). 
The final velocity profiles of this merged cluster are significantly 
different from  those of simulation S4; in particular, the mean radial 
velocity is not close to zero.
Values of $V_r\simeq 50$ km /sec persist up to 
$ r\simeq 800$ kpc. Similarly, the circular velocity $V_c$ is as high as 
 $V_c\simeq 400$ km/sec within $ r\la 50$ kpc.

These values suggest that for this merger a fully relaxed status 
has not yet been  achieved  at $t=10$ Gyr. To verify this 
possibility we have continued the simulation until  $t=11$ Gyr.
 The velocity profiles corresponding to this epoch are shown 
in the middle panels of Figure  \ref{fig:advmrb} as solid black lines, 
 and they clearly show lower velocities.

\begin{figure*}
\centering
\includegraphics[width=0.95\textwidth]{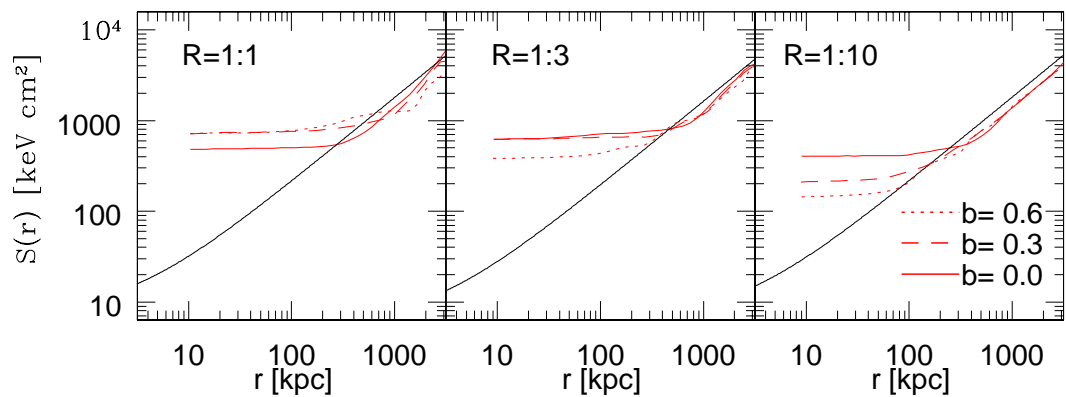}
\caption{Final entropy profiles for the nine
non-radiative merging runs at $t=10$ Gyr (red lines).
The panel presentation and line styles are identical to 
 Figure~\ref{fig:adrho}, solid black lines represent the initial entropy
	profile of the primary.
\label{fig:adentr}}
\end{figure*}

Figure \ref{fig:adentr} shows the final entropy profiles of the merged clusters. 
Note that the astrophysical entropy parameter $S \equiv kT / n_e^{2/3}$ is not the physical entropy and is not an extensive, additive quantity.
Thus, in averaging $S$ over spherical shells, $S$ was converted into the physical entropy (which is proportional to $\ln S$), and this was averaged over the spherical shell.
Then, the average physical entropy was converted back to the entropy parameter $S$.

The presence of an entropy core is common to all of the profiles, with 
its level and extent depending on the mass ratio and impact parameter of the 
simulation. A comparison with the corresponding Figure 24 of \citetalias{ZuH11} shows a 
substantial agreement in the radial behavior of the profiles, with differences
in the central levels of core entropy which can be reconciled in light
of the previous discussions.

In particular, at any specified radius and for a given mass ratio, the hierarchy
of the entropy profiles as a function of the impact parameters is strictly reproduced.
Following \citetalias{ZuH11}, differences in the various levels of entropy profiles 
can be interpreted in terms of the different amounts of entropy mixing
taking place during the mergers.

The core entropy of the primary increases during the merger owing to
the mixing of low- with high-entropy gas. This high-entropy gas is made 
available  by the secondary as it falls through the ICM of the primary and 
is ram-pressure stripped. The gas of the secondary is then efficiently
mixed with that of the primary through the development  
of Kelvin--Helmholtz instabilities. This scenario has been confirmed by
various authors in several merging simulations \citep{Tak05,Mi09,ZuH11}.

Following this line of argument, the level of core heating of the primary should 
depend on the impact parameter $b$ of the simulation. 
The higher the impact parameter, the lower is the amount of mixing.
{ This follows because in off-center collisions, the amount of gas stripped from the
secondary depends on  the ram pressure it encounters, and in turn on  the initial
mass ratio and angular momentum of the merger.
For a given mass ratio the quantity of stripped material, which is available in
the inner regions
of the primary to raise core entropy through mixing, is then expected to
depend sensitively on the orbit traced by the secondary.}

In accord with this scenario,
the third panel of Figure \ref{fig:adentr}
shows an increase in the central level of final entropy as the impact parameter
decreases.  However, this behavior is clearly seen 
for the mergers with mass ratio 1:10 (third panel) but is progressively 
less pronounced as the mass ratio $R$ becomes higher. In fact, 
 for the 1:1 mass ratio case the dependency of the final core entropy
 level on the impact parameter $b$ is reversed, i.e. the first panel of
 Figure \ref{fig:adentr} shows that 
 simulation   $R01b06$ has an higher level of central entropy than $R01b00$.  

The likely origin for this difference with
the results from the  $R=$ 1:10 merger is that in the equal-mass mergers,
a significant amount of core heating is provided dynamically by the secondary during the final merging with the primary.
This effect is almost completely absent in the $R=$ 1:10 
cases, in which the mass of the secondary is small with respect that
of the primary, and for off-axis mergers
the secondary
is totally stripped
by instabilities  before coalescing with the primary.

For a better understanding of this scenario, in Section \ref{sec:entgen}
we  present a thorough  discussion of how entropy is generated during the 
merging process. For several merging runs, we investigate in detail the time
evolution of entropy and other related quantities,
in order
to demonstrate
how the final entropy profile of the merged clusters depends critically 
on the mass ratio and angular momentum of the collision.

\subsection{Stability issues }\label{sec:stab}

As a convergence test, we compared the final entropy profiles for several merger runs with
simulations in which we varied 
the numerical resolution and/or the adopted initial conditions.

\begin{figure*}
\centering
\includegraphics[width=0.95\textwidth]{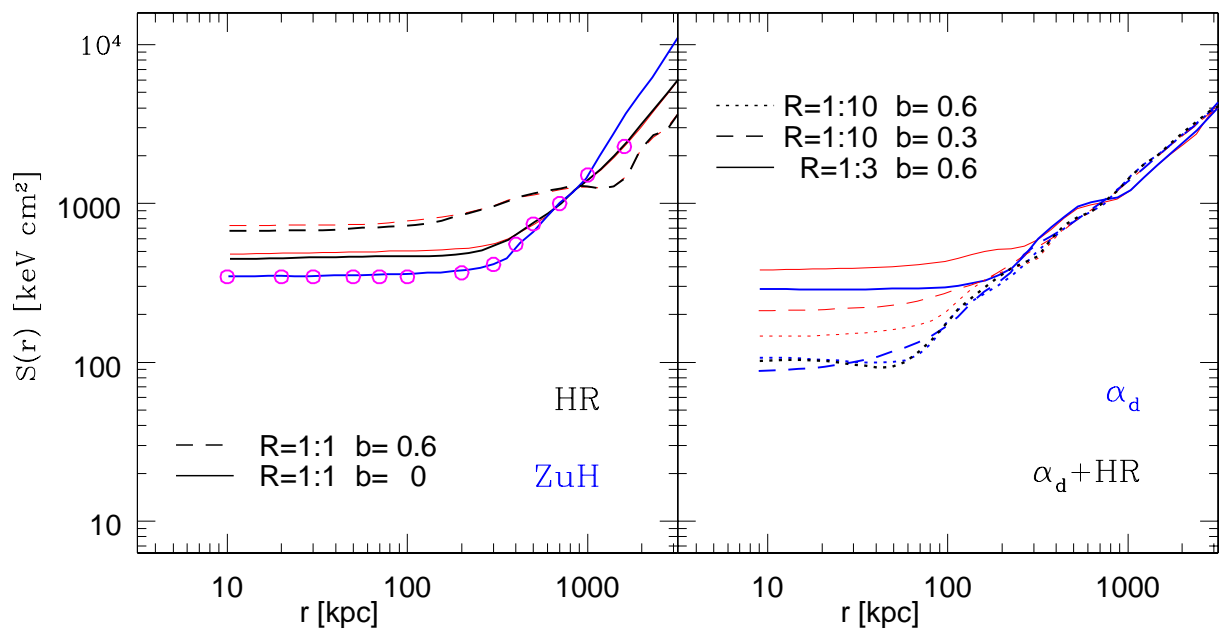}
\caption{For several merging runs, final entropy profiles 
from Figure  \ref{fig:adentr} (red lines) are 
contrasted against the corresponding profiles extracted 
from merging simulations with different numerical parameters.
Left panel: black lines labeled HR show the entropy profiles
 of high-resolution simulations, performed by using a number of particles
 about four times higher than in the baseline runs. 
 The solid blue line is the entropy profile extracted from a merging run 
 with $R=1:1$ and $b=0$, but with its initial entropy profile identical to that in \citetalias{ZuH11}.
Open circles are taken from the entropy
profile of the corresponding simulation S1, as given in Figure 24 of \citetalias{ZuH11}.
Right panel:  blue lines ($\alpha_{damp})$ refer to merging runs in which the friction term is switched off at a simulation time $t>0$ which depends
on the impact parameter $b$. 
For the merging simulation $R=1:10$ and $b=0.6$, the black dots ($\alpha_d+HR$) are from a high-resolution run that also had the damping term switched off.
 \label{fig:adHS}}
\end{figure*}

For two  equal-mass  mergers ($b=0$ and $b=0.6$), in Figure \ref{fig:adHS}
 (left panel) we show 
the final entropy profiles together with those extracted from 
the corresponding higher-resolution runs (HR, black lines). These simulations were 
performed 
by adopting the same initial conditions as the baseline runs, but with the
particle masses reduced by a factor $\sim$4. 
The plots show a radial behavior of HR profiles which is in excellent agreement 
with the corresponding standard resolution profiles, a result which leads us 
 to conclude that the simulations presented here are numerically converged.

Similarly, the entropy profile (blue line) of the ZuH run is contrasted 
with its parent simulation S1 (open circles, the points are taken 
from Figure 24 of \citetalias{ZuH11}).
There is a significant agreement between the two profiles, the only exception being 
the outermost point ($ r\simeq 1800$ kpc) for which 
 the entropy of the ZuH simulation  is higher than that of S1. 
This result is interpreted in light of the 
steepening  of the ZuH density profile at large radii
 (Figure \ref{fig:adrho}). As already outlined, this
 outer
 behavior follows from
 the  adopted initial conditions and the lack of an external buffer 
surrounding the SPH particles.

This strict agreement between the entropy profiles of two independent
simulations is very significant and it has a number of implications. 
Firstly, for a given merging configuration, it definitively shows 
that the only parameter which determines the thermodynamic structure of the 
final merged cluster is the initial entropy profile. 
The other direct consequence is that the numerical scheme used here produces,
 for the same initial conditions, a final entropy profile which is identical 
to that obtained by \citetalias{ZuH11} using the adaptative mesh refinement (AMR)  
code FLASH.
This is a non trivial issue, and consistency between 
hydrodynamical test cases  performed using Lagrangian SPH schemes and 
 mesh-based codes has been the subject of many investigations.

Specifically, \citet{Ag07}  found that the standard formulation of SPH 
\citep[SSPH,][]{Price2012} fails to reproduce the results of several hydrodynamic 
test cases, when contrasted against those obtained from Eulerian mesh based codes.
In particular, non-radiative SSPH simulation  of galaxy clusters exhibit 
entropy profiles with a power-law behavior. 
This is in sharp contrast with the constant entropy cores produced in
 Eulerian mesh simulations \citep{Mi09}.
These discrepancies are due, in part, to the intrinsic difficulty SSPH has in 
modeling density gradients around contact discontinuities, 
which in turn implies that there is a surface tension effect that
inhibits the growth of fluid instabilities \citep{Ag07}.

To address these problems, several solutions have been proposed 
\citep[][and references cited therein]{Ho15}.
In particular a possible solution is to add  a dissipative term
to the SPH thermal equation, with the purpose of smoothing the thermal 
energy at fluid interfaces \citep{pr08,wa08}. The presence of this 
 AC term has the effect of smoothing entropy transitions at 
contact discontinuities, thus  enforcing pressure continuity and in turn 
removing the artificial surface tension effect that suppresses
 the growth of the instabilities at fluid interfaces.

The SPH scheme employed here is based on this AC formulation, but the 
adopted signal velocity (equation~\ref{vsgv.eq}) is different from that originally 
proposed by \citet{pr08} and it is  better suited when gravity is present 
\citep{wa08,VA12}.
Other formulations of SPH aimed at solving these issues are the 
 SPH scheme proposed by \citet[][SPHS]{Read12}, which is based on the use of 
a high order dissipation switch, and the density-independent scheme
of \citet[][DISPH]{Sa16}.
To validate these numerical schemes,    
it is important to assess the degree of 
consistency between the level and radial extent of the  core entropies 
produced  by these codes in cluster simulations.
To this end, radial entropy profiles extracted from galaxy cluster simulations 
 can be contrasted with the corresponding ones 
obtained from their AMR counterparts.

On this issue, the results reported in the literature show the absence of a 
general agreement between the various final entropy profiles.
From DISPH simulations of galaxy clusters, \citet{Sa16} obtain final levels of 
core entropies  which are intermediate between results from SSPH and those from 
AMR codes. Their entropy levels are also in accord with those obtained using 
 the moving mesh scheme AREPO \citep{Sp10}, or the meshless code GIZMO 
\citep{Ho15}.  These schemes are both based on Riemann solvers.

\citet{Semb16} carried out a systematic comparison between the final entropy 
profiles extracted from a suite of simulations of an individual cluster. Their
simulation set is constructed by using different codes. Their results showed
that a flat inner entropy profile, such as that obtained using the AMR code 
RAMSES \citep{Tey02}, is similarly formed  in cluster simulations  
produced using SPH variants which are based on some form of 
artificial dissipation. In particular, both  the improved SPH code of 
 \citet{Be10} and the SPHS scheme \citep{Read12} give entropy profiles 
 in accord with mesh-based results.
The AC implementation of the former SPH scheme is very similar to the 
one employed here, thus reinforcing the consistency 
between our ZuH entropy profile and that of the corresponding S1 run of \citetalias{ZuH11}.

It must be stressed that the core entropy level and
size of the core are mainly regulated by the maximum value $\alpha_{MAX}^C$ 
of the AC particle parameter $\alpha^C_{i}$. 
For the simulations presented, here we set $\alpha_{MAX}^C=1.5$, 
this upper limit  being derived from the consistency of self-gravity 
tests with mesh results \citep{VA12}.
This limiting value is also  in accord
with the DISPH runs of \citet{Sa16}, who concluded that a core entropy is
established when $\alpha_{MAX}^C\simgt1$.

\citet{pow14} criticized the AC formulation of SPH;
based on the results from \citet{wa08},
they suggest that the AC scheme may not always achieve numerical convergence.
However, the HR profiles of Figure \ref{fig:adHS} are 
fully converged and  do not support this view. We  argue that it is the
adopted method to estimate gradients using a matrix inversion  
that is more relevant in this context. As demonstrated in \citetalias{V16}, 
our scheme is seen to exhibit excellent convergence properties.

Finally, it must be emphasized the strict agreement between the final 
entropy profile of our ZuH test run with the corresponding S1 profile of \citetalias{ZuH11}
does not imply that the produced core entropy levels are correct.
It just demonstrates that the two codes consistently obtain the same results, 
when adopting the same initial conditions. It remains unclear which is the 
correct core entropy level in these sort of simulations,
with SSPH lacking of any mixing process 
and Eulerian codes having the tendency 
to overestimate mixing effects because of numerical diffusion \citep{Sp10}.

In the right panel of Figure \ref{fig:adHS}, 
we analyze the consistency of our setup procedure
for several merger configurations.
Specifically, the friction parameter $\alpha_{damp}$ 
 introduced in Section \ref{subsec:icgas}  
is switched off when the two cluster
orbits have reached the initial conditions of \citetalias{ZuH11} ($t=0$).
This friction term is introduced to maintain
a stable 
realization of the CC entropy profile before the occurrence of the
cluster collision. However, in merging simulations with large angular momentum, 
the time interval between $t=0$ and the direct collision can be large 
($\ga 5 $ Gyr).  For these mergers, a certain amount of numerical 
heating can modify the core of the original entropy profile before the collision 
(see Figure \ref{fig:ichaloa}),  
thereby increasing  the final level of core entropy.

In order to assess the impact of this effect on the final entropy profiles,
we ran three additional merger simulations.
Among the simulation suite. we have chosen three merging configurations with the 
criterion of having the highest angular momentum. 
These are the two merging simulations 
with mass ratio $R=1:10$ and impact parameters $b=0.6$ and $b=0.3$, and 
 the simulation with $R=1:3$ and $b=0.6$. For these simulations, the damping 
parameter $\alpha_{damp}$ is switched off at a simulation time $t_{\rm hit}>0$ 
and not at $t=0$. 
This procedure guarantees that the initial entropy profile of the primary 
cluster maintains its form for a certain period of time after $t=0$.
Henceforth, we will generically refers to these simulations as 
 $\alpha_{d}$.

The choice of the time $t_{\rm hit}$ is a compromise between the need to avoid the possible numerical heating of the primary cluster core, 
and at the same time, to not damp significantly 
the primary core's heating due to entropy mixing driven by dynamical interactions with the secondary. 
As outlined in Section \ref{subsec:icstab},
we define $t_{\rm hit}$ approximately as the  epoch
when, after the first pericenter passage, 
 the distance between the two clusters center of mass becomes
 { smaller than $\simeq r^1_{200}$}.   

Our estimates give $t_{\rm hit}\simeq9$ Gyr for the 
merging simulation $R10b06$,  and $t_{\rm hit}\simeq5$ Gyr
in the case of the $R10b03$ and $R03b06$ merging runs. 
Clearly, a higher value 
of $t_{\rm hit}$ leads to a longer time required for the damping term to keep the entropy 
profile stable.

The first value of $t_{\rm hit}$ corresponds to the maximum required
period of damping, whereas in the case of off-axis mergers
$t_{\rm hit}\simeq5$ Gyr constitutes an approximate lower limit to $t_{\rm hit}$.
This choice of different off-axis merger cases allows us to assess the
impact of numerical heating on the entropy profile of the final merger
remnant.

By comparing the final entropy
profiles of these simulations ($\alpha_d$, blue lines) in the right panel of Figure \ref{fig:adHS} with the corresponding 
ones in Figure \ref{fig:adentr}, we see that a certain amount of numerical
heating is always present.  All of the $\alpha_{d}$ profiles  have core 
entropy levels  systematically smaller than their standard counterpart. 
For instance, in the case of $R03b06$  the final central entropy of the 
$\alpha_{d}$  ran is smaller by about $ \simeq 30 \%$. For the other two 
 merging simulations the difference is even higher,
being almost a factor of two in the case of 
$R10b03$. 
Note that the core entropy levels  of the $\alpha_d$ runs are now in 
better agreement with the corresponding ones displayed in Figure 24 of \citetalias{ZuH11}.

Moreover, to demonstrate that the final profiles are numerically 
converged, for the simulation $R10b06$,  we also run a high resolution
simulation ($\alpha_d$+HR, black dots). In Figure \ref{fig:adHS},
 it can be seen that the entropy profiles of the two simulations are 
almost coincident, thus confirming that our simulations are not affected by insufficient 
resolution. 
 
These results demonstrate that our setup procedures as described in Section 
\ref{subsec:icsetp} are not entirely free of relaxation effects, 
 with some amount of numerical heating being present in the final
entropy profiles of the merged clusters. 
The result of the  $\alpha_{d}$  run with $R=1:3$ and $b=0.6$ suggests
 that this effect leads to  an overestimate 
of the final core entropy level by $ \simeq 30 \%$. 
For the merging run $R10b06$ the increase is similar ($ \simeq 50 \%$),
whereas for the merging simulation $R10b03$ the difference in the central
entropy values is larger,
$\sim 100$ keV cm$^2$.

Thus, we conclude that with the setup procedure adopted here, 
in merger simulations with high angular momentum and a $1:$10 mass ratio there
is the tendency to overestimate final entropy in cluster cores by about
  $ \simeq 30 \%$.
This relaxation effect can be compensated for by switching 
off the damping term at a later time $t_{\alpha_d}>0$; however, the correct
implementation of this correction requires a careful choice of 
 $t_{\alpha_d}$ to avoid over-correction leading to artificially lower entropy levels.

\subsection{How is central entropy generated  during mergers? }
\label{sec:entgen}

In this Section, we examine the time evolution of entropy and other related quantities for several merger simulations.
The analysis is aimed at investigating the origin and amount of final
central entropy as a function of the initial merging parameters. 

To this end, we construct radial profiles of gas density
and entropy at given times. 
The  radial profile of a given quantity is  calculated 
for each radial bin by averaging grid values computed on spherical shells,
defined by  a set of $(\theta,\phi) = 40 \times 40$
grid points uniformly spaced
in $\cos \theta$ and $\phi$.
Unlike in Section \ref{sec:ad}, the origin of the shells 
is centered at the gas density peak of the primary, tracked at run time. 
For equal-mass mergers we
adopt the convention of defining as primary the cluster on the left in the 
$ \{x,y\}$ plane  of the collision.

\begin{figure*}
\centering
\includegraphics[width=0.95\textwidth]{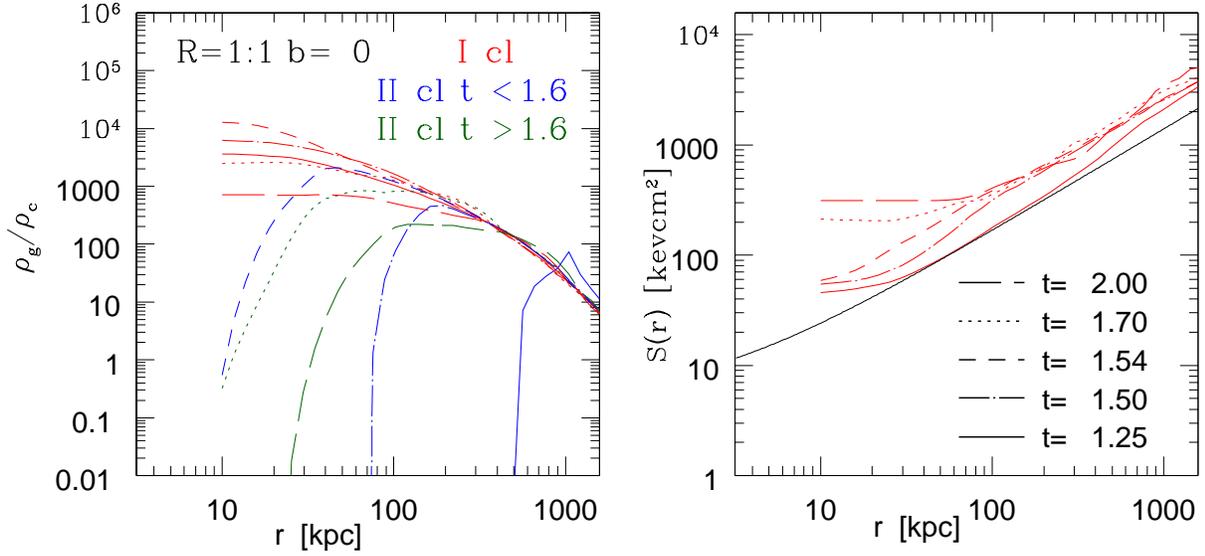}
\caption{For the head-on $1:1$ mass ratio case, we show radial profiles 
of gas density (entropy) in the left (right) panels.
Different line styles refer to different epochs, and the time is in Gyr.
In the left panel, red lines indicate the  gas density profiles of the 
primary (I), whilst blue and green lines  those of the secondary (II).
At each epoch the profiles of both clusters are evaluated in a frame centered on
the peak of the gas density of the primary. For the secondary blue (green)
	lines are used to indicate profiles extracted at times before (after) 
 	$t\simeq 1.6$ Gyr, an epoch which is approximately identified as that 
	when the distance between the two cluster centers-of-mass reaches its first
	minimum.
	Right panels shows the entropy profiles of the primary at different
	epochs. The solid black line refers to the initial profile.
\label{fig:ptav_01_00}}
\end{figure*}

We first discuss the head-on, $1:1$ mass ratio case. This is the most energetic 
event  and its dynamics are relatively simple.
In the left panel of Figure \ref{fig:ptav_01_00}, we show the time evolution of
the gas density radial profile of the primary at different epochs. Similarly, 
 the evolution of the entropy radial profile is shown in the right panel.
Additionally, we also show in the left panel the density profile of the secondary; 
this quantity has been evaluated in the frame of the primary. The gas profiles of each cluster member are constructed separately by culling from the set of SPH
particles the corresponding gas particles. These are
tagged at the start of the simulation according to their membership.

We have chosen to evaluate the profiles at five different times:
$t=1.25,~1.50,~1.54,~1.70,$ and $2$ Gyr.
These are centered around $ t_{p}\sim 1.6$ Gyr, the epoch
at which the distance between the centers-of-mass of the two colliding cores
attains its first minimum. In other mergers with non-zero impact 
parameters, this epoch is identified as the first pericenter passage and 
we denote it as $ t_{p}$.

From the time evolution of the profiles of Figure \ref{fig:ptav_01_00}  we
conclude that most of the final core entropy  is generated at the time $ t_p$ 
of the first core collision. From the right panel it can be seen that the 
level of central entropy is already at $S(0)\sim 300$ keV cm$^2$ at $ t\simeq 2$ Gyr. 
This entropy level is about $\sim 70\%$ of its final value,
as can be inferred from  the right panel of Figure \ref{fig:adentr}.

\begin{figure*}
\centering
\includegraphics[width=0.95\textwidth]{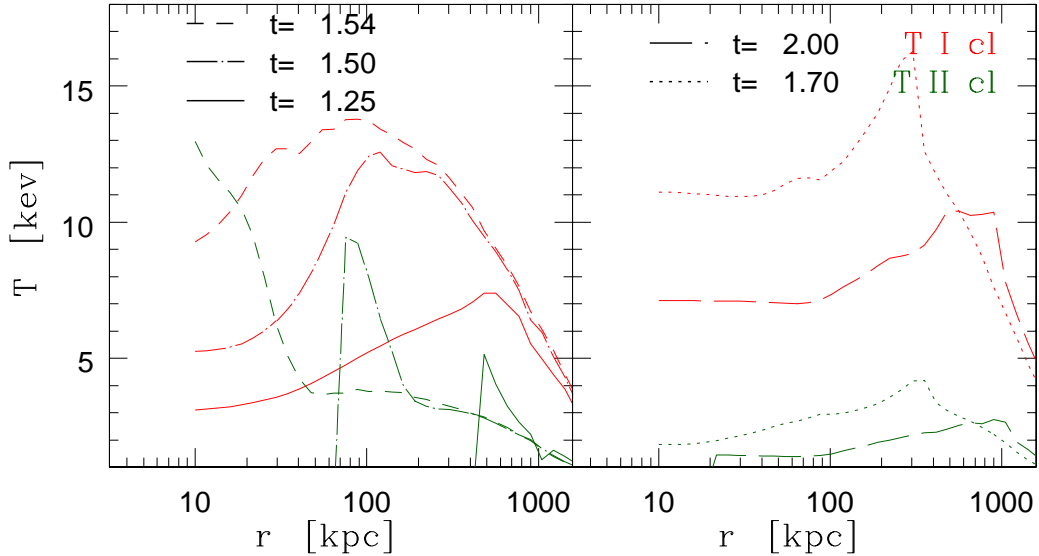}
\caption{
For  the same merger as in Figure \ref{fig:ptav_01_00},
we show here the evolution of the gas temperature profiles,
 red lines are for the primary and green lines refer to the secondary.
For the sake of clarity, in the left
panels we show profiles evaluated at $t< 1.6$ Gyr , when the two
clusters are approaching each other, and in the right panels, the
profiles are evaluated at  $t> 1.6$ Gyr, after the first pericenter
passage. As in  Figure \ref{fig:ptav_01_00}, at each epoch radial profiles
are evaluated in the primary frame, with  its origin being defined as the gas
density peak of the primary tracked at run time.
\label{fig:ptbav_01_00}}
\end{figure*}

This scenario is also confirmed by the radial behavior of the temperature profiles
displayed in Figure \ref{fig:ptbav_01_00}.
The two panels show
the temperature profiles of both primary and secondary at different times.
As in Figure \ref{fig:ptav_01_00},
the secondary profiles are
evaluated in the primary frame. To avoid overcrowding, we have divided the 
profiles in two categories: before (left panels) and after (right panels) 
the epoch $t_p=1.6$ Gyr.

The left panel of  Figure \ref{fig:ptbav_01_00} shows how 
the temperature profile of the primary  increases progressively,
as the secondary approaches the primary and the gas is shock heated.
This increase is first characterized by a peak in the primary outskirts, 
which moves inward and gets wider at late times ($t\rightarrow t_p$).
The temperature of the secondary increases too, in fact at $t=1.54$ Gyr both
the two core temperatures approach the same level.
These findings consistently support the view that for head-on mergers  most of the final
 core entropy is generated during the first core collision, with the remainder of
  the high-entropy gas being accreted later during the final phases of the
   merging.

 A different scenario of entropy generation emerges when analyzing mergers with an 
 initial angular momentum (AM).
  We have chosen to  discuss first the case of the off-axis merger with $1:1$ 
  mass ratio and $b=0.6$. Among the  equal mass mergers this   
  has the highest AM, so its study is particularly interesting 
  in order to analyze how entropy is generated during these collisions.
 
\begin{figure*}
\centering
\includegraphics[width=0.95\textwidth]{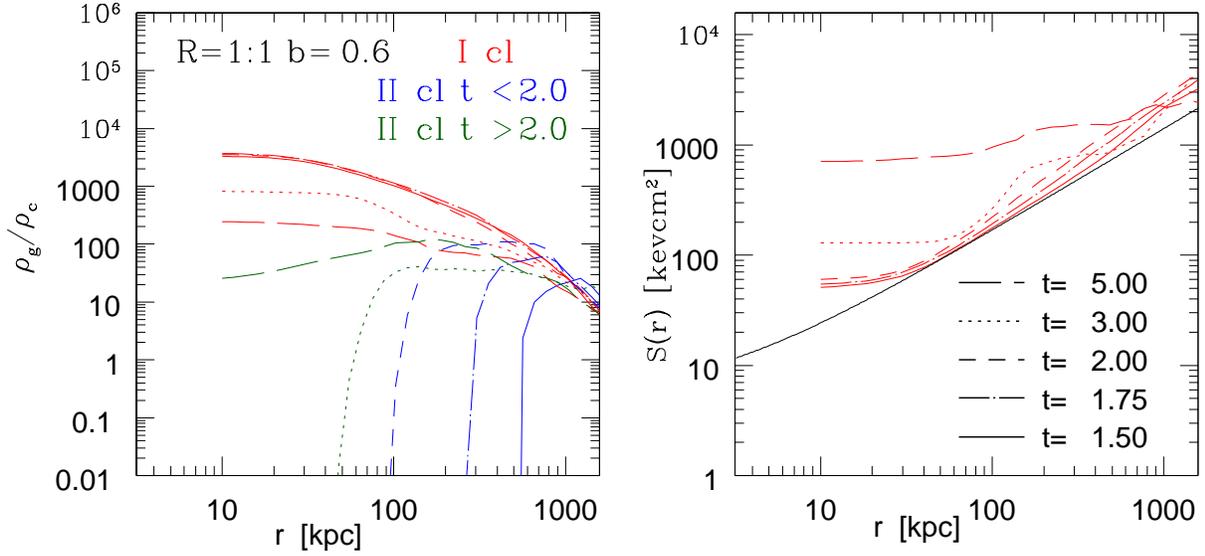}
\caption{Identical to Figure \ref{fig:ptav_01_00}, but  for the $R=1:1$ off-center merger 
	with $b=0.6$. The first pericenter passage  occurs approximately 
	at $t_p\simeq 2$ Gyr.
\label{fig:ptav_01_06}}
\end{figure*}
 
  The two panels of Figure \ref{fig:ptav_01_06} depict density and entropy
  profiles as done in Figure \ref{fig:ptav_01_00}, but for the  $R01b06$  merging run. 
  In particular,  the first pericenter passage  occurs at  $ t_{p}\sim 2$ Gyr, and 
  we show profiles extracted at five different times centered around this epoch.

From the entropy profiles displayed in the right panel of 
  Figure \ref{fig:ptav_01_06}  one can recognize that, unlike the head-on case, most 
  of the core entropy of the primary
   is generated well after the pericenter passage, between $t\sim3$ and $t\sim5$ Gyr.
   This last epoch corresponds to when the secondary has passed the apocenter and is 
   falling back onto the primary.
   Note that in the left panel of Figure 
   \ref{fig:ptav_01_06}, the density profile of the secondary at $t=5$ Gyr extends closer 
   to the core of the primary than at  $t=3$ Gyr.

   At $t\simgt 3$ Gyr, core heating of the primary proceeds as previously discussed
   for the head-on merger, with the secondary coalescing with the primary and the 
   low-entropy gas in the core being mixed with the high-entropy gas  
    generated during the core collision. This suggests that we can decompose the 
    generation of entropy in the central regions of the primary into 
    two distinct phases:  a first one when the primary has a grazing encounter 
    with the secondary,  and a second phase when the secondary finally collapses 
    onto the primary. We now investigate how entropy generation proceeds 
    during the first
    phase, when the secondary first approaches the primary.

\begin{figure*}
\centering
\includegraphics[width=0.95\textwidth]{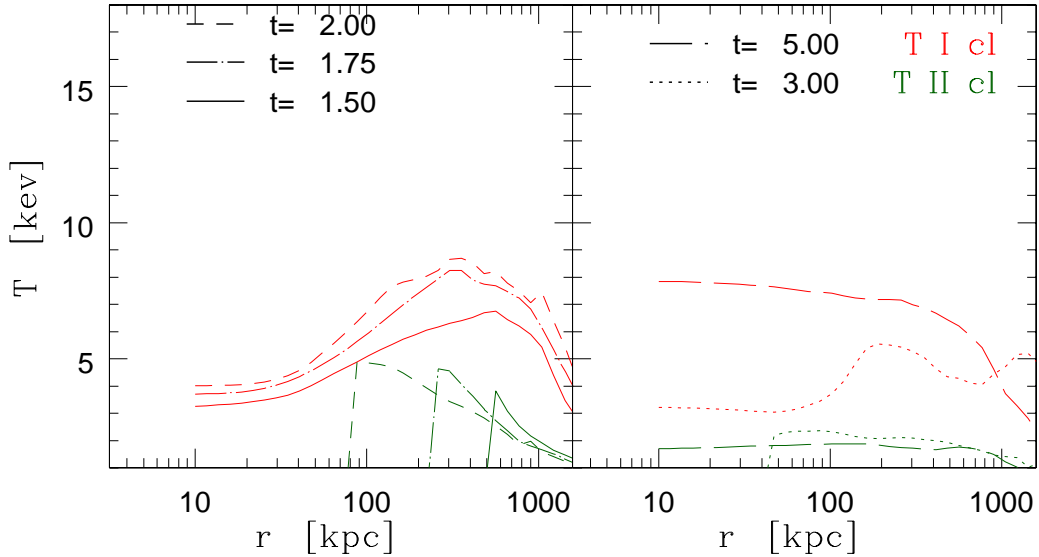}
\caption{
As in Figure \ref{fig:ptbav_01_00}, for the merging case  of Figure  
        \ref{fig:ptav_01_06} ($R=1:1$, $b=0.6$),
        we show here the time evolution
        of  the gas temperature profiles for both the primary and
        secondary cluster.
\label{fig:ptbav_01_06}}
\end{figure*}

    To this end, we first show in the two panels of Figure \ref{fig:ptbav_01_06} the
    time evolution of the primary and secondary temperature profiles.
        Their time variations exhibit a behavior
        in line with that seen with the corresponding profiles
      of Figure \ref{fig:ptav_01_00}, with a significant increase in the gas
      temperature of the primary at late epochs.
     
      However, the right panel of  Figure \ref{fig:ptav_01_06} shows a jump in
      the  entropy of the primary between $t=2$ and $t=3$ Gyr. We argue that
       this entropy increase is a consequence of  a transfer
        of AM between the secondary and the primary as the secondary is getting closer.
        This  in turn is due to tidal torques
        that become significant as the two clusters reach  their
        closest approach along their orbits.
        
\begin{figure*}
\includegraphics[width=0.95\textwidth]{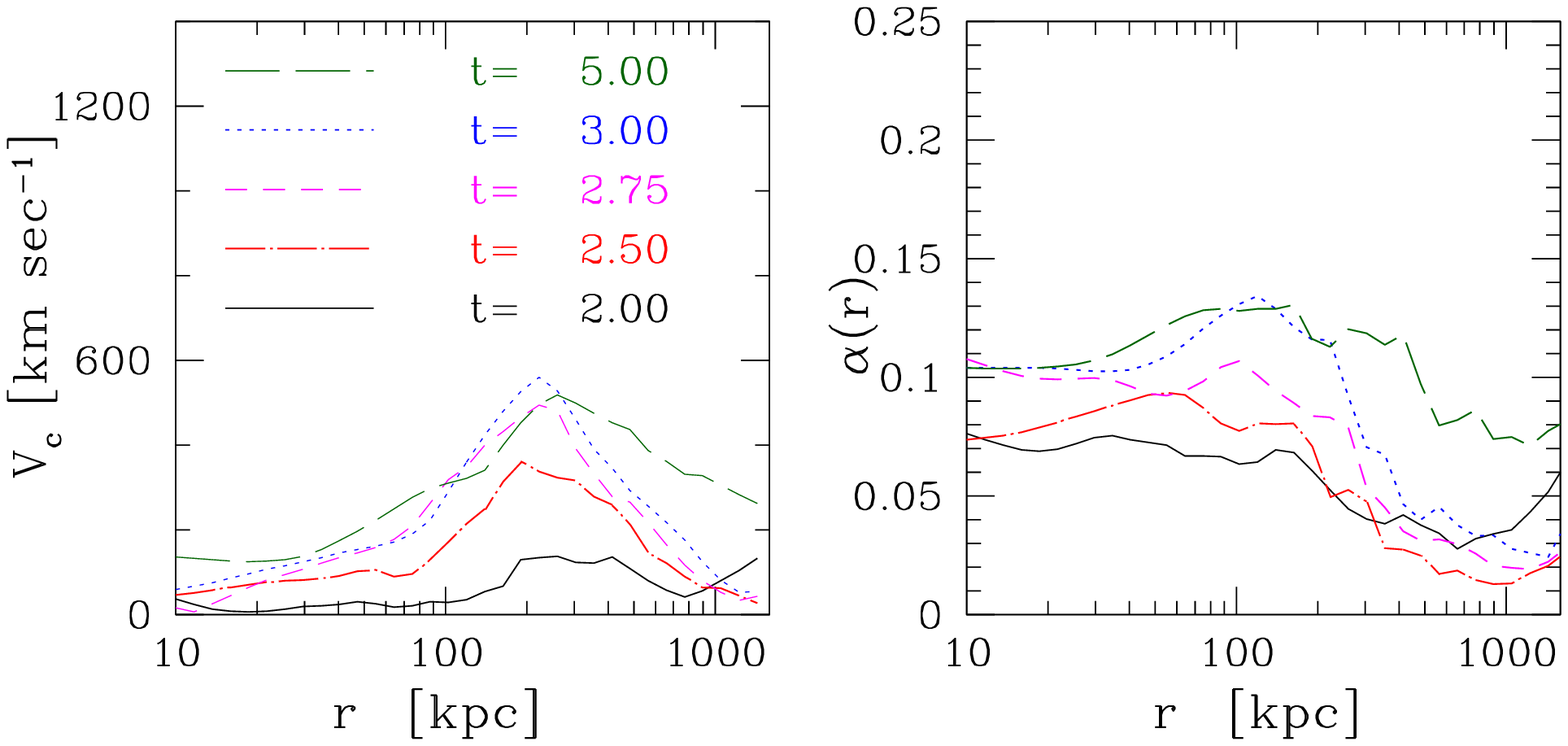}
        \caption{
 Average radial profiles of the mean circular velocity $V_c(r)$
    (left panel)
and artificial viscosity parameter $\alpha(r)$ (right panel) are shown at  different epochs
        for the primary cluster of the merging run of Figure \ref{fig:ptav_01_06}.
        The  center-of-mass  velocity of the primary has been subtracted from the
         mean gas velocities before evaluating the velocity profiles.
The $\alpha(r)$ profiles have been extracted
from the AV parameters $\alpha_i$ of the gas particles of the primary cluster,
        in the same way in which the radial profiles of other gas properties
        (e.g., temperature) have been derived.
        Time is in Gyr.
\label{fig:ptdav_01_06}}
\end{figure*}

                This transfer of AM  leads to an increase of the gas
        circular velocity of the primary and, subsequently, to the
        development of instabilities and entropy mixing.
        To better quantify this point,
 we show in Figure \ref{fig:ptdav_01_06} the radial profiles of
the  mean circular velocity $V_c(r)$ (left panel)   and artificial viscosity parameter
$\alpha(r)$ (right panel) for the primary cluster.
The latter is constructed  by radial averaging  the AV parameters $\alpha_i$  in a
manner similar to that adopted   to calculate the other radial profiles.
        The   mean  velocity $V_c(r)$  is evaluated  by subtracting  the
        center-of-mass  velocity of the primary from the gas velocities.
The profiles have been extracted
at the following epochs: $t=2$,  $t=2.5$, $t=2.75$, $t=3$ and $t=5$ Gyr.

   The time evolution of the mean circular velocity profiles reveals several important
   features.  In particular, there is a progressive  increase in the amplitude of
   the  profiles  as $t\rightarrow 3$ Gyr. All of the profiles at
   different epochs have the tendency to reach their peak values around
  $\sim 100-200$ kpc, these are of the order of
         $\sim$600 km s$^{-1}$  at $t\simgt 3$ Gyr.

 This behavior is shared by the corresponding artificial viscosity profiles
  $\alpha(r)$, which at epochs $t\simgt 3$ Gyr exhibit peak values of $\sim 0.12-0.15$ in the same radial
  range.
  This clearly
  shows evidence of significant gradients in the flow velocity 
  at $\sim 100-300$ kpc, induced by the strong increase in the gas rotational motions.
  
    These motions will generate local instabilities in the medium
  which  in turn will lead to the development of turbulence,  thus driving
  the diffusion of entropy. This is indirectly confirmed by the radial
 behavior of the AC  profiles $\alpha^C(r)$, which are constructed  from the
  AC parameters $\alpha^C_{i}$  in the same way as we did  for the
  AV profiles $\alpha(r)$. The qualitative behavior of the profiles $\alpha^C(r)$
  mirrors very closely that of the corresponding $\alpha(r)$, thus showing
  the presence of diffusive processes associated with the appearance of rotational
  motions.

     We argue that the  increase in  core entropy seen  between $t=2$ and $t=3$ Gyr
     is then a consequence of the rotational gas motions, induced by the passage of the
   secondary at the pericenter.  However, to validate this picture,  the  local
   diffusion time scale must be of the same order or lower than the
   estimated time span ($\sim 1$ Gyr) over which entropy undergoes its changes.
 To further elucidate this issue, we now try to assess the diffusion
 time scale associated  with the development of  Kelvin-Helmholtz instabilities
    (KHI).

\begin{figure*}
\centering
\includegraphics[width=0.95\textwidth]{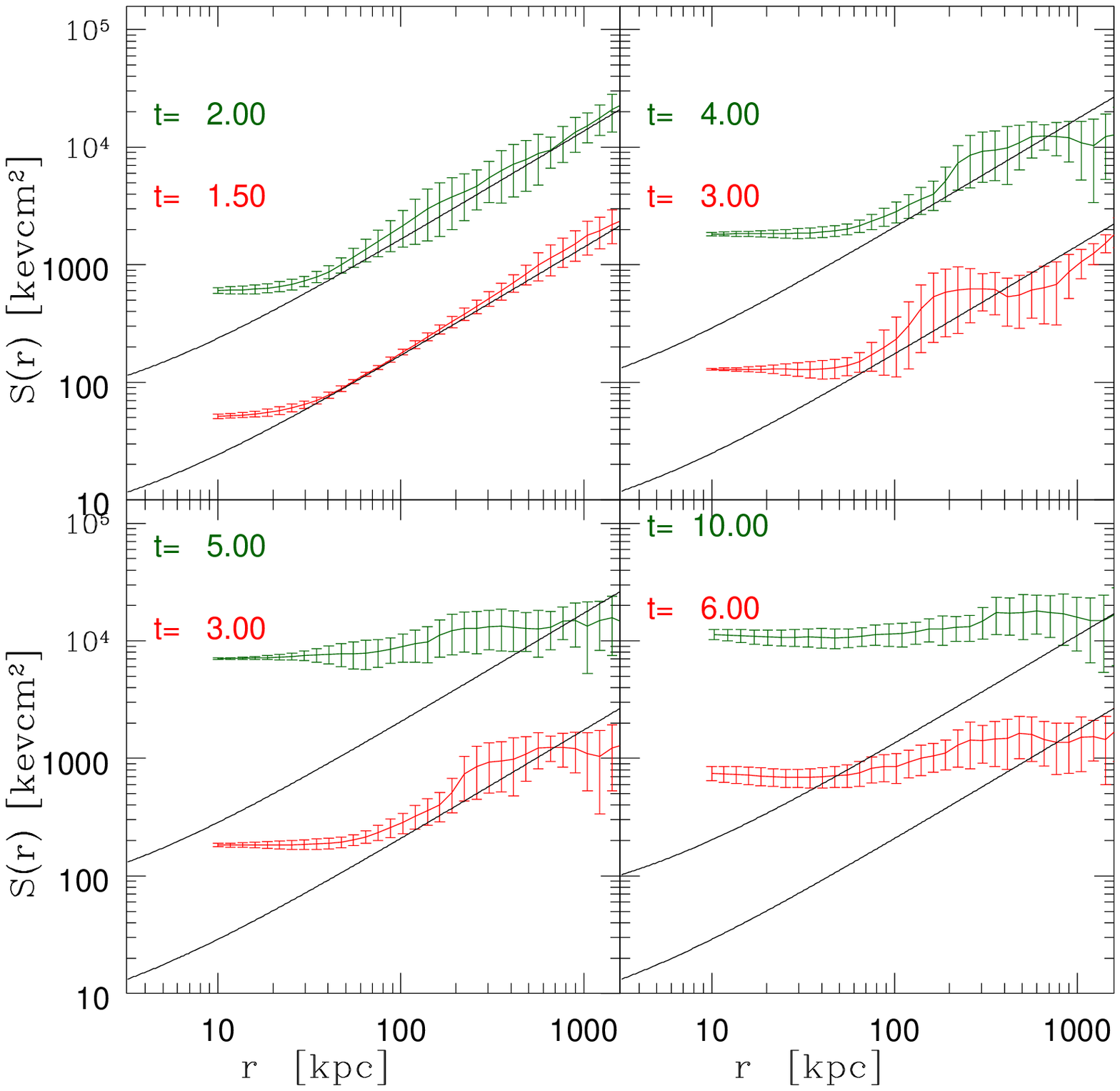}
\caption{For the equal-mass merger with $b=0.6$  of  Figure \ref{fig:ptav_01_06},
 we show here the entropy profiles of the primary at different epochs.
 At a given time, we compute the entropy dispersion (the error bars) by 
	averaging over angular grid values of the shell for each radial bin (see text). As in 
	Figure \ref{fig:ptav_01_06} the solid black lines refer to the initial profile.
	For the sake of clarity, in each panel the entropy profile
	referring to the later epoch has 
	been shifted upward by a factor 10.
\label{fig:ptavc_01_06}}
\end{figure*}

    The generation of KHI leads to the development of
turbulent motions and to the formation of eddies at different spatial scales.
 The size of these eddies  is expected to be significant, with scales of
   $\lambda \sim 100-300$ kpc  \citep{Tak05,Sub06}.
  These estimates are also in accord  with length scales 
  found in   simulations aimed at studying turbulent properties of the ICM 
    \citep[][see in particular  Figure 9 of the first authors]{Vaz12,V19}.

\begin{figure*}
\centering
\includegraphics[width=0.95\textwidth]{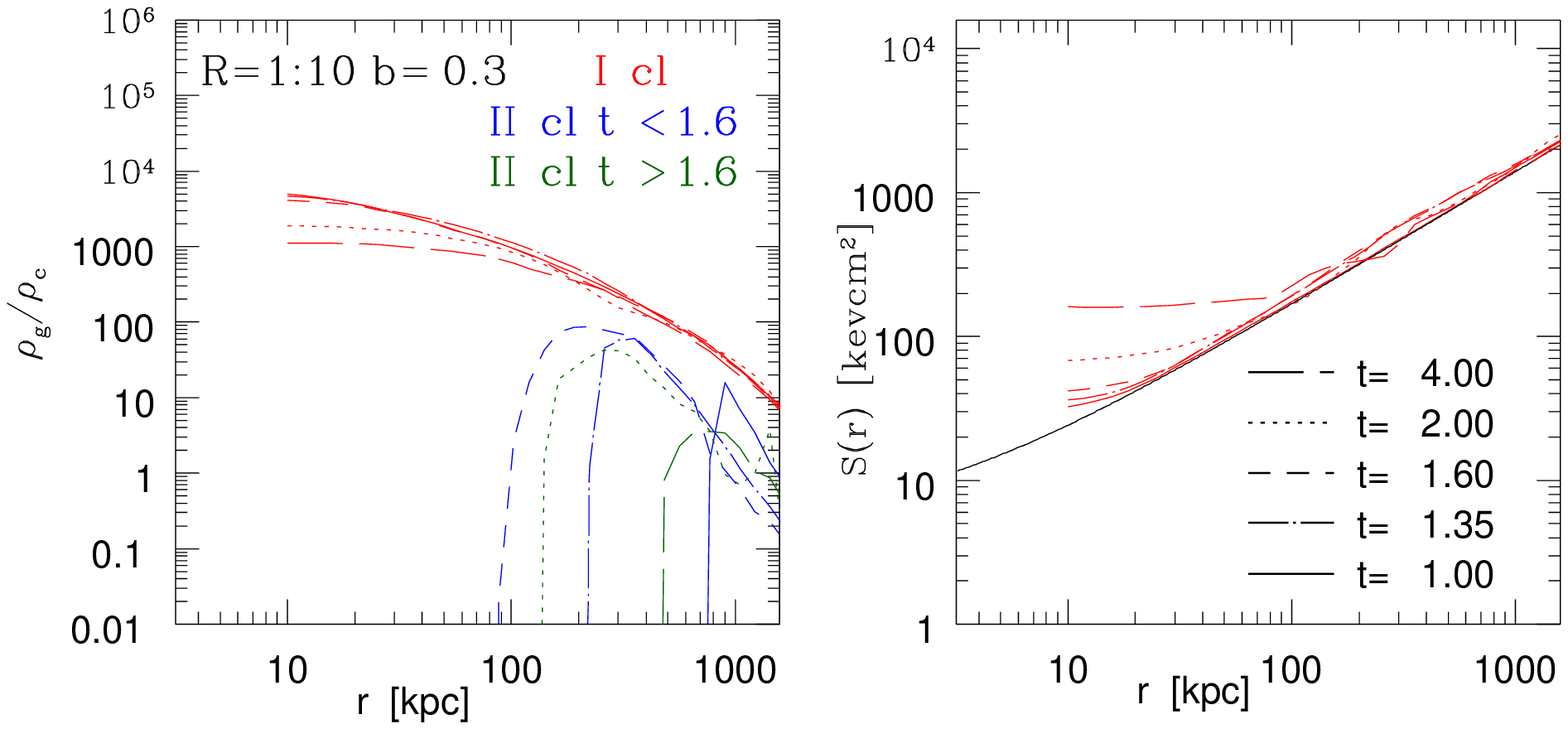}
\caption{The same as Figure \ref{fig:ptav_01_00}, but  for the $R=1:10$ off-center
	merger with $b=0.3$. The first pericenter passage is estimated
	 to occur at $t_p \simeq 1.6$ Gyr.
\label{fig:ptav_10_03}}
\end{figure*}

 From  Eq. 8 of \citet{Vaz12}, we estimate the coefficient  for turbulent diffusion  
 as given by
\begin {equation}
 D_{turb}\simeq 0.1 {\lambda}  \sigma \simeq 3 \times 10^{30} 
 \left(\frac{\lambda}{100 \, {\textrm{kpc}} }\right) \left(\frac{\sigma}{1000 \, {\textrm{km/sec}}}\right)
 \, {\textrm{cm}}^2 \, {\textrm{s}}^{-1} \, ,
\label{diff.eq}
\end{equation}
and the corresponding diffusion time scale by 
\begin {equation}
 \tau_D\simeq {R^2}  /D_{turb} \sim 3 \left(\frac{R}{100 \, {\textrm{kpc}} }\right)^2 
 \frac{1}{D_{30} } \, {\textrm{Gyr}} \, ,
\label{taud.eq}
\end{equation}
where $\sigma$ is the gas velocity dispersion, $D_{30} \equiv D / ( 10^{30}$ cm$^2$ s), and $R$ is the considered scale. 

\begin{figure*}
\centering
\includegraphics[width=0.95\textwidth]{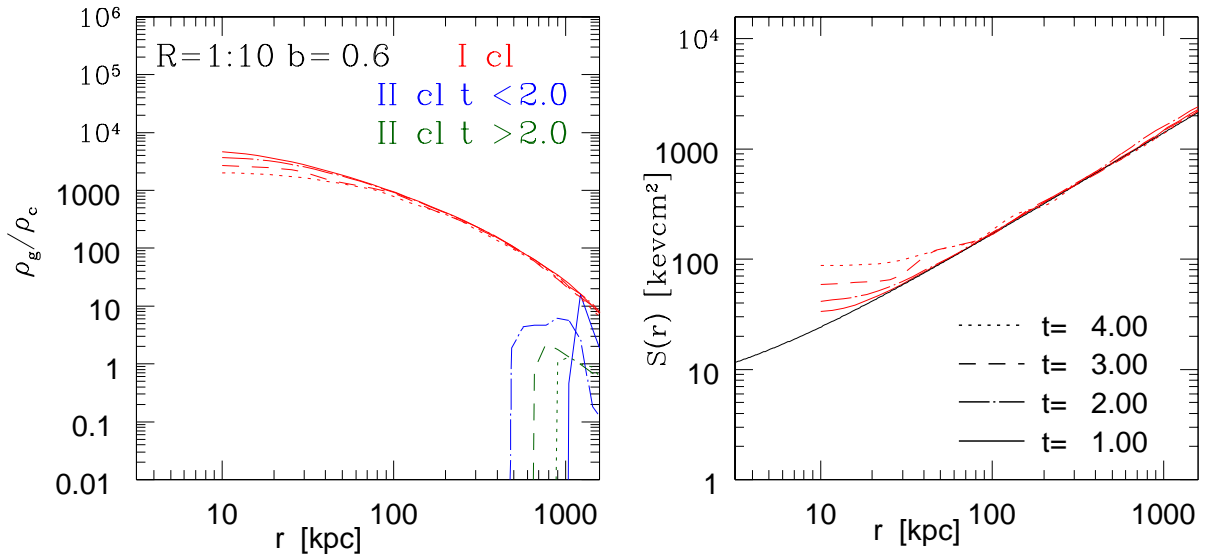}
\caption{ The same as  in Figure \ref{fig:ptav_10_03}, but  for the  off-center
	case with $b=0.6$. Here the first pericenter passage is around
	  $t_p \simeq 2$ Gyr.
\label{fig:ptav_10_06}}
\end{figure*}

\begin{figure*}
\centering
\includegraphics[width=0.95\textwidth]{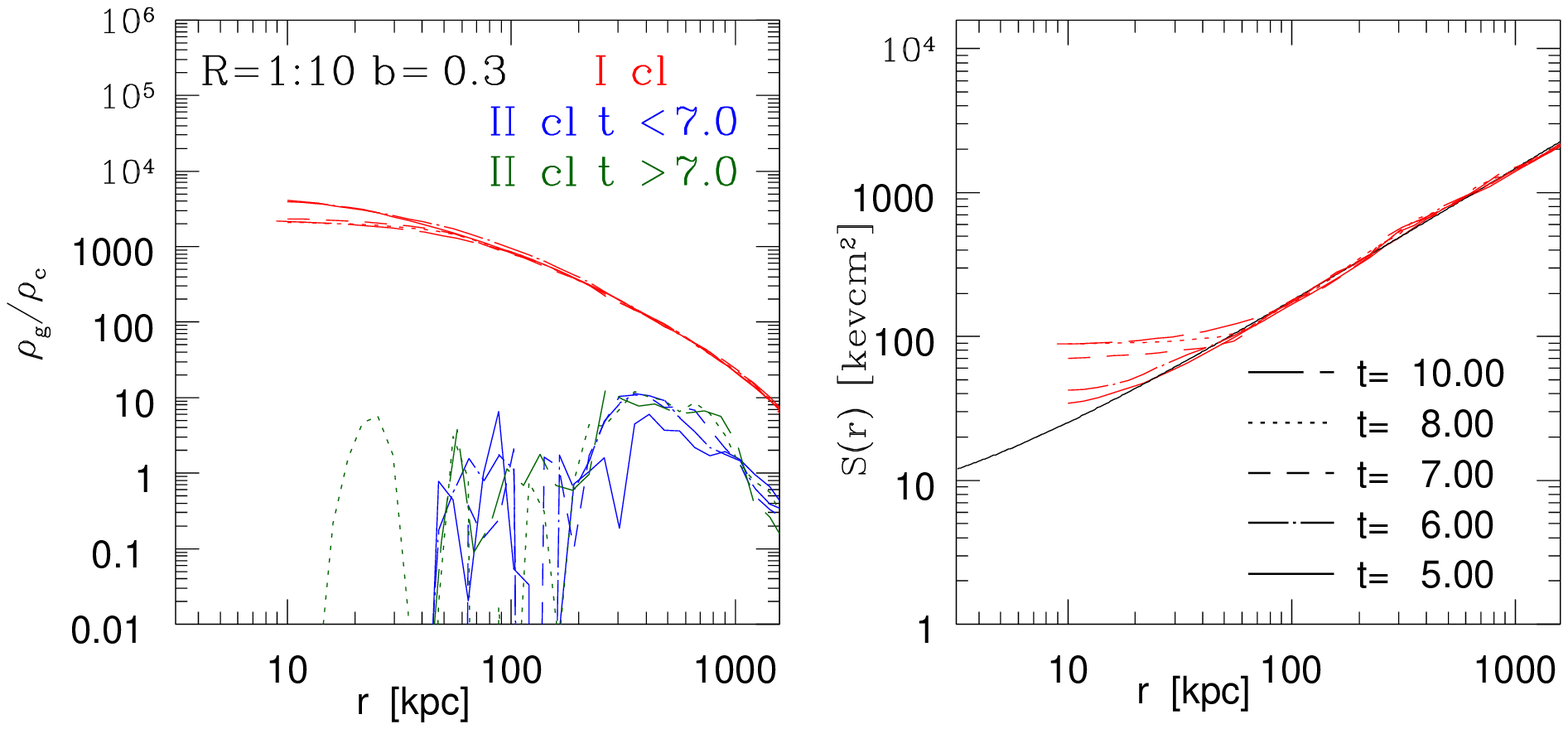}
\caption{ The same as for the merging case shown in Figure \ref{fig:ptav_10_03}, 
but for the  $\alpha_{d}$  simulation of Section \ref{sec:stab} (see text), in which the 
parameter $\alpha_{damp}$ is switched off at a simulation time 
	$t_{\rm hit}=5$
Gyr. The first pericenter passage is between $t=1$ and $t=2$ Gyr,  and 
after it the secondary falls back  onto the primary shortly after $t=7$ Gyr.
\label{fig:rptav_10_03}}
\end{figure*}

  Between $t=2$ Gyr and $t=3$ Gyr the gas velocity dispersion of the primary
  is found to change drastically from $ \sim 800$ km/sec down to $ \sim 200$ km/sec
 within  radii  $r \le 100$ kpc.
 By assuming an upper limit of $\lambda\sim 300$ kpc for KHI eddies,
 in the given time interval we then obtain  values of $D_{30}$ lying in the range
 $7 \simlt D_{30}\simlt  1.8$.
 This correspondingly gives
 $0.4 \simlt \tau_D / \, ( R/100 \, {\textrm{kpc}})^2 \, {\textrm{Gyr}} \simlt1 $
 between $t=2$ Gyr and $t=3$ Gyr,
 showing  the consistency of the diffusion time scale with the increase  in
 central entropy seen between $t=2$ and $t=3$ Gyr at $r \le 100$ kpc.

 To further elucidate this critical point, we show in 
Figure \ref{fig:ptavc_01_06} the entropy profiles of the primary  at different epochs.
Unlike in the right panel of Figure \ref{fig:ptav_01_06}, we show the radial
entropy profiles extracted from a wider range of time frames; moreover,
for each radial bin we also evaluate the entropy dispersion $\sigma_S(r)$.
As noted in the text discussing Figure \ref{fig:adentr} above, the average $S$ over a spherical shell is determined by converting the entropy parameter $S$ into a physical entropy, averaging this, and then converting the mean physical entropy back to $S(r)$.
Similarly, the dispersion of the physical entropy is added and subtracted from the mean physical entropy, and these two values are converted back to $S$ to give the upper and lower limits of the error bars in Figure \ref{fig:ptavc_01_06}.

Within each panel of Figure \ref{fig:ptavc_01_06}, we show the entropy profiles 
 at two distinct epochs; for the sake of better understanding  the profile referring to 
 the latest epoch is shifted upward by one  order of magnitude.
 The time evolution of these entropy profiles clearly illustrates a progressive
  increase in entropy at outer radii, with a subsequent propagation toward the inner 
  regions.
  If we now adopt the reasonable assumption that the entropy dispersion
  $\sigma_S(r)$ can be taken as a metric  for assessing the amount of mixing present 
  at that radius, $\sigma_S(r)$ (and presumably mixing) are initially very small  
    at radii $ r \simlt 500$ kpc. 
   The entropy dispersion becomes progressively wider in the outer regions of the primary as 
   the secondary approaches ($t\rightarrow t_p \approx 2$  Gyr), but with an amplitude
    which is still negligible within $ r \simlt 100$ kpc at $t=2$ Gyr. 
  At $t=3$ Gyr,  there is a  widening of  $\sigma_S(r)$, which is
    significant at radii  $ r \simgt 100$ kpc.
    { However, it is only when $t\simgt 5$ Gyr that the entropy dispersion is
approximately constant across all the cluster  and the gas has now a higher 
degree of mixing.}
To summarize, we conclude that for the considered merging configuration,
   part 
  of the final core entropy   owes its origin to
  rotational motions  induced by tidal torques  occurring  during the collision.

   We now discuss the two off-axis mergers for the $1:10$ mass ratio case. 
 Figure \ref{fig:ptav_10_03} is the analogue of Figure 
  \ref{fig:ptav_01_00}  but for  the $R=1$:10 and $b=0.3$ 
    ($ t_p\sim 1.6$  Gyr) merging, and   Figure \ref{fig:ptav_10_06} for $b=0.6$
    ($ t_p\sim 2$  Gyr).
   In both cases the level of central entropy at late times is
    around $\sim 100$ keV cm$^2$, while in the left panel of the Figures the time
    evolution of the secondary density profiles shows  a negligible
    interaction with the primary's core. In fact, for the $b=0.6$ merging the 
    impact of the secondary on the density profile of the primary 
    at $t=4 $ Gyr can be considered completely absent.

As already discussed in Section \ref{sec:stab}, in merging simulations with
 $1:$10 mass ratio and AM, numerical heating effects can modify the level of core entropy
 before the two clusters merge together. For instance,  one can assume that for the
$b=0.6$ run, all of the core entropy increase at $t=4 $ Gyr is due to this effect.
 This level can be contrasted with that expected  in an isolated halo 
 when the damping term is absent. From the GL run in the right panel of Figure \ref{fig:ichaloa}, 
 we obtain an entropy value of 
 $S/S_{500}\simeq 0.15 $ at $r/r_{500}=0.1$ at $t=5$ Gyr. This translates into $S\sim 130$ at 
 $\sim 115$ kpc, where we have taken $S_{500}$ and $r_{500}$ from 
 Table ~\ref{clparam.tab}.  This implies that at the same epoch, for the $b=0.3$ merging,
  at most $\sim 30\%$ of the entropy core level is due to merging effects. 

  To illustrate  the impact of this effect, in  Figure \ref{fig:rptav_10_03} we depict density 
 and entropy profiles for the  $R=1:10$ and $b=0.3$  merger  
    at various epochs, but extracted from   the corresponding 
 $\alpha_{d}$ simulation presented in Section \ref{sec:stab}.
 For this simulation, the damping 
parameter $\alpha_{damp}$ was switched off at a simulation time $t_{\rm hit}=5$ Gyr 
in order to stably maintain the initial entropy profile until $t=t_{\rm hit}$.
The density profiles are time-centered around $t=7$ Gyr, an epoch at which the secondary
begins to coalesce with the primary.

The final entropy profiles depicted in the right panel of Figure \ref{fig:rptav_10_03}
show a core entropy level of $\sim 100$ keV cm$^2$ at $t=10$ Gyr. This level  is
in accord with previous findings and demonstrates that in mergers with 
low mass ratios and AM, the bulk of core heating occurs at late stages.
A similar level of core entropy at $t=10$ Gyr is obtained from the
 $\alpha_{d}$ simulation with $b=0.6$ (Figure \ref{fig:adHS}).

To summarize, our findings indicate that heating of the core in off-axis mergers
depends critically on the initial merging mass ratio as well as on the AM of the
system. 
For equal-mass mergers and high AM, a significant contribution to the central entropy level  
is sourced by instabilities 
generated by tidal torques, as
the secondary first reaches its pericenter.
On the contrary, for unequal-mass mergers with $1:$10  mass ratio,
 the secondary is progressively disrupted along its orbit by
ram pressure and the
 development of hydrodynamical instabilities,  and core heating 
  becomes significant only during the late merging phases.

\subsection{Radiative runs}\label{sec:cr}

We now investigate the heating of gas cores and the survival of CCs in a more 
realistic set of merging simulations. In this section, we present results extracted
from simulations where the physics of the gas 
includes  cooling, star formation,  and energy feedback following supernova explosions. 

The computational cost of these simulations is much higher than that of
their adiabatic counterparts. Because of cooling, the development of short 
cooling times and large central densities  during the simulations requires
very small timesteps. For this reason, we refrain from resimulating all of 
 the merging cases previously discussed and perform radiative simulations
 only for several of them. Additionally, we also present results from 
 a merging run with a new initial condition set-up (see later).

\begin{figure*}
\centering
\includegraphics[width=0.95\textwidth]{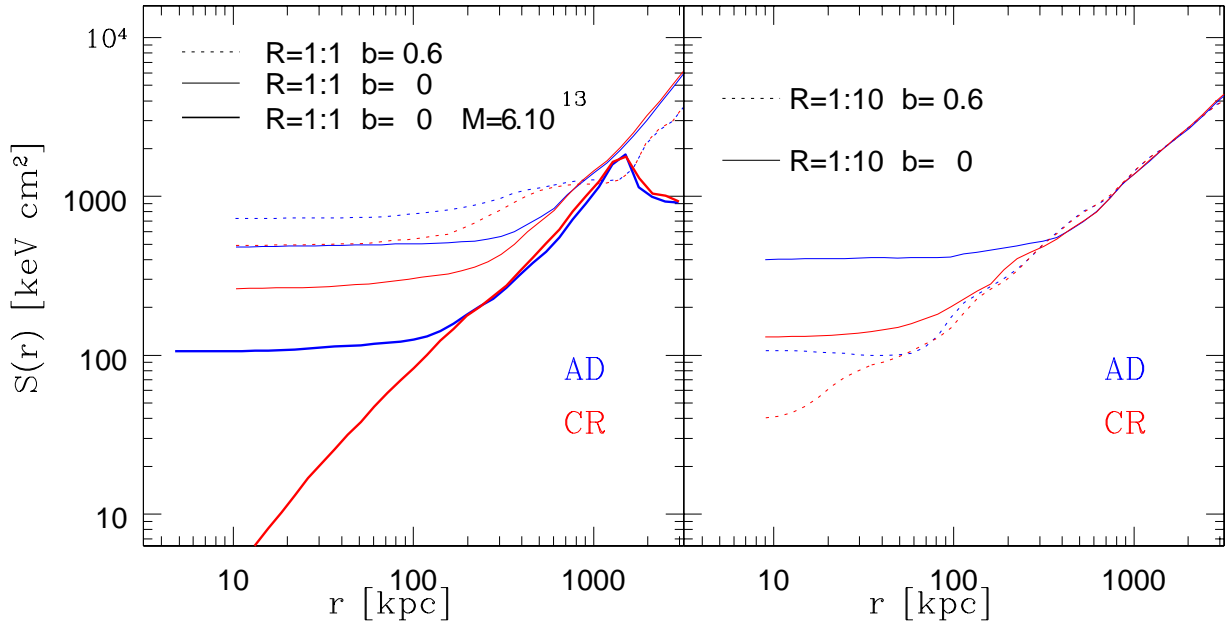}
\caption{Final entropy profiles extracted from adiabatic (AD: blue lines)
 and cooling simulations (CR: red lines).
The left panel shows some runs with the mass ratio $R=1:1$ and
the right panels refers to merging simulations with 
 mass ratio $R=1:10$. In the left panel, thick lines indicate an head-on merging 
	with cluster masses $M_1=M_2= 6 \times 10^{13} \msun$.
 \label{fig:crRS}}
\end{figure*}

As described in Section \ref{sec:stab}, merging simulations with cooling are 
initialized as in the adiabatic case. 
The cooling term $Q_R$ in equation  (\ref{aen.eq})  is switched on  at times $t>0$, 
so that both adiabatic and
radiative  simulations start at $t=0$ with the same profiles.
For the chosen cases we show in Figure \ref{fig:crRS} the final entropy 
profiles of the radiative merging simulations. These are contrasted with the 
profiles of the corresponding adiabatic runs.

In these simulations, the final level of core entropy, and as a consequence the CC
 ability to maintain its integrity, will depend on the various processes which have contributed to the core heating
  during the merging. We thus expect shock
  heating and entropy mixing to be counteracted in part by radiative cooling in shaping
  the central entropy profile of the merged remnant.

  A discriminant criterion to assess whether or not the CC is destroyed during 
  the merger is the level to which the central entropy  is raised during the
  collision because of the heating processes. If this amount of entropy is high
  enough to raise the central cooling time, let's say above several Gyrs, then
  the CC cannot be re-established in a Hubble time. This implies  the survival of CCs
  to be strictly related to the energetics of the collisions, i.e. to 
  the mass ratio and initial orbit of the merger.

  While the inclusion of gas cooling in the simulations will lead to the development
  of large gas core densities, the growth of KHI and the 
  degree of gas mixing are not expected to be modified in a significant way 
  when compared to the adiabatic runs. 
   Damping of KHI  in the presence of cooling will occur  whenever
$\tau_c < \tau_{s}$ \citep{Vietri97}, where
 $\tau_{s}$  is the sound crossing time of the perturbation.  We estimate $\tau_{s}$ 
  as $\lambda/c_s$:
\begin {equation}
 \tau_{s}\simeq 0.2 \frac{\lambda}{100 \, {\rm kpc} } 
\frac{1 }{ \sqrt {kT/ {\rm keV}} }\, {\rm Gyr} \, .
\label{taus.eq}
\end{equation}

From the previous discussion in Section \ref{sec:entgen}, we adopt 
  a lower limit of  $\lambda\simgt 100$ kpc for the eddy size.
The lower value of $\tau_c$ is about $\sim 1$ Gyr in
 the core (Figure \ref{fig:ichalob}), and it reaches  $\sim 10$ Gyr  at $ r \sim 100$ kpc.
From the profiles of Figures \ref{fig:ptbav_01_00} and \ref{fig:ptbav_01_06},
 we estimate gas temperatures  in the 
range  of few keV in the cluster central regions. Therefore, we conclude that  the impact of cooling 
 on the growth of KHI  can be considered negligible, in line
 with previous findings \citep{ZuH10}. 

The final entropy 
 profiles (red lines) of  several equal-mass radiative merger simulations are displayed in the left panel of  Figure \ref{fig:crRS}.
We first consider  the two merging cases with impact parameter $b=0$ and $b=0.6$,
previously investigated in Section \ref{sec:entgen}. For the sake of comparison their
adiabatic counterparts are also shown (blue lines).

In the head-on case we expect the core entropy to  undergo a very steep increase
 because of the strong shock following the collision of the cores, with a subsequent
 decrease due to radiative cooling.  The difference at $t=10$ Gyr 
 between the core entropy of the adiabatic simulation and the radiative one (solid lines,
left panel of  Figure \ref{fig:crRS}) is $\Delta S(0) \simeq 200$ keV cm$^2$.

\begin{figure*}
\centering
\includegraphics[width=0.95\textwidth]{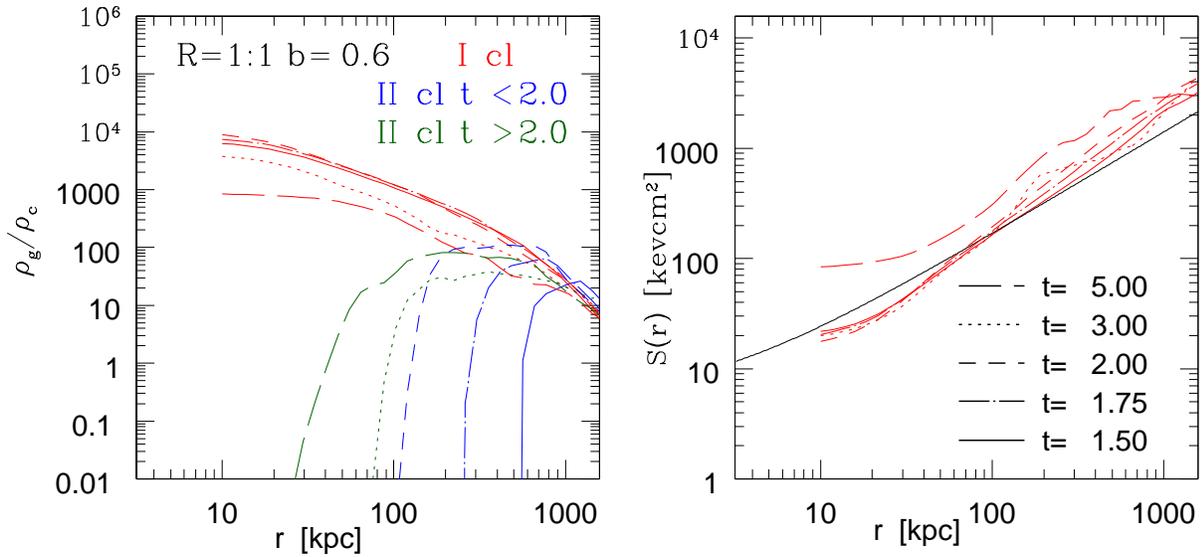}
\caption{As in  Figure \ref{fig:ptav_01_06}, we show here the
 gas density and entropy  radial profiles
  for the $R=1:1$ off-center merger with $b=0.6$.
        At variance with the run of Figure \ref{fig:ptav_01_06},
        the simulation here incorporates radiative cooling.
\label{fig:ptcr_01_06}}
\end{figure*}

The off-axis merger (dot lines) exhibits  a similar difference in final core 
entropies.  In Figure \ref{fig:ptcr_01_06} 
 we show  the time evolution of density and entropy profiles  for this run, 
 as we did in  Figure \ref{fig:ptav_01_06} for the adiabatic case. A comparison 
 between the two sets of entropy  profiles  shows 
a much more modest increase in entropy for the cooling run.

In fact, at $t=5$ Gyr the level of core entropy is about 
$S_{cr}(0) \simeq 100$ keV cm$^2$, while in the adiabatic run it is 
about $S_{ad}(0) \simeq 800$ keV cm$^2$. The difference is subsequently
reduced soon after  $t=5$ Gyr, as the secondary finally merges with the primary 
and the process raises the central entropy up to $S_{cr}(0) \simeq 500$ keV cm$^2$
at $t=10$ Gyr. 

This is in accord with \citet{Poole08}. From their merging simulations 
the authors argue that to 
re-establish a CC it is necessary  for the remnant to be relaxed  for a significant 
fraction of the cooling time $\tau_c$. For the considered merger, we estimate 
$n_e(0) \sim 10^{-2}$ cm$^{-3}$ at 
 $t=5$ Gyr and  
 $\tau_c(0) \sim 6 
 \left ( S(0)/100 \, {\rm keV} \, {\rm cm}^2\right) ^{1/2}$ Gyr $ \simeq 6$ Gyr, 
 from Equation \ref{tc.eq}.
 This time scale is larger than that set by the final collision of the secondary,
  which raises again the central entropy and in turn $\tau_c$ to above $ \sim 10$ Gyr.

  Finally, note that in the left panel of Figure \ref{fig:ptcr_01_06}  the density
  profile of the secondary is significantly reduced 
  in the cluster inner regions, when contrasted with the adiabatic case at $t=5$ Gyr. 
  Similarly, in the same regions the core gas density of the primary is higher
  by a factor $\sim 5$. These differences follow because of radiative cooling, 
  with the primary developing a steeper profile
  than the secondary.

In the right panel  of Figure \ref{fig:crRS}  we show the entropy profiles of two unequal mass mergers with the mass ratio $R=1:10$. The dynamics of the head-on case
mirrors that of the $R=1:$1 merging, but to a lesser extent because of the 
 reduced mass of the secondary. There is a smaller increase in entropy as the two
 core collide, with radiative losses subsequently reducing entropy down to
$S_{cr}(0) \simeq 100$ keV cm$^2$  at $t=10$ Gyr. In contrast to the adiabatic
case, this level of core entropy is a factor $\sim$ two smaller.

In the most off-center ($b=0.6$) merger, the impact  of the secondary on the core is 
negligible. As in the adiabatic case, low-mass subclumps with large AM are 
progressively disrupted by instabilities before being able to significantly 
shock-heat the primary's core. However, in the adiabatic run there is a certain
amount ($\Delta S(0) \simeq 100$ keV cm$^2$) of core heating taking place
during the late phases ($t\sim9-10$ Gyr) of the merger. This small entropy jump
is now absent because of radiative losses, thus allowing the CC to survive.

Additionally, we ran another  head-on radiative merging simulation with the mass ratio
$1:1$. In contrast to the merging case previously discussed, for the two
clusters we now adopt a halo mass of  $M_{200} =6 \times 10^{13} \msun$.
The initial condition set-up being the same as described in Section \ref{subsec:icsetp}.

For this simulation we show in the right panel of Figure \ref{fig:crRS} the 
final entropy profile of the merger remnant  (solid thick line), together
with the profile of its adiabatic counterpart. The entropy profile of the radiative
run exhibits a near power-law behavior and, unlike the head-on $R=1:10$ merging case,
the 
CC has been rapidly re-established. This happens because now the mass of the primary
is an order of magnitude smaller, thus  the collision with the secondary  
(at $t\sim 2$ Gyr) 
is able to raise the central entropy only up to $ S(0) \simeq 80$ keV cm$^2$.
This is a factor $\sim$ two smaller than in the the head-on $R=1:10$ merger, so 
that the cooling time is $\sim 5$ Gyr and the CC is  soon restored.

{ These findings demonstrate that radiative cooling dominates the final ICM core
properties, with physical processes governed by timescales much shorter than those
set by diffusion \citep{Bif15}.}

To summarize, the results of this Section demonstrate that the 
final level of core entropy, and thus the resiliency of CCs to disruption,
depends critically on the merging mass ratio and initial orbit.
Specifically, CCs are destroyed in head-on high-mass mergers, but can survive low-mass 
mergers or off-axis low mass ratio mergers.
This suggests that the merging AM is a key
parameter which determines the final remnant  core entropy. 
In merging systems with high AM, a CC is re-established after the final 
collision as long as the cooling time is shorter than the core free-fall time.
This condition depends on the merging mass  ratio as well.

Our results are, partially, in agreement with \citet{Ha17}. The authors argue that 
AM is a fundamental quantity to determine whether  a CC can survive a 
cluster merger or not. They found CC disruption to occur in major mergers with low AM,
but it is absent if the AM is high. This is in contrast with our findings, for which the
disruption of CC in
major mergers (mass ratio higher than $1:5$) of massive clusters
(i.e. with a primary mass 
$M_1 \simgt 6 \times 10^{14} \msun$)
  occurs regardless of whether the merging is head-on or 
  off-axis. Note also that the CC is not disrupted, and the AM becomes unimportant, 
  when the primary mass is small ($M_1 \simlt  10^{14} \msun$).

  Finally, our findings are in contrast with those of \citetalias{ZuH11}. For the same set of merging 
  initial conditions, that paper finds final levels of core entropy high enough
  to erase CCs, regardless of the considered merging case. This is at variance 
  with the results of this Section, in which the merging simulations now include 
  gas cooling. This shows that a realistic physical modeling of the simulations
  is crucial to address the issue of CC  survival in merging 
  clusters.

\section{Conclusions}
\label{sec:discuss}
In this paper we have presented results extracted from a suite of idealized
binary cluster merger simulations, realized using an N-body/hydro  code
which employs an improved SPH scheme. Each merging cluster simulation was
performed by constructing two isolated gas+DM halos in equilibrium; initial
positions and velocities of the halos are then assigned according to the
specific orbital trajectory. We purposely adopted the same range of initial
mass ratios and impact parameters as in a previous paper (\citetalias{ZuH11}), so as to
consistently compare our results with previous findings.

Our simulations are aimed at investigating how the heating of the gas core leads to an increase of its entropy and to
 the disruption of the original CC profile during cluster mergers.
 In order to assess the impact
 of different physical processes on the survival of CC systems, we consider adiabatic
 as well as radiative merging simulations. The latter incorporate cooling,
  star formation and supernovae feedback. Our main conclusions can be summarized as
  follows.
  
   i) For adiabatic simulations and head-on mergers, the dominant source of entropy
  is the  shocking of the gas at the epoch of the first collision,
  with core heating being later driven by mixing processes.

  ii) In the case of off-axis mergers, core heating
depends critically on the initial merging mass ratio, as well as on the angular momentum (AM) of the
system.  For equal-mass mergers
   part of the final core entropy   owes its origin to the transfer of AM,
   induced by tidal torques  occurring  during the  first encounter of the
   two clusters along their orbit. 
The corresponding increase in the primary circular
   velocity at this epoch (see Section \ref{sec:entgen}) generates instabilities
   in the cluster inner regions,
    and in turn an increase in its core entropy.

iii)  For mergings with low mass ratios the previous effect is negligible or absent
and the bulk of core heating occurs at late stages, when the secondary accretes onto
the primary. The final increase in core entropy can be modest because,
before the final merging with the primary, most of the
secondary mass has been stripped along its orbit by
instabilities and ram pressure.

iv) In general, our results from adiabatic simulations are in accord
  with previous findings \citepalias{ZuH11}. The initial CC profiles do not survive
   the various merger cases we considered, and because of the
   different physical processes occurring during cluster merging,
    high-entropy gas is always present in the cluster core after the merger.

    v) From a numerical point of view, it worth noting the good hydrodynamical
    behavior of  the ISPH  code presented here.
    For a specific run (see Figure \ref{fig:adHS}), we find the final entropy profile
    in good accord with the corresponding one shown in \citetalias{ZuH11}.
     This is a non trivial result since it demonstrates how the ISPH scheme, based on
     a Lagrangian formulation,
     can be considered competitive with Eulerian-based AMR codes
     in terms of hydrodynamical performance.

     The scenario outlined above changes in several respects when cooling
     is incorporated in the simulations. The most important differences are:

    i) The increase in core entropy during cluster merging is now counteracted in part by
     radiative cooling, thus leading to lower levels
      of final entropy in the cluster inner regions than in the corresponding adiabatic case. CCs are found to
      survive if the merger is only able to raise the central entropy to
      $S(0) \la 80$ keV cm$^2$.
This implies that the cooling time is shorter than the Hubble time and the CC is restored.

      ii) For high mass mergers, CCs are destroyed in major merger, but are resilient to off-axis mergers
      with low mass ratios. This suggests that the survival of CCs depends both on the
      initial mass ratio and AM, and is thus characterized by a two-parameter dependency.
      This can be considered as the most important result of this paper.

      iii) Finally, this dependence on AM  tends to disappear as one considers low-mass
      cluster mergers. We ran a head-on merger with a primary mass an order of magnitude
      smaller ($M_{200} =6 \times 10^{13} \msun $) than in the baseline simulations,
       and the final core entropy is found to be low and cooling-dominated.

    Overall, these findings support the observational evidence \citep{Pratt10,Chon12}
    of a correlation between the CC/NCC core morphology and  cluster
    mergers \citep[but see][for a different viewpoint]{Barnes18}.
    The results presented here are also in broad agreement with
    previous works \citep{RT02,Bu08,Poole08,Ha17}, aimed at investigating the impact of
    merging clusters on core properties.

    In particular \citet{Ha17} argued that CCs can survive major mergers with large AM.
    While our merging simulations also clearly indicate a significant role of
     AM in determining the status of the final core remnant, for the considered
    merging case we find that CCs are destroyed in high mass major mergers even 
    in the case large AM.  We suggest that this discrepancy is not significant and is
   of statistical origin.  Our results are obtained from
    binary cluster mergers  realized in isolation with specific initial
    conditions, whereas the \citet{Ha17} sample is comprised of only ten clusters
   extracted from a previous cosmological simulation.
   
{ A possibility which has been left open  by our study of merging clusters occurs 
  when the merger is between a NCC and a CC cluster. In such a case, 
  if the CC of the secondary is able to survive the merger process, it might settle 
  in the center of the primary, leading to a transition of the primary from a NCC to 
  a CC state.  However, we  argue that this scenario is unlikely to occur. 

  This is justified  by the chosen initial condition set up of our merger 
  simulations, in which both the primary and the secondary are initialized as CC 
  clusters. Our previous findings indicate that the CC of the secondary does not 
  survive ram pressure and shock heating as it enters the atmosphere of the primary, 
   regardless of the initial mass ratio and AM of the binary system. This result
   is valid for a primary CC cluster, but we expect it to be little affected
   by the level of core entropy of the primary.}

   The validity of our simulation results depends both on the numerical resolution
   of our simulations as well as on the adopted physical modeling  of the gas.
   For several merging runs, in Section  \ref{sec:stab} we contrasted the final
   entropy profiles against parent merging simulations performed using a higher
   resolution. The stability of the corresponding profiles is shown in Figure
 \ref{fig:adHS}, 
indicating  that our simulations can be considered free of resolution
 effects and are numerically converged.

 In our simulations the physical modeling of the ICM is based on a number of
 simplifying assumptions.  In particular, the most relevant is  the absence of
  a subgrid model for the energy injection from AGNs, which can offset
  radiative cooling in cluster cores. For the purpose of the present
  investigation, it is then important to assess the impact of AGN feedback on
  the results presented in Section \ref{sec:cr}.

  \citet{Ras15} argued that CC thermal properties are affected by AGN feedback, and that
  its absence renders CCs more resilient against late-time mergers. This is in contrast
  with the findings of \citet{Ha17}, for whom CC stability is not affected  by
  incorporating AGN feedback, regardless of the adopted feedback parameters.
  
     \citet{Ha17} suggest that this discrepancy is directly connected to the
    hydrodynamic codes used in the two sets of simulations. The authors
    performed their simulations using an Eulerian AMR code,  while \citet{Ras15}
     employed an improved SPH scheme
  \citep[see][]{Semb16}.
   \citet{Ha17}  argued that the treatment of thermal diffusion in the two codes
   is critical in determining  thermal properties of the simulated CCs.

   This topic has been discussed  at length in Section \ref{sec:stab} and, as
   mentioned in point v) above, there is a full  consistency between the final entropy
   profile of our $R=1:$1 $b=0$  merging simulation with the parent one of \citetalias{ZuH11}.
    This strongly suggests that the artificial diffusion parameters of our
    hydrodynamical scheme are correctly calibrated. We thus assume that the inclusion
    of AGN feedback in radiative merging simulations will mirror the thermal behavior
    of ICM seen by \citet{Ha17}, and in turn  should be of limited impact on the findings
    of Section \ref{sec:cr}.

    However, the entropy profile of the head-on low-mass radiative merger depicted
    in the left panel of Figure \ref{fig:crRS} clearly exhibits an overcooling
    behavior in its inner regions. Incorporating thermal AGN
    feedback in this simulation  will avoid runaway cooling and will bring the final level of core entropy
    to higher values. Without a dedicated simulation including AGN feedback, it is
    difficult to assess the fate of the CC at the end of this merger.

    From the parameters of cluster C3 ($M_{200} =6 \times 10^{13} \msun $) given
    in Table ~\ref{clparam.tab} and the entropy profile   (\ref{sprof.eq}), we
    estimate an initial entropy value of $\simeq 12\sev$ at $r=10$ kpc.
    Assuming that AGN feedback during the collision will maintain the core entropy of
    the primary at approximately this level, we require that the
     gain in core entropy be limited to $ \simlt 50-60$ keV cm$^2$ during the collision  to keep the integrity of the CC.
     In this case, we expect the final central entropy level of the
 merger remnant  to be below the CC threshold ($ \simeq 80$ keV cm$^2$) previously
 given.

 An upper limit to the level of core entropy achieved by the primary during the
 collision can be inferred by looking at the adiabiatic merging simulation.
  From Figure \ref{fig:crRS} we obtain a final core entropy $ \simeq 100$ keV cm$^2$,
  but in the radiative simulations cooling effects will reduce this level to lower
  values. We thus conclude that for this specific merging case, the inclusion of
  thermal AGN feedback will change the ICM thermal state of the core.
  The final level of core entropy is likely to be close to the threshold
  above which the CC will be destroyed.
  
    To summarize, our findings support the scenario in which the observed CC/NCC
  dichotomy is driven by cluster mergers. The difference in the disruption histories
   of the CCs, between adiabatic and radiative merging simulations, demonstrates
   that a realistic modeling of ICM physics is crucial in order to properly
   investigate the behavior of core morphology  during the merging phase.
   We argue that the inclusion in our simulations of AGN thermal feedback
    is unlikely to impact most of our findings, at least in merging simulations
    in which the final cluster remnant has a virial mass
 $M_{200} \simgt6 \times 10^{14} \msun $.
 
  However, it must be stressed that in our high mass merging simulations, CCs survive only in mergers with low mass ratios and high AM. This leaves open the problem
  if such a result is consistent with the observed fraction
 of CC/NCC clusters  at the present epoch \citep{Barnes18}.
 
  This issue can only be addressed  in a cosmological framework, in which the
 evolution of simulated clusters can be followed self-consistently in a
 cosmological volume. Our idealized merging simulations are performed in isolation,
 so that environment effects are absent. In a cosmological simulation, these effects
  are automatically taken into account, and one expects merging environments to
  be affected.

On the other hand, our
 results indicate that the majority of core heating occurs when the
  secondary  enters the innermost regions of the primary cluster. We thus
  suggest  that environmental effects will be of limited impact on our findings.
  
\section*{Acknowledgments}
RV gratefully acknowledges J.A. ZuHone for clarifying comments on
the setup of the initial conditions described in Section 2.2.
CLS was supported in part by  NASA {\it Chandra} Grants GO7-18122X and GO8-19106X.
{ The computations of this paper were carried out  using the Ulisse cluster at SISSA
and the Marconi cluster at CINECA (Italy), under a SISSA-CINECA agreement.}
\section*{Data availability}
The data underlying this article will be shared on reasonable request
to the corresponding author.

%
\bibliographystyle{mn2e}
\bibliography{clmerge.bib}

\begin{thebibliography}{}

\bibitem[\protect\citeauthoryear{{Agertz}, {Moore}, {Stadel} \& { et
  al.}}{{Agertz} et~al.}{2007}]{Ag07}
{Agertz} O.,  {Moore} B.,  {Stadel} J.,    { et al.} 2007, \mnras, 380, 963

\bibitem[\protect\citeauthoryear{{Arth}, {Donnert}, {Steinwandel}, {B{\"o}ss},
  {Halbesma}, {P{\"u}tz}, {Hubber} \& {Dolag}}{{Arth} et~al.}{2019}]{Ar19}
{Arth} A.,  {Donnert} J.,  {Steinwandel} U.,  {B{\"o}ss} L.,  {Halbesma} T.,
  {P{\"u}tz} M.,  {Hubber} D.,    {Dolag} K.,  2019, arXiv e-prints, p.
  arXiv:1907.11250

\bibitem[\protect\citeauthoryear{Balsara}{Balsara}{1995}]{ba95}
Balsara D.,  1995, Journal of Computational Physics, 121, 357

\bibitem[\protect\citeauthoryear{{Barnes}, {Vogelsberger}, {Kannan},
  {Marinacci}, {Weinberger}, {Springel}, {Torrey}, {Pillepich}, {Nelson},
  {Pakmor}, {Naiman}, {Hernquist} \& {McDonald}}{{Barnes}
  et~al.}{2018}]{Barnes18}
{Barnes} D.~J.,  {Vogelsberger} M.,  {Kannan} R.,  {Marinacci} F.,
  {Weinberger} R.,  {Springel} V.,  {Torrey} P.,  {Pillepich} A.,  {Nelson} D.,
   {Pakmor} R.,  {Naiman} J.,  {Hernquist} L.,    {McDonald} M.,  2018, \mnras,
  481, 1809

\bibitem[\protect\citeauthoryear{{Beck}, {Murante}, {Arth} \& {et al.}}{{Beck}
  et~al.}{2016}]{Be10}
{Beck} A.~M.,  {Murante} G.,  {Arth} A.,    {et al.} 2016, \mnras, 455, 2110

\bibitem[\protect\citeauthoryear{{Bekki}, {Owers} \& {Couch}}{{Bekki}
  et~al.}{2010}]{Bek10}
{Bekki} K.,  {Owers} M.~S.,    {Couch} W.~J.,  2010, \apjl, 718, L27

\bibitem[\protect\citeauthoryear{{Biffi} \& {Valdarnini}}{{Biffi} \&
  {Valdarnini}}{2015}]{Bif15}
{Biffi} V.,  {Valdarnini} R.,  2015, \mnras, 446, 2802

\bibitem[\protect\citeauthoryear{{Binney} \& {Tremaine}}{{Binney} \&
  {Tremaine}}{1987}]{Bin87}
{Binney} J.,  {Tremaine} S.,  1987, {Galactic dynamics}

\bibitem[\protect\citeauthoryear{{Buote}}{{Buote}}{2002}]{Buote02}
{Buote} D.~A.,  2002, {X-Ray Observations of Cluster Mergers: Cluster
  Morphologies and Their Implications}.
pp 79--107

\bibitem[\protect\citeauthoryear{{Burns}, {Hallman}, {Gantner}, {Motl} \&
  {Norman}}{{Burns} et~al.}{2008}]{Bu08}
{Burns} J.~O.,  {Hallman} E.~J.,  {Gantner} B.,  {Motl} P.~M.,    {Norman}
  M.~L.,  2008, \apj, 675, 1125

\bibitem[\protect\citeauthoryear{{Cavagnolo}, {Donahue}, {Voit} \&
  {Sun}}{{Cavagnolo} et~al.}{2009}]{Cav09}
{Cavagnolo} K.~W.,  {Donahue} M.,  {Voit} G.~M.,    {Sun} M.,  2009, \apjs,
  182, 12

\bibitem[\protect\citeauthoryear{{Chon}, {B{\"o}hringer} \& {Smith}}{{Chon}
  et~al.}{2012}]{Chon12}
{Chon} G.,  {B{\"o}hringer} H.,    {Smith} G.~P.,  2012, \aap, 548, A59

\bibitem[\protect\citeauthoryear{Cullen \& Dehnen}{Cullen \&
  Dehnen}{2010}]{cul10}
Cullen L.,  Dehnen W.,  2010, \mnras, 408, 669

\bibitem[\protect\citeauthoryear{{Diehl}, {Rockefeller}, {Fryer}, {Riethmiller}
  \& {Statler}}{{Diehl} et~al.}{2015}]{Diehl12}
{Diehl} S.,  {Rockefeller} G.,  {Fryer} C.~L.,  {Riethmiller} D.,    {Statler}
  T.~S.,  2015, PASA, 32, e048

\bibitem[\protect\citeauthoryear{{Donnert}}{{Donnert}}{2014}]{Donnert14}
{Donnert} J.~M.~F.,  2014, \mnras, 438, 1971

\bibitem[\protect\citeauthoryear{{Donnert}, {Beck}, {Dolag} \&
  {R{\"o}ttgering}}{{Donnert} et~al.}{2017}]{Donnert17}
{Donnert} J.~M.~F.,  {Beck} A.~M.,  {Dolag} K.,    {R{\"o}ttgering} H.~J.~A.,
  2017, \mnras, 471, 4587

\bibitem[\protect\citeauthoryear{{Drakos}, {Taylor} \& {Benson}}{{Drakos}
  et~al.}{2017}]{Drakos17}
{Drakos} N.~E.,  {Taylor} J.~E.,    {Benson} A.~J.,  2017, \mnras, 468, 2345

\bibitem[\protect\citeauthoryear{{Feretti}, {Giovannini}, {Govoni} \&
  {Murgia}}{{Feretti} et~al.}{2012}]{Feretti12}
{Feretti} L.,  {Giovannini} G.,  {Govoni} F.,    {Murgia} M.,  2012, \aapr, 20,
  54

\bibitem[\protect\citeauthoryear{{Fujita}, {Takizawa}, {Nagashima} \&
  {Enoki}}{{Fujita} et~al.}{1999}]{Fu99}
{Fujita} Y.,  {Takizawa} M.,  {Nagashima} M.,    {Enoki} M.,  1999, \pasj, 51,
  L1

\bibitem[\protect\citeauthoryear{{Garc{\'{\i}}a-Senz}, {Cabez{\'o}n} \&
  {Escart{\'{\i}}n}}{{Garc{\'{\i}}a-Senz} et~al.}{2012}]{ga12}
{Garc{\'{\i}}a-Senz} D.,  {Cabez{\'o}n} R.~M.,    {Escart{\'{\i}}n} J.~A.,
  2012, \aap, 538, A9

\bibitem[\protect\citeauthoryear{{Gastaldello}, {Buote}, {Humphrey},
  {Zappacosta}, {Bullock}, {Brighenti} \& {Mathews}}{{Gastaldello}
  et~al.}{2007}]{gast07}
{Gastaldello} F.,  {Buote} D.~A.,  {Humphrey} P.~J.,  {Zappacosta} L.,
  {Bullock} J.~S.,  {Brighenti} F.,    {Mathews} W.~G.,  2007, \apj, 669, 158

\bibitem[\protect\citeauthoryear{{Ghirardini}, {Eckert}, {Ettori},
  {Pointecouteau}, {Molendi}, {Gaspari}, {Rossetti}, {De Grandi}, {Roncarelli}
  \& {Bourdin}}{{Ghirardini} et~al.}{2019}]{Ghi19}
{Ghirardini} V.,  {Eckert} D.,  {Ettori} S.,  {Pointecouteau} E.,  {Molendi}
  S.,  {Gaspari} M.,  {Rossetti} M.,  {De Grandi} S.,  {Roncarelli} M.,
  {Bourdin} H.,  2019, \aap, 621, A41

\bibitem[\protect\citeauthoryear{{G{\'o}mez}, {Loken}, {Roettiger} \&
  {Burns}}{{G{\'o}mez} et~al.}{2002}]{GO02}
{G{\'o}mez} P.~L.,  {Loken} C.,  {Roettiger} K.,    {Burns} J.~O.,  2002, \apj,
  569, 122

\bibitem[\protect\citeauthoryear{{Groener}, {Goldberg} \& {Sereno}}{{Groener}
  et~al.}{2016}]{gr16}
{Groener} A.~M.,  {Goldberg} D.~M.,    {Sereno} M.,  2016, \mnras, 455, 892

\bibitem[\protect\citeauthoryear{{Hahn}, {Martizzi}, {Wu}, {Evrard}, {Teyssier}
  \& {Wechsler}}{{Hahn} et~al.}{2017}]{Ha17}
{Hahn} O.,  {Martizzi} D.,  {Wu} H.-Y.,  {Evrard} A.~E.,  {Teyssier} R.,
  {Wechsler} R.~H.,  2017, \mnras, 470, 166

\bibitem[\protect\citeauthoryear{{Halbesma}, {Donnert}, {de Vries} \&
  {Wise}}{{Halbesma} et~al.}{2019}]{Ha19}
{Halbesma} T.~L.~R.,  {Donnert} J.~M.~F.,  {de Vries} M.~N.,    {Wise} M.~W.,
  2019, \mnras, 483, 3851

\bibitem[\protect\citeauthoryear{{Herant}}{{Herant}}{1994}]{he94}
{Herant} M.,  1994, Mem. Soc. Astron. Ital., 65, 1013

\bibitem[\protect\citeauthoryear{{Hopkins}}{{Hopkins}}{2015}]{Ho15}
{Hopkins} P.~F.,  2015, \mnras, 450, 53

\bibitem[\protect\citeauthoryear{{Johnson}, {Ponman} \& {Finoguenov}}{{Johnson}
  et~al.}{2009}]{Joh09}
{Johnson} R.,  {Ponman} T.~J.,    {Finoguenov} A.,  2009, \mnras, 395, 1287

\bibitem[\protect\citeauthoryear{{Kazantzidis}, {Magorrian} \&
  {Moore}}{{Kazantzidis} et~al.}{2004}]{Kaz04}
{Kazantzidis} S.,  {Magorrian} J.,    {Moore} B.,  2004, \apj, 601, 37

\bibitem[\protect\citeauthoryear{{Kim}, {Peter} \& {Wittman}}{{Kim}
  et~al.}{2017}]{Kim17}
{Kim} S.~Y.,  {Peter} A. H.~G.,    {Wittman} D.,  2017, \mnras, 469, 1414

\bibitem[\protect\citeauthoryear{{Kuijken} \& {Dubinski}}{{Kuijken} \&
  {Dubinski}}{1994}]{KD94}
{Kuijken} K.,  {Dubinski} J.,  1994, \mnras, 269, 13

\bibitem[\protect\citeauthoryear{{Landau} \& {Lifshitz}}{{Landau} \&
  {Lifshitz}}{1980}]{Lan80}
{Landau} L.~D.,  {Lifshitz} E.~M.,  1980, {Statistical Physics. (3d ed.;
  Oxford:Pergamon) }

\bibitem[\protect\citeauthoryear{{Machado} \& {Lima Neto}}{{Machado} \& {Lima
  Neto}}{2013}]{Mac13}
{Machado} R. E.~G.,  {Lima Neto} G.~B.,  2013, \mnras, 430, 3249

\bibitem[\protect\citeauthoryear{{Mansheim}, {Lemaux}, {Tomczak}, {Lubin},
  {Rumbaugh}, {Wu}, {Gal}, {Shen}, {Dawson} \& {Squires}}{{Mansheim}
  et~al.}{2017}]{Man17}
{Mansheim} A.~S.,  {Lemaux} B.~C.,  {Tomczak} A.~R.,  {Lubin} L.~M.,
  {Rumbaugh} N.,  {Wu} P.~F.,  {Gal} R.~R.,  {Shen} L.,  {Dawson} W.~A.,
  {Squires} G.~K.,  2017, \mnras, 469, L20

\bibitem[\protect\citeauthoryear{{Markevitch}, {Gonzalez}, {Clowe},
  {Vikhlinin}, {Forman}, {Jones}, {Murray} \& {Tucker}}{{Markevitch}
  et~al.}{2004}]{Mar04}
{Markevitch} M.,  {Gonzalez} A.~H.,  {Clowe} D.,  {Vikhlinin} A.,  {Forman} W.,
   {Jones} C.,  {Murray} S.,    {Tucker} W.,  2004, \apj, 606, 819

\bibitem[\protect\citeauthoryear{{Markevitch} \& {Vikhlinin}}{{Markevitch} \&
  {Vikhlinin}}{2007}]{Mark07}
{Markevitch} M.,  {Vikhlinin} A.,  2007, \physrep, 443, 1

\bibitem[\protect\citeauthoryear{{Mastropietro} \& {Burkert}}{{Mastropietro} \&
  {Burkert}}{2008}]{Mas08}
{Mastropietro} C.,  {Burkert} A.,  2008, \mnras, 389, 967

\bibitem[\protect\citeauthoryear{{McCarthy}, {Bower}, {Balogh}, {Voit},
  {Pearce}, {Theuns}, {Babul}, {Lacey} \& {Frenk}}{{McCarthy}
  et~al.}{2007}]{MC07}
{McCarthy} I.~G.,  {Bower} R.~G.,  {Balogh} M.~L.,  {Voit} G.~M.,  {Pearce}
  F.~R.,  {Theuns} T.,  {Babul} A.,  {Lacey} C.~G.,    {Frenk} C.~S.,  2007,
  \mnras, 376, 497

\bibitem[\protect\citeauthoryear{{McDonald}, {Benson}, {Vikhlinin}, {Stalder},
  {Bleem}, {de Haan}, {Lin}, {Aird}, {Ashby} \& {Bautz}}{{McDonald}
  et~al.}{2013}]{McD13}
{McDonald} M.,  {Benson} B.~A.,  {Vikhlinin} A.,  {Stalder} B.,  {Bleem} L.~E.,
   {de Haan} T.,  {Lin} H.~W.,  {Aird} K.~A.,  {Ashby} M.~L.~N.,    {Bautz}
  M.~W.,  2013, \apj, 774, 23

\bibitem[\protect\citeauthoryear{{Mitchell}, {McCarthy}, {Bower}, {Theuns} \&
  {Crain}}{{Mitchell} et~al.}{2009}]{Mi09}
{Mitchell} N.~L.,  {McCarthy} I.~G.,  {Bower} R.~G.,  {Theuns} T.,    {Crain}
  R.~A.,  2009, \mnras, 395, 180

\bibitem[\protect\citeauthoryear{{Molnar}}{{Molnar}}{2016}]{Molnar16}
{Molnar} S.,  2016, Frontiers in Astronomy and Space Sciences, 2, 7

\bibitem[\protect\citeauthoryear{{Molnar} \& {Broadhurst}}{{Molnar} \&
  {Broadhurst}}{2018}]{Molnar18}
{Molnar} S.~M.,  {Broadhurst} T.,  2018, \apj, 862, 112

\bibitem[\protect\citeauthoryear{{Molnar}, {Hearn} \& {Stadel}}{{Molnar}
  et~al.}{2012}]{Molnar12}
{Molnar} S.~M.,  {Hearn} N.~C.,    {Stadel} J.~G.,  2012, \apj, 748, 45

\bibitem[\protect\citeauthoryear{Monaghan}{Monaghan}{1997}]{mo97}
Monaghan J.~J.,  1997, Journal of Computational Physics, 136, 298

\bibitem[\protect\citeauthoryear{{Monaghan} \& {Price}}{{Monaghan} \&
  {Price}}{2006}]{Mo06}
{Monaghan} J.~J.,  {Price} D.~J.,  2006, \mnras, 365, 991

\bibitem[\protect\citeauthoryear{{Nagasawa}, {Nakamura} \& {Miyama}}{{Nagasawa}
  et~al.}{1988}]{Nag88}
{Nagasawa} M.,  {Nakamura} T.,    {Miyama} S.~M.,  1988, \pasj, 40, 691

\bibitem[\protect\citeauthoryear{{Navarro}, {Frenk} \& {White}}{{Navarro}
  et~al.}{1997}]{NA97}
{Navarro} J.~F.,  {Frenk} C.~S.,    {White} S. D.~M.,  1997, \apj, 490, 493

\bibitem[\protect\citeauthoryear{{Pakmor}, {Edelmann}, {R{\"o}pke} \&
  {Hillebrand t}}{{Pakmor} et~al.}{2012}]{Pakmor12}
{Pakmor} R.,  {Edelmann} P.,  {R{\"o}pke} F.~K.,    {Hillebrand t} W.,  2012,
  \mnras, 424, 2222

\bibitem[\protect\citeauthoryear{{Planelles} \& {Quilis}}{{Planelles} \&
  {Quilis}}{2009}]{Pla09}
{Planelles} S.,  {Quilis} V.,  2009, \mnras, 399, 410

\bibitem[\protect\citeauthoryear{{Poole}, {Babul}, {McCarthy}, {Sand erson} \&
  {Fardal}}{{Poole} et~al.}{2008}]{Poole08}
{Poole} G.~B.,  {Babul} A.,  {McCarthy} I.~G.,  {Sand erson} A.~J.~R.,
  {Fardal} M.~A.,  2008, \mnras, 391, 1163

\bibitem[\protect\citeauthoryear{{Poole}, {Fardal}, {Babul}, {McCarthy},
  {Quinn} \& {Wadsley}}{{Poole} et~al.}{2006}]{Poole06}
{Poole} G.~B.,  {Fardal} M.~A.,  {Babul} A.,  {McCarthy} I.~G.,  {Quinn} T.,
  {Wadsley} J.,  2006, \mnras, 373, 881

\bibitem[\protect\citeauthoryear{{Power}, {Read} \& {Hobbs}}{{Power}
  et~al.}{2014}]{pow14}
{Power} C.,  {Read} J.~I.,    {Hobbs} A.,  2014, \mnras, 440, 3243

\bibitem[\protect\citeauthoryear{{Pratt}, {Arnaud}, {Piffaretti},
  {B{\"o}hringer}, {Ponman}, {Croston}, {Voit}, {Borgani} \& {Bower}}{{Pratt}
  et~al.}{2010}]{Pratt10}
{Pratt} G.~W.,  {Arnaud} M.,  {Piffaretti} R.,  {B{\"o}hringer} H.,  {Ponman}
  T.~J.,  {Croston} J.~H.,  {Voit} G.~M.,  {Borgani} S.,    {Bower} R.~G.,
  2010, \aap, 511, A85

\bibitem[\protect\citeauthoryear{{Price}}{{Price}}{2008}]{pr08}
{Price} D.~J.,  2008, Journal of Computational Physics, 227, 10040

\bibitem[\protect\citeauthoryear{{Price}}{{Price}}{2012}]{Price2012}
{Price} D.~J.,  2012, Journal of Computational Physics, 231, 759

\bibitem[\protect\citeauthoryear{{Price} \& {Monaghan}}{{Price} \&
  {Monaghan}}{2007}]{Price07}
{Price} D.~J.,  {Monaghan} J.~J.,  2007, \mnras, 374, 1347

\bibitem[\protect\citeauthoryear{{Price} Daniel, Wurster, {Tricco} \& {et
  al.}}{{Price} et~al.}{2018}]{Price18}
{Price} Daniel J.,  Wurster J.,  {Tricco} T.~S.,    {et al.} 2018, PASA, 35,
  e031

\bibitem[\protect\citeauthoryear{{Rasia}, {Borgani}, {Murante}, {Planelles},
  {Beck}, {Biffi}, {Ragone-Figueroa}, {Granato}, {Steinborn} \&
  {Dolag}}{{Rasia} et~al.}{2015}]{Ras15}
{Rasia} E.,  {Borgani} S.,  {Murante} G.,  {Planelles} S.,  {Beck} A.~M.,
  {Biffi} V.,  {Ragone-Figueroa} C.,  {Granato} G.~L.,  {Steinborn} L.~K.,
  {Dolag} K.,  2015, \apjl, 813, L17

\bibitem[\protect\citeauthoryear{{Raskin} \& {Owen}}{{Raskin} \&
  {Owen}}{2016}]{raskin16}
{Raskin} C.,  {Owen} J.~M.,  2016, \apj, 820, 102

\bibitem[\protect\citeauthoryear{{Read} \& {Hayfield}}{{Read} \&
  {Hayfield}}{2012}]{Read12}
{Read} J.~I.,  {Hayfield} T.,  2012, \mnras, 422, 3037

\bibitem[\protect\citeauthoryear{{Reinhardt} \& {Stadel}}{{Reinhardt} \&
  {Stadel}}{2017}]{Rein17}
{Reinhardt} C.,  {Stadel} J.,  2017, \mnras, 467, 4252

\bibitem[\protect\citeauthoryear{{Ricker} \& {Sarazin}}{{Ricker} \&
  {Sarazin}}{2001}]{Ricker01}
{Ricker} P.~M.,  {Sarazin} C.~L.,  2001, \apj, 561, 621

\bibitem[\protect\citeauthoryear{{Ritchie} \& {Thomas}}{{Ritchie} \&
  {Thomas}}{2002}]{RT02}
{Ritchie} B.~W.,  {Thomas} P.~A.,  2002, \mnras, 329, 675

\bibitem[\protect\citeauthoryear{{Robertson}, {Massey} \& {Eke}}{{Robertson}
  et~al.}{2017}]{Rob17}
{Robertson} A.,  {Massey} R.,    {Eke} V.,  2017, \mnras, 465, 569

\bibitem[\protect\citeauthoryear{{Roediger}, {Bruggen}, {Owers}, {Ebeling} \&
  {Sun}}{{Roediger} et~al.}{2014}]{Roe14}
{Roediger} E.,  {Bruggen} M.,  {Owers} M.~S.,  {Ebeling} H.,    {Sun} M.,
  2014, \mnras, 443, L114

\bibitem[\protect\citeauthoryear{{Roettiger}, {Burns} \& {Loken}}{{Roettiger}
  et~al.}{1996}]{Ro96}
{Roettiger} K.,  {Burns} J.~O.,    {Loken} C.,  1996, \apj, 473, 651

\bibitem[\protect\citeauthoryear{{Rosswog}}{{Rosswog}}{2015}]{Ross15}
{Rosswog} S.,  2015, \mnras, 448, 3628

\bibitem[\protect\citeauthoryear{{Rosswog} \& {Price}}{{Rosswog} \&
  {Price}}{2007}]{Ross07}
{Rosswog} S.,  {Price} D.,  2007, \mnras, 379, 915

\bibitem[\protect\citeauthoryear{{Saitoh} \& {Makino}}{{Saitoh} \&
  {Makino}}{2016}]{Sa16}
{Saitoh} T.~R.,  {Makino} J.,  2016, \apj, 823, 144

\bibitem[\protect\citeauthoryear{{Sarazin}}{{Sarazin}}{2002}]{Sar02}
{Sarazin} C.~L.,  2002, {The Physics of Cluster Mergers}.
pp 1--38

\bibitem[\protect\citeauthoryear{{Schmidt}, {Byrohl}, {Engels}, {Behrens} \&
  {Niemeyer}}{{Schmidt} et~al.}{2017}]{Sch17}
{Schmidt} W.,  {Byrohl} C.,  {Engels} J.~F.,  {Behrens} C.,    {Niemeyer}
  J.~C.,  2017, \mnras, 470, 142

\bibitem[\protect\citeauthoryear{{Sembolini}, {Yepes}, {Pearce} \& {et
  al.}}{{Sembolini} et~al.}{2016}]{Semb16}
{Sembolini} F.,  {Yepes} G.,  {Pearce} F.~R.,    {et al.} 2016, \mnras, 457,
  4063

\bibitem[\protect\citeauthoryear{{Soker}}{{Soker}}{2016}]{Soker16}
{Soker} N.,  2016, NewAR, 75, 1

\bibitem[\protect\citeauthoryear{{Springel}}{{Springel}}{2010}]{Sp10}
{Springel} V.,  2010, \mnras, 401, 791

\bibitem[\protect\citeauthoryear{{Springel} \& {Farrar}}{{Springel} \&
  {Farrar}}{2007}]{Spr07}
{Springel} V.,  {Farrar} G.~R.,  2007, \mnras, 380, 911

\bibitem[\protect\citeauthoryear{{Stroe}, {Sobral}, {Paulino-Afonso}, {Alegre},
  {Calhau}, {Santos} \& {van Weeren}}{{Stroe} et~al.}{2017}]{St17}
{Stroe} A.,  {Sobral} D.,  {Paulino-Afonso} A.,  {Alegre} L.,  {Calhau} J.,
  {Santos} S.,    {van Weeren} R.,  2017, \mnras, 465, 2916

\bibitem[\protect\citeauthoryear{{Subramanian}, {Shukurov} \&
  {Haugen}}{{Subramanian} et~al.}{2006}]{Sub06}
{Subramanian} K.,  {Shukurov} A.,    {Haugen} N. E.~L.,  2006, \mnras, 366,
  1437

\bibitem[\protect\citeauthoryear{{Sun}, {Voit}, {Donahue}, {Jones}, {Forman} \&
  {Vikhlinin}}{{Sun} et~al.}{2009}]{Sun09}
{Sun} M.,  {Voit} G.~M.,  {Donahue} M.,  {Jones} C.,  {Forman} W.,
  {Vikhlinin} A.,  2009, \apj, 693, 1142

\bibitem[\protect\citeauthoryear{{Takizawa}}{{Takizawa}}{2005}]{Tak05}
{Takizawa} M.,  2005, \apj, 629, 791

\bibitem[\protect\citeauthoryear{{Teyssier}}{{Teyssier}}{2002}]{Tey02}
{Teyssier} R.,  2002, \aap, 385, 337

\bibitem[\protect\citeauthoryear{{Turner}, {Chapman}, {Bhattal}, {Disney},
  {Pongracic} \& {Whitworth}}{{Turner} et~al.}{1995}]{Turner95}
{Turner} J.~A.,  {Chapman} S.~J.,  {Bhattal} A.~S.,  {Disney} M.~J.,
  {Pongracic} H.,    {Whitworth} A.~P.,  1995, \mnras, 277, 705

\bibitem[\protect\citeauthoryear{{Valdarnini}}{{Valdarnini}}{2006}]{VA06}
{Valdarnini} R.,  2006, \na, 12, 71

\bibitem[\protect\citeauthoryear{{Valdarnini}}{{Valdarnini}}{2012}]{VA12}
{Valdarnini} R.,  2012, \aap, 546, A45

\bibitem[\protect\citeauthoryear{{Valdarnini}}{{Valdarnini}}{2016}]{V16}
{Valdarnini} R.,  2016, \apj, 831, 103 (V16)

\bibitem[\protect\citeauthoryear{{Valdarnini}}{{Valdarnini}}{2019}]{V19}
{Valdarnini} R.,  2019, \apj, 874, 42

\bibitem[\protect\citeauthoryear{{Vazza}, {Roediger} \& {Br{\"u}ggen}}{{Vazza}
  et~al.}{2012}]{Vaz12}
{Vazza} F.,  {Roediger} E.,    {Br{\"u}ggen} M.,  2012, \aap, 544, A103

\bibitem[\protect\citeauthoryear{Vela, Sanchez \& Geiger}{Vela
  et~al.}{2018}]{vela18}
Vela L.~V.,  Sanchez R.,    Geiger J.,  2018, CPC, 224, 186

\bibitem[\protect\citeauthoryear{Vietri, Ferrara \& Miniati}{Vietri
  et~al.}{1997}]{Vietri97}
Vietri M.,  Ferrara A.,    Miniati F.,  1997, The Astrophysical Journal, 483,
  262

\bibitem[\protect\citeauthoryear{{Vitvitska}, {Klypin}, {Kravtsov}, {Wechsler},
  {Primack} \& {Bullock}}{{Vitvitska} et~al.}{2002}]{Viv02}
{Vitvitska} M.,  {Klypin} A.~A.,  {Kravtsov} A.~V.,  {Wechsler} R.~H.,
  {Primack} J.~R.,    {Bullock} J.~S.,  2002, \apj, 581, 799

\bibitem[\protect\citeauthoryear{{Voit}}{{Voit}}{2005}]{Voit05}
{Voit} G.~M.,  2005, Reviews of Modern Physics, 77, 207

\bibitem[\protect\citeauthoryear{{Wadsley}, {Veeravalli} \&
  {Couchman}}{{Wadsley} et~al.}{2008}]{wa08}
{Wadsley} J.~W.,  {Veeravalli} G.,    {Couchman} H.~M.~P.,  2008, \mnras, 387,
  427

\bibitem[\protect\citeauthoryear{{Wang} \& {White}}{{Wang} \&
  {White}}{2007}]{Wa07}
{Wang} J.,  {White} S. D.~M.,  2007, \mnras, 380, 93

\bibitem[\protect\citeauthoryear{{Zemp}, {Moore}, {Stadel}, {Carollo} \&
  {Madau}}{{Zemp} et~al.}{2008}]{zemp08}
{Zemp} M.,  {Moore} B.,  {Stadel} J.,  {Carollo} C.~M.,    {Madau} P.,  2008,
  \mnras, 386, 1543

\bibitem[\protect\citeauthoryear{{Zhang}, {Yu} \& {Lu}}{{Zhang}
  et~al.}{2014}]{Zh14}
{Zhang} C.,  {Yu} Q.,    {Lu} Y.,  2014, \apj, 796, 138

\bibitem[\protect\citeauthoryear{{Zhang}, {Yu} \& {Lu}}{{Zhang}
  et~al.}{2015}]{Zh15}
{Zhang} C.,  {Yu} Q.,    {Lu} Y.,  2015, \apj, 813, 129

\bibitem[\protect\citeauthoryear{{Zhang}, {Yu} \& {Lu}}{{Zhang}
  et~al.}{2018}]{Zh18}
{Zhang} C.,  {Yu} Q.,    {Lu} Y.,  2018, \apj, 855, 36

\bibitem[\protect\citeauthoryear{{ZuHone} J.}{{ZuHone}}{2011}]{ZuH11}
{ZuHone} J. A.,  2011, \apj, 728, 54 (Z11)

\bibitem[\protect\citeauthoryear{{ZuHone}, {Markevitch} \& {Johnson}}{{ZuHone}
  et~al.}{2010}]{ZuH10}
{ZuHone} J.~A.,  {Markevitch} M.,    {Johnson} R.~E.,  2010, \apj, 717, 908

\bibitem[\protect\citeauthoryear{{ZuHone}, {Ricker}, {Lamb} \& {Karen
  Yang}}{{ZuHone} et~al.}{2009}]{ZuH09}
{ZuHone} J.~A.,  {Ricker} P.~M.,  {Lamb} D.~Q.,    {Karen Yang} H.~Y.,  2009,
  \apj, 699, 1004

\bibitem[\protect\citeauthoryear{{ZuHone}, {Zavala} \& {Vogelsberger}}{{ZuHone}
  et~al.}{2019}]{ZuH19}
{ZuHone} J.~A.,  {Zavala} J.,    {Vogelsberger} M.,  2019, \apj, 882, 119

\bibitem[\protect\citeauthoryear{{Zurek} \& {Benz}}{{Zurek} \&
  {Benz}}{1986}]{Zu86}
{Zurek} W.~H.,  {Benz} W.,  1986, \apj, 308, 123

\end{thebibliography}
%


\bsp
\label{lastpage}
\end{document}